\documentclass[a4paper,fleqn,usenatbib]{mnras}

\usepackage{newtxtext,newtxmath}
\usepackage[T1]{fontenc}
\usepackage{ae,aecompl}
\usepackage{url}

\usepackage{graphicx}   % Including figure files
\usepackage{amsmath}    % Advanced maths commands
\usepackage{longtable}
\usepackage[hyphenbreaks]{breakurl} % line breaks in long URLs

\newcommand{\orcid}[1]{$^{\rm \href{https://orcid.org/#1}{\includegraphics[height=0.6em]{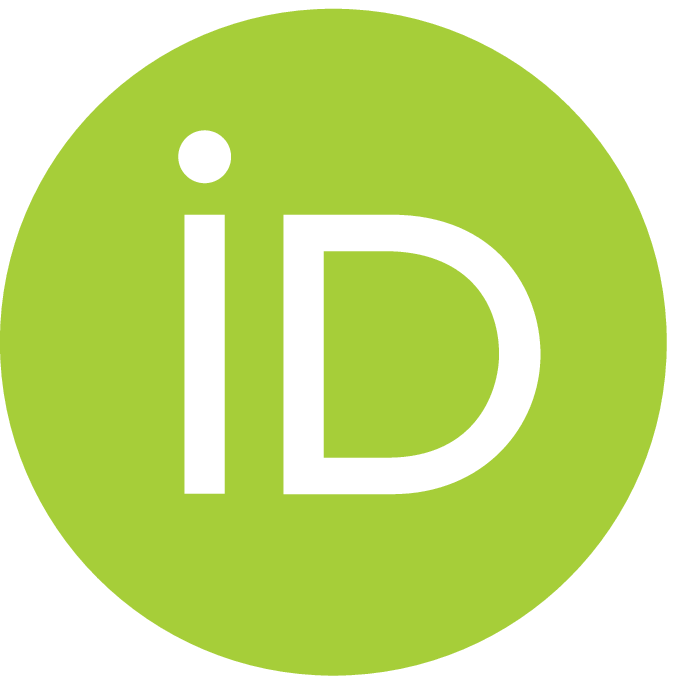}}}$}

\newcommand{\nustarenergylow}{3.0}
\newcommand{\nustarenergyhigh}{30}

\newcommand{\nustarobsid}{90701321002}
\newcommand{\nustarobsstartutc}{2021-06-23 11:24}
\newcommand{\nustarobsstoputc}{2021-06-24 10:24}
\newcommand{\nustarexpks}{39}

\newcommand{\nustarcaldbver}{20220706}
\newcommand{\nustarheasoftver}{V6.30.1}

\newcommand{\nustargroupmincounts}{25}

\newcommand{\nustarlcchiprob}{0.0003}
\newcommand{\nustarorbitalperiodsec}{5804}
\newcommand{\nustarmeancountrateperfpm}{0.065}

\newcommand{\nova}{V1674\,Her}

\newcommand{\fermilat}{\emph{Fermi}-LAT}

\newcommand{\opticalpeaklum}{$2.3 \times 10^{39}$\,erg\,s$^{-1}$}
\newcommand{\opticalnustareplum}{$9.6 \times 10^{37}$\,erg\,s$^{-1}$}
\newcommand{\latpeaklum}{$3.2 \times 10^{36}$\,erg\,s$^{-1}$}
\newcommand{\latnustareplum}{$< 1 \times 10^{36}$\,erg\,s$^{-1}$}
\newcommand{\nustarlum}{$1.0 \times 10^{34}$\,erg\,s$^{-1}$}
\newcommand{\nustarlumextrapolated}{$1.4 \times 10^{34}$\,erg\,s$^{-1}$}

\newcommand{\tnustarep}{$t_0+11.3$\,d}

\def\jaavso{JAAVSO}

\title[Multi-wavelength view of shocks in \nova{}]{The multi-wavelength view of shocks in the fastest nova \nova{}}

\author[K.~V.~Sokolovsky et al.]{
\parbox{\textwidth}{
K.~V.~Sokolovsky\orcid{0000-0001-5991-6863},$^{1,2}$ %\thanks{E-mail: kirx@kirx.net (KVS)}
T.~J.~Johnson\orcid{0000-0002-2771-472X},$^{3}$
S.~Buson\orcid{0000-0002-3308-324X},$^{4}$
P.~Jean\orcid{0000-0002-1757-9560},$^{5,6}$
C.~C.~Cheung\orcid{0000-0002-4377-0174},$^{7}$ %\\ % "Cheung, Chi (Teddy) CIV USN NRL (7655) Washington DC (USA)" <teddy.cheung@nrl.navy.mil>
K.~Mukai\orcid{0000-0002-8286-8094},$^{8}$
L.~Chomiuk\orcid{0000-0002-8400-3705},$^{2}$
E.~Aydi\orcid{0000-0001-8525-3442},$^{2}$
B.~Molina,$^{2}$
A.~Kawash\orcid{0000-0003-0071-1622},$^{2}$
J.~D.~Linford\orcid{0000-0002-3873-5497},$^{9}$
A.~J.~Mioduszewski,$^{9}$
M.~P.~Rupen,$^{10}$
J.~L.~Sokoloski,$^{11}$
M.~N.~Williams,$^{9}$
E.~Steinberg\orcid{0000-0003-0053-0696},$^{12}$
I.~Vurm\orcid{0000-0003-1336-4746},$^{13}$
B.~D.~Metzger\orcid{0000-0002-4670-7509},$^{11,14}$
K.~L.~Page\orcid{0000-0001-5624-2613},$^{15}$
M.~Orio\orcid{0000-0003-1563-9803},$^{16,17}$
R.~M.~Quimby\orcid{0000-0001-9171-5236},$^{18,19}$
A.~W.~Shafter\orcid{0000-0002-1276-1486},$^{18}$
H.~Corbett\orcid{0000-0002-6339-6706},$^{20}$
S.~Bolzoni\orcid{0000-0003-0114-1318},$^{21,22}$
J.~DeYoung,$^{21}$
K.~Menzies,$^{21}$
F.~D.~Romanov\orcid{0000-0002-5268-7735},$^{21,22}$
M.~Richmond,$^{23}$
J.~Ulowetz,$^{24}$
T.~Vanmunster,$^{21,25,26}$
G.~Williamson,$^{21}$
D.~J.~Lane\orcid{0000-0002-6097-8719},$^{27,21}$
M.~Bartnik,$^{2}$
M.~Bellaver,$^{2}$
E.~Bruinsma,$^{2}$
E.~Dugan,$^{2}$
J.~Fedewa,$^{2}$
C.~Gerhard,$^{2}$
S.~Painter,$^{2}$
D.-M.~Peterson,$^{2}$
J.~E.~Rodriguez\orcid{0000-0001-8812-0565},$^{2}$
C.~Smith,$^{2}$
H.~Sullivan,$^{2}$
S.~Watson$^{2}$}
\vspace{0.4cm}\\
\parbox{\textwidth}{
$^{1}$Department of Astronomy, University of Illinois at Urbana-Champaign, 1002 W. Green Street, Urbana, IL 61801, USA\\
$^{2}$Department of Physics and Astronomy, Michigan State University, 567 Wilson Rd, East Lansing, MI 48824, USA\\
$^{3}$College of Science, George Mason University, Fairfax, VA 22030, resident at Naval Research Laboratory, Washington, DC 20375, USA\\
$^{4}$Lehrstuhl f\"ur Astronomie, Universit\"at W\"urzburg, Emil-Fischer-St. 31, W\"urzburg, 97074, Germany\\
$^{5}$CNRS, IRAP, F-31028 Toulouse cedex 4, France\\
$^{6}$GAHEC, Universit\'e de Toulouse, UPS-OMP, IRAP, Toulouse, France\\
$^{7}$Space Science Division, Naval Research Laboratory, Washington, DC 20375-5352, USA\\
$^{8}$CRESST and X-ray Astrophysics Laboratory, NASA/GSFC, Greenbelt, MD 20771, USA\\
$^{9}$National Radio Astronomy Observatory, Domenici Science Operations Center, 1003 Lopezville Road, Socorro, NM 87801, USA\\
$^{10}$National Research Council, Herzberg Astronomy and Astrophysics, 717 White Lake Rd, PO Box 248, Penticton, BC V2A 6J9, Canada \\
$^{11}$Department of Physics and Columbia Astrophysics Laboratory, Columbia University, New York, NY 10027, USA\\
$^{12}$Racah Institute of Physics, The Hebrew University, 9190401 Jerusalem, Israel\\
$^{13}$Tartu Observatory, University of Tartu, T\~oravere, 61602 Tartumaa, Estonia\\
$^{14}$Center for Computational Astrophysics, Flatiron Institute, 162 5th Ave, New York, NY 10010, USA\\
$^{15}$School of Physics \&  Astronomy, University of Leicester, LE1 7RH, UK\\
$^{16}$Department of Astronomy, University of Wisconsin-Madison, 475 N. Charter Street, Madison, WI 53706, USA\\
$^{17}$INAF-Padova, vicolo Osservatorio, 5, I-35122 Padova, Italy\\
$^{18}$Department of Astronomy and Mount Laguna Observatory, San Diego State University, San Diego, CA 92182, USA\\
$^{19}$Kavli Institute for the Physics and Mathematics of the Universe (WPI), The University of Tokyo Institutes for Advanced Study, The University of Tokyo, Kashiwa, Chiba 277-8583, Japan\\
$^{20}$Department of Physics and Astronomy, University of North Carolina at Chapel Hill, Chapel Hill, NC 27599-3255, USA\\
$^{21}$AAVSO Observer\\
$^{22}$Remote observer of Burke-Gaffney Observatory, Abbey Ridge Observatory and Mini-Robotic Observatory, Canada\\
$^{23}$School of Physics and Astronomy, Rochester Institute of Technology, 84 Lomb Memorial Drive, Rochester, NY 14623, USA\\
$^{24}$Center for Backyard Astrophysics Illinois, Northbrook Meadow Observatory, 855 Fair Ln, Northbrook, IL 60062, USA\\
$^{25}$Center for Backyard Astrophysics Belgium, Walhostraat 1a, B-3401 Landen, Belgium\\
$^{26}$Center for Backyard Astrophysics Extremadura, e-EyE Astronomical Complex, ES-06340 Fregenal de la Sierra, Spain\\
$^{27}$Burke-Gaffney Observatory, Saint Mary's University, 923 Robie Street, Halifax, NS B3H 3C3, Canada
} 
}

\date{Accepted 2023 March 14. Received 2023 February 03; in original form 2023 February 03}

\pubyear{2023}

\begin{document}
\label{firstpage}
\pagerange{\pageref{firstpage}--\pageref{lastpage}}
\maketitle

\begin{abstract}
%
% MNRAS Abstract word limit: 250 words for Main Journal papers or 200 words for Letters. 
%
% I'm thinking of A&A-style structured abstract while writing a
% non-structured abstract for MNRAS.
%
%Context.
Classical novae are shock-powered multi-wavelength transients triggered by
a thermonuclear runaway on an accreting white dwarf. 
%Aims.
\nova{} 
is the fastest nova ever recorded (time to declined by two magnitudes is $t_2=1.1$\,d)
that challenges our understanding of shock formation in novae.
%Methods.
We investigate the physical mechanisms behind nova emission from GeV $\gamma$-rays to cm-band radio 
using coordinated \fermilat{}, {\em NuSTAR}, {\em Swift} and VLA observations supported by optical photometry.
%Results.
\fermilat{} detected short-lived (18\,h) 
0.1-100\,GeV emission from \nova{} that appeared 6\,h after the eruption began; 
this was at a level of $(1.6 \pm 0.4) \times 10^{-6}$\,photons\,cm$^{-2}$\,s$^{-1}$.
Eleven days later, simultaneous {\em NuSTAR} and {\em Swift} X-ray observations 
revealed optically thin thermal plasma shock-heated to ${\rm k} T_{\rm shock} = 4$\,keV.
The lack of a detectable 6.7\,keV Fe~K$\alpha$ emission suggests super-solar CNO abundances. % in the ejecta. 
The radio emission from \nova{} was consistent with thermal emission at early times and synchrotron at late times.
The radio spectrum 
steeply rising with frequency may be a result 
of either free-free absorption of synchrotron and thermal emission by unshocked outer
regions of the nova shell or the Razin-Tsytovich effect attenuating synchrotron emission in dense plasma.
%Conclusions.
The development of the shock inside the ejecta is unaffected
by the extraordinarily rapid evolution and the intermediate polar host of this nova. 
\end{abstract}

\begin{keywords}
% based on the list https://academic.oup.com/DocumentLibrary/mnras/keywords.pdf
transients: novae -- stars: novae, cataclysmic variables -- stars: white dwarfs -- stars: individual: \nova{}
\end{keywords}

%%%%%%%%%%%%%%%%%%%%%%%%%%%%%%%%%%%%%%%%%%%%%%%%%%

%%%%%%%%%%%%%%%%% BODY OF PAPER %%%%%%%%%%%%%%%%%%

%\linenumbers

\section{Introduction}
\label{sec:intro}

%\subsection{Novae as shock-powered multi-wavelength transients}
%\label{sec:innovashock}

% Nova overview
Novae are multiwavelength transients powered by a sudden ignition of
thermonuclear fusion at the bottom of a hydrogen-rich shell accreted by 
a white dwarf from its binary companion \citep[e.g.][]{2008clno.book.....B,2016PASP..128e1001S,2020ApJ...895...70S,2020A&ARv..28....3D}. % last one is on observations
The ignition leads to a dramatic expansion and ejection of the white dwarf atmosphere at 
typical velocities of $\sim$500-5000\,km/s -- a hallmark feature of 
the nova phenomenon recognized since the earliest days of spectroscopy \citep{1895Obs....18..436P,1956VA......2.1477M,2020ApJ...905...62A}. 
The expanded atmosphere leads to a dramatic, albeit temporary, increase in the optical brightness of the host 
binary system by $\sim 8$--15\,mag \citep{1990ApJ...356..609V,2008clno.book.....W,2021ApJ...910..120K}, 
reaching absolute magnitudes of $-4$ to $-10$\,mag \citep{2017ApJ...834..196S,2009ApJ...690.1148S,2022MNRAS.517.6150S}.
While the optical continuum light of a nova fades on a time-scale of days to 
months, the warm ejected envelope remains the source of optical line and radio continuum 
emission for months and years after the eruption \citep{2010AJ....140...34S,2021ApJS..257...49C}.
About 30 such events occur in the Galaxy each year, 
with only 10 events per year typically observed while others remain hidden 
by dust extinction \citep{2017ApJ...834..196S,2021ApJ...912...19D,2022arXiv220614132K,2022arXiv220705689R}.

Novae are prominent sources of X-rays. As the eruption progresses, a nova goes 
through the following phases of X-ray lightcurve development \citep{2017PASP..129f2001M,2010AN....331..169H}:
\begin{enumerate}
\item `fireball' phase - a bright soft ($<0.1$\,keV) thermal X-ray flash seen hours before the
optical rise \citep{2022Natur.605..248K,2022ApJ...935L..15K};
\item shock-dominated phase - hard ($\sim1$--$10$\,keV) thermal X-ray emission of plasma 
heated by shocks within the nova ejecta
\citep{1994MNRAS.271..155O,2014MNRAS.442..713M,2014ASPC..490..327M,2020ApJ...895...80O,2021ApJ...910..134G};
\item super-soft ($<0.5$\,keV, SSS) thermal X-rays that
appear when the ejecta clears, revealing the white dwarf heated by 
the ongoing thermonuclear reactions
\citep{2011ApJS..197...31S,2013A&A...559A..50N,2018ApJ...862..164O}; 
\item accretion-powered hard (typically $>1$\,keV) X-rays produced by
shocked plasma at the interface between the stream of accreting material and the white dwarf surface --
similar to non-nova accreting white dwarf binaries
\citep{2020AdSpR..66.1097B,2020AdSpR..66.1209D,2020MNRAS.499.3006S}.
\end{enumerate}

The discovery that shocks are playing an essential role in energy transport within the nova shell has led to 
a renewed interest in novae \citep[for a recent review see][]{2021ARA&A..59..391C}.
The role of shocks was revealed by the initial detection of continuum GeV $\gamma$-ray emission from novae with
\fermilat{} \citep{2010Sci...329..817A,2014Sci...345..554A}, 
followed by observations of shocks contributing to nova optical light  
\citep{2017NatAs...1..697L,2020NatAs...4..776A} 
and recent very high-energy (TeV) detections of the recurrent nova RS\,Oph
\citep{2022NatAs...6..689A,2022Sci...376...77A,2022ApJ...935...44C}.

Novae can serve as laboratories for studying astrophysical shocks, 
which may power the emission of diverse 
transients \citep{2020ApJ...904....4F} including Type~IIn and 
super-luminous supernovae \citep[e.g.][]{2014ApJ...788..154O,2018SSRv..214...27C}, 
tidal disruption events \citep{2015ApJ...806..164P}, 
stellar mergers or `Luminous Red Novae' \citep{2017MNRAS.471.3200M} 
and neutron star mergers \citep{2018ApJ...858...53L}. 
Understanding particle acceleration efficiency at shocks \citep{2014ApJ...783...91C,2018MNRAS.479..687S} 
and the prospects of detecting neutrinos from a nearby ($\sim 1$\,kpc) nova eruption 
are also of interest \citep{2010PhRvD..82l3012R,2016MNRAS.457.1786M,2020ApJ...904....4F,2022arXiv220904873G,2022arXiv221206810A}. 
Finally, as the nova envelope swells to encompass the binary star, it may remain marginally bound to the system 
\citep[e.g.][]{2016MNRAS.461.2527P}. Thus, 
each nova eruption serves as a test of common envelope evolution
\citep{1991ApJ...374..623S,2021ApJ...914....5S}
-- a poorly understood evolutionary stage passed by all interacting 
binaries \citep{1976IAUS...73...75P,1988ApJ...329..764L,2010MmSAI..81..849R,2013A&ARv..21...59I}. 
The angular momentum loss during nova eruption may be the key to understanding 
white dwarf binaries evolution \citep{1998MNRAS.297..633S,2016MNRAS.455L..16S,2021ApJ...923..100M,2022MNRAS.510.6110P}. 

% Here is a quote from 1986ApJ...310..222P regarding the heat-driven ejection
% "After the shock wave has deposited a large amount of
% (kinetic) energy in the outer part of the envelope, but not
% nearly enough as to cause mass ejection, another energy
% source becomes dominant : nuclear energy, released by the
% decay of beta-unstable nuclei produced by the hot CNO cycle in
% the hydrogen burning shell and carried outward by convection."
%
Multiple physical mechanisms, including 
hydrodynamic pressure supported by heat from nuclear reactions 
\citep{1969ApJ...156..569S,1978ApJ...226..186S,1986ApJ...310..222P}, 
radiation pressure \citep{1976MNRAS.175..305B,1978ApJ...220.1063S,1994ApJ...437..802K,2001MNRAS.326..126S}, 
and interactions with a binary companion
\citep{1985ApJ...294..263M,1990LNP...369..342L,1990ApJ...356..250L}, 
have long been recognized as potential causes of envelope ejection in novae.
The `slow torus -- fast bipolar wind' scenario of nova eruption outlined 
by \cite{1990LNP...369..342L,2014Natur.514..339C,2021ARA&A..59..391C,2019PhT....72k..38M}, and \cite{2022ApJ...938...31S} can be summarized as follows.
Thermonuclear reactions heat the white dwarf atmosphere that expands engulfing the binary. 
Little (if any) material is ejected 
as the result of the sudden explosive onset of the nuclear burning.
The weight of the expanded atmosphere would prevent the formation of fast radiation-driven wind from the white dwarf 
until most of the atmosphere is ejected via the common envelope interaction.
The velocity of the wind is expected to be close to the escape 
velocity at the distance from the white dwarf centre where the wind forms.
Without a close companion that would disrupt the expanded atmosphere, 
the wind would launch farther away from the white dwarf's centre and at a slower speed \citep{2022ApJ...938...31S}.
The ejected common envelope produces the slow equatorial
flow - the presumed target for the fast white dwarf wind to shock. 
We put this scenario to the test with the observations of \nova{}.

%
%\subsection{Scope of this work}
%\label{sec:thispaper}

We examine observations of \nova{} in the GeV $\gamma$-ray (0.1--300\,GeV from \fermilat{}; \S~\ref{sec:latobs}),
hard (3--78\,keV from {\em NuSTAR}; \S~\ref{sec:nustarobs}), and soft X-ray 
(0.3-10\,keV from {\em Swift}/XRT), and ultraviolet ({\em Swift}/UVOT; \S~\ref{sec:swiftobs}), 
as well as radio (Karl G. Jansky Very Large Array -- VLA; \S~\ref{sec:vla}) bands, putting them in the context of its optical lightcurve. 
In \S~\ref{sec:discussion} we discuss how the observed high-energy and radio behaviour results 
from shock waves mediating energy transport within the expanding nova shell 
and compare \nova{} to other novae, specifically the ones previously observed by {\em NuSTAR}. 
We make concluding remarks in \S~\ref{sec:conclusions}. 

% Yes, we report 95% cl for Fermi
Throughout this paper we report uncertainties at the $1 \sigma$ level, unless stated otherwise.
For hypothesis testing we adopt a significance level $\alpha_{\rm lim} = 0.05$ 
(the probability of rejecting the null hypothesis when it is true),
which is equivalent to the confidence level $(1-\alpha_{\rm lim}) = 0.95$ (or $2 \sigma$).
% The following is a clumsy attempt to explain when large p value is good and when it's bad
Note, that when reporting $p$-values in relation to the variability and 
periodicity detection in \S~\ref{sec:nustarvar}, \ref{sec:swiftobs}, \ref{sec:latobs} 
the null hypothesis is the absence of the effect ($p>\alpha_{\rm lim}$ means non-detection), 
while in the X-ray spectral fitting discussion (\S~\ref{sec:nustarspec} and \ref{sec:swiftobs}) 
we follow the \textsc{XSPEC} \citep{1996ASPC..101...17A} convention of the null hypothesis
being that `the adopted spectral model is true' ($p>\alpha_{\rm lim}$ means we have a good model).
For power law spectra, we use the positively-defined spectral index $\alpha$: $F_\nu \propto \nu^\alpha$ 
where $F_\nu$ is the flux density and $\nu$ is the frequency; 
the corresponding index in the distribution of the number of photons 
as a function of energy  is ${\rm d}N(E)/dE \propto E^{-\Gamma}$, 
where $\Gamma$ is the photon index and $\Gamma = 1 - \alpha$.
The same power law expressed in spectral energy distribution units 
(SED; \citealt{1997NCimB.112...11G}) is $\nu F_\nu \propto \nu^{\alpha + 1} \propto \nu^{-\Gamma + 2}$.

\section{\nova{} -- Nova Herculis 2021}
\label{sec:thisnova}

The eruption of \nova{} (also known as Nova Herculis 2021, TCP\,J18573095$+$1653396, ZTF19aasfsjq) was discovered
on 2021-06-12.5484\,UTC by Seiji Ueda and reported via 
the Central Bureau for Astronomical Telegrams' Transient Objects Confirmation Page\footnote{\url{http://www.cbat.eps.harvard.edu/unconf/followups/J18573095+1653396.html}} 
\citep{2021CBET.4976....1U,2021CBET.4977....1K}. 
The transient was spectroscopically confirmed as a classical nova by 
\cite{2021ATel14704....1M}, \cite{2021ATel14710....1A} and \cite{2021CBET.4976....1U}. 
The All-Sky Automated Survey for Supernovae \citep[ASAS-SN][]{2014ApJ...788...48S,2017PASP..129j4502K} detected 
\nova{} on 2021-06-12.1903 (8.4~hours before discovery) at $g = 16.62$. The final pre-eruption ASAS-SN observation of the field without 
a detection was on 2021-06-10.9660, which places the start of the eruption between these two dates. 
Throughout this paper we adopt the date of the first ASAS-SN detection (the first available observation 
of \nova{} above the quiescence level) as the eruption start time $t_0 = {\rm JD(UTC)}2459377.6903$.

\begin{figure*}
%                                                        left  bottom right  top
        \includegraphics[width=0.48\linewidth,clip=true,trim=0.15cm 0.5cm 0.8cm 0.2cm,angle=0]{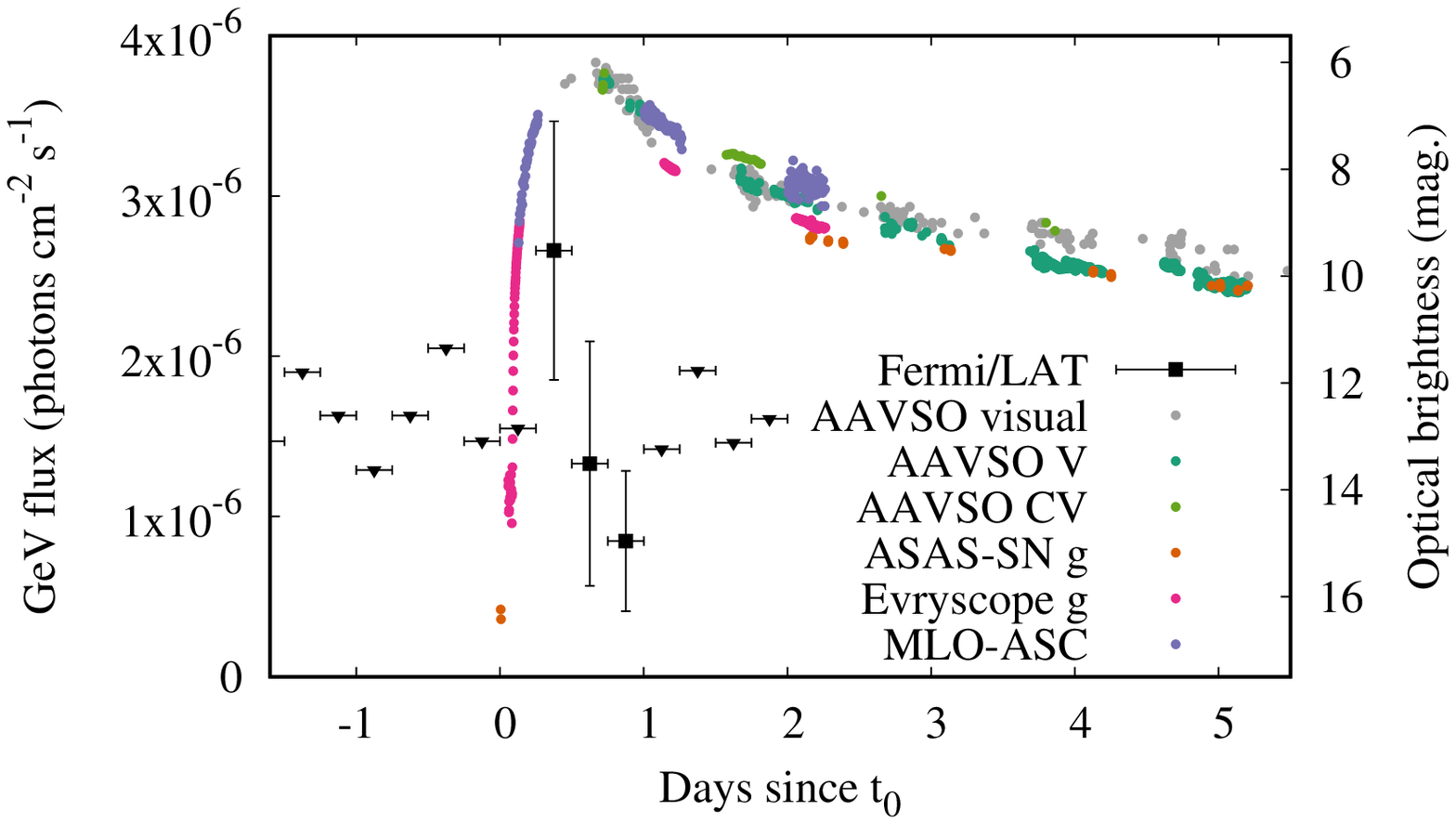}~~~~
        \includegraphics[width=0.48\linewidth,clip=true,trim=0.15cm 0.5cm 0.8cm 0.2cm,angle=0]{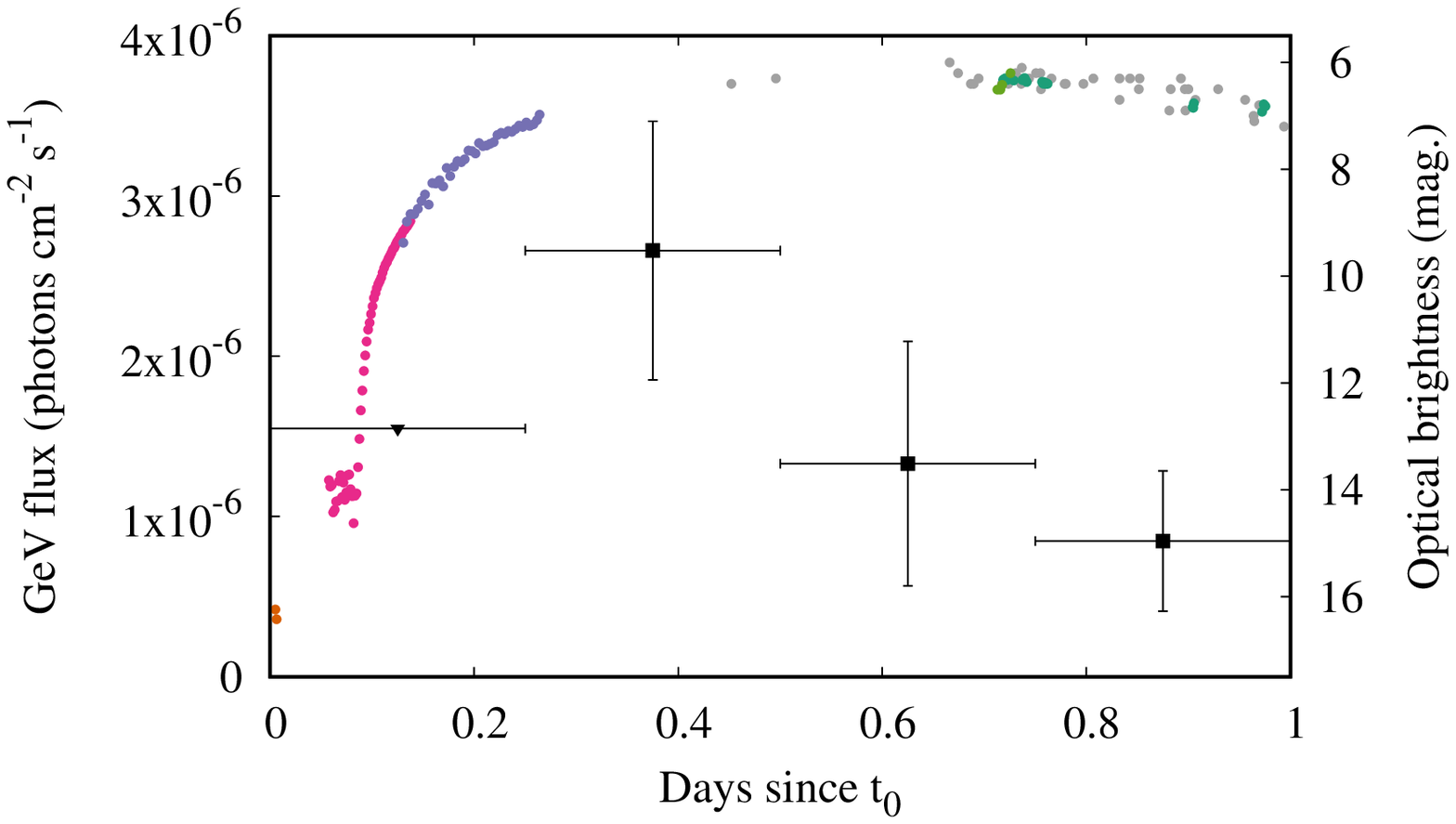}
\caption{The optical and $\gamma$-ray lightcurves of \nova{}.
The \fermilat{} detections %in the 6\,h bins 
are shown as black squares while the 95\,per~cent upper limits are marked with black triangles.
The left panel shows the full duration of the \fermilat{} 6\,h binned lightcurve.
The right panel zooms into the first day of the eruption.}
    \label{fig:latlc}
\end{figure*}

% So many details about the MLO-ASC camera here because you told me I can't have a separate "optical photometry" section.
\cite{2021RNAAS...5..160Q} report photometry of \nova{} on the rise to maximum light using 
Evryscope \citep[$g$ band;][]{2014SPIE.9145E..0ZL} and the Mount Laguna Observatory All-Sky Camera 
(MLO-ASC; an unfiltered monochrome camera based on a blue-sensitive Panasonic MN34230 CMOS chip) that
is normally used for cloud cover monitoring. \cite{2021RNAAS...5..160Q} used a custom code based on \textsc{astropy} and \textsc{photutils} 
to perform photometry on the MLO-ASC images that 
was calibrated using {\em Gaia} $G$ magnitudes of nearby field stars. 
The MLO-ASC data cover the near-peak time when the nova was saturated for Evryscope. 
We reproduce these observations in Fig.~\ref{fig:latlc}, 
combining them with ASAS-SN \citep[$g$ band;][]{2014ApJ...788...48S,2017PASP..129j4502K}
data, as well as $V$ band and $CV$ (unfiltered observations with $V$ magnitude zero-point)
photometry and visual brightness estimates collected by the AAVSO observers \citep{AAVSODATA}. 

The lightcurve of \nova{} is presented at Fig.~\ref{fig:latlc}. 
The nova experienced a pre-maximum halt at $g \sim 14$ lasting for at least
three hours \citep{2021RNAAS...5..160Q}. 
The halt was followed by a steep rise to the peak around the visual magnitude of 6 on 2021-06-12.856 according to the AAVSO photometry \citep{AAVSODATA}
and the measurements reported by \cite{2021CBET.4976....1U} and \cite{2021CBET.4977....1K}.
\nova{} rapidly declined from the peak, fading by two magnitudes ($t_2$) 
in 1.1 \citep{2021RNAAS...5..160Q} 
or 1.2 days \citep{2021PZ.....41....4S} making it one of the fastest novae ever observed 
\citep{2021ApJ...922L..10W,2022RNAAS...6..124W}. 
The colour of novae near peak brightness changes rapidly 
\citep[due to changing photospheric temperature and development of emission lines;][]{1987A&AS...70..125V}, 
resulting in slight differences in decline rates between the bands.
The uncertainty in the maximum light epoch and magnitude may have also
contributed to the difference between the reported $t_2$ estimates for \nova{}.
Regardless of the exact value of $t_2$, \nova{} is clearly among the fastest
novae observed, leading its nearest competitors 
U\,Sco \citep[$t_2=1.2$\,d;][]{2010ApJS..187..275S}, 
V838\,Her (\citealt{2010AJ....140...34S} report $t_2=1$\,d 
whereas \citealt{1996MNRAS.282..563V} quote $t_2\sim2$\,d),
M31N\,2008-12a ($t_2=1.6$\,d; \citealt{2016ApJ...833..149D}),
V1500\,Cyg ($t_2=2$\,d),
V4160\,Sgr ($t_2=2$\,d), 
V4739\,Sgr \citep[$t_2=2$\,d;][]{2010AJ....140...34S}, and
V392\,Per \citep[$t_2=2$\,d;][]{2022MNRAS.514.6183M}; see also Table~5 of
\cite{2016ApJ...833..149D}.

Spectroscopic observations revealed shell expansion velocities that are some of the fastest observed in novae. 
\citet{2021ATel14704....1M} report P~Cygni profiles of Balmer and 
\ion{Fe}{II} with absorption troughs blueshifted by 3000\,km\,s$^{-1}$ less than a day after the discovery. 
\citet{2021ATel14710....1A} noted dramatic changes in the line profiles over the course of a day --- in addition to 
the initial 3000\,km\,s$^{-1}$ absorption components, faster components (P~Cygni absorptions with troughs at blueshifted 
velocities $>5000$\,km\,s$^{-1}$) appeared in less than a day. 
\citet{2021ATel14710....1A} interpreted these two velocity components in the context of multiple outflows described in \citet{2020ApJ...905...62A}. 
NIR spectroscopic observations were reported by \cite{2021ApJ...922L..10W} showing the emergence of coronal lines as early
as $t_0 + 11$\,d, the earliest onset yet observed for any classical nova. 
Based on late time optical spectroscopic follow up taken more than 300 days after eruption, 
\cite{2022RNAAS...6..124W} suggested that the eruption is over. They also report
P~Cygni-like profile of H$\alpha$, suggesting the presence of a wind emanating from the binary system.

GeV $\gamma$-ray emission from \nova{} was detected by \fermilat{} as reported
by \cite{2021ATel14705....1L,2021ATel14707....1L,2022MNRAS.517L..97L}, see Sec.~\ref{sec:latobs} for our independent analysis.
Along with the dedicated X-ray observations \citep[][and Sec.~\ref{sec:nustarobs}]{2021ATel14747....1P,2021ApJ...922L..42D},
\nova{} was detected in the course of the SRG/eROSITA survey \citep{2021AstL...47..587G}. 
Radio emission from \nova{} was detected using the VLA 
%Karl G. Jansky Very Large Array (VLA) 
\citep[][\S~\ref{sec:vla}, \S~\ref{sec:radiodiscussion}]{2021ATel14731....1S}.
The very long baseline interferometry (VLBI) observation with e-EVN on $t_0+10$\,d resulted in 
an upper limit \citep{2021ATel14758....1P}.

A remarkable feature of \nova{} is the emergence of orbital 
\citep[3.67\,h $=$ 0.153\,d;][]{2021PZ.....41....4S,2021JAVSO..49..257S} 
and white dwarf spin periods \citep[8.36\,min $=$ 0.00580\,d;][]{2021ATel14856....1P} 
shortly after the eruption. The two periods are seen in X-rays in addition to optical data 
\citep{2021ATel14776....1M,2021ATel14798....1P,2022ATel15317....1P,2022ApJ...932...45O,2022MNRAS.517L..97L}. 
The spin period was present before the eruption according to the Zwicky Transient Facility
photometry reported by \cite{2021ATel14720....1M}. 
The spin period change may be caused by some combination of 
magnetic coupling between the rotating white dwarf and the ejecta, 
non-rigid rotation or substantial radial expansion of the heated white dwarf. 
Following the spin-down associated with the eruption, a spin-up in the post-eruption 
phase is reported by \citep{2022ApJ...940L..56P} on the basis of optical photometry, 
while the presence of changes in the X-ray derived period deserves further
investigation \citep{2021ApJ...922L..42D,2022ApJ...932...45O}.

%%% ZTF photometry https://lasair-ztf.lsst.ac.uk/object/ZTF19aasfsjq/
%%% ASAS-SN photometry https://asas-sn.osu.edu/sky-patrol/coordinate/381a105f-acc8-4c04-a87c-bb362e84df68

The 3.67\,h orbital period firmly identifies the donor star as a dwarf: 
an evolved donor would not fit in such a compact orbit \citep{2011ApJS..194...28K}. 
The white dwarf spin period is substantially shorter than the orbital period, 
revealing the system as an intermediate polar (IP). 
IPs host white dwarfs with magnetic fields strong enough to 
disrupt the inner part of the accretion disc and redirect the accreting
matter to the magnetic poles. As the white dwarf rotates, the magnetic poles
come in and out of view modulating the X-ray and optical light of the system  
%The optical and X-ray properties of IPs have been reviewed by
(see the reviews by \citealt{1994PASP..106..209P}, \citealt{2000NewAR..44...63B}, \citealt{2017PASP..129f2001M}).

Astrometric measurements during the eruption of \nova{} \citep[e.g.][]{2021ATel14710....1A}
allow the identification of the nova progenitor as a $G = 19.95 \pm 0.02$ star 
{\em Gaia}~DR3~4514092717838547584 located at R.A. and Dec. 
\begin{verbatim}
18:57:30.98324 +16:53:39.5895
\end{verbatim}
equinox J2000.0, mean epoch 2016.0;
with the positional uncertainty of 0.6 and 0.8\,mas 
and the proper motion of  
$-4.1 \pm 0.7$ and $-4.7 \pm 0.9$\,mas\,yr$^{-1}$ in R.A. and Dec. directions,
respectively; there is no measured parallax 
\citep{2016A&A...595A...1G,2022arXiv220800211G}.
The detection of the white dwarf spin period in the pre-eruption Zwicky Transient Facility
\citep{2019PASP..131a8003M} photometry by \cite{2021ATel14720....1M}
unambiguously confirms the progenitor identification.

\cite{2021ATel14704....1M} report $E(B-V)=0.55$\,mag based on the \cite{1997A&A...318..269M} relation between extinction and 
the equivalent width of the \ion{K}{I}~7699\,\AA{} line.
For the standard value of $\frac{A_V}{E(B-V)}=3.1$, this corresponds to $A_V = 1.70$\,mag.
We use $A_V$ to estimate the expected Galactic X-ray absorbing column to \nova{} following \cite{2009MNRAS.400.2050G}: 
\begin{equation}
N_\mathrm{H} = 2.21 \times 10^{21}\,{\rm cm}^{-2} \times A_V = 3.77 \times 10^{21}\,{\rm cm}^{-2}
\end{equation}
-- this is the value we use throughout this paper. 
The total line of sight neutral hydrogen column density in this direction, as derived from the 21\,cm line observations of
\cite{2005A&A...440..775K}, %\footnote{\url{https://www.astro.uni-bonn.de/hisurvey/euhou/LABprofile/}}, 
is $N_{\rm HI} = 2.99 \times 10^{21}\,{\rm cm}^{-2}$, 
lower than the above optical reddening-based estimate. 
The similar value $N_{\rm HI} = 2.95 \times 10^{21}\,{\rm cm}^{-2}$ is listed in 
the \ion{H}{I}\,4\,$\pi$ survey data \citep[HI4PI;][]{2016A&A...594A.116H}.

\begin{figure}
%                                                        left  bottom right  top
        \includegraphics[width=1.0\linewidth,clip=true,trim=0cm 5.5cm 0cm 0cm,angle=0]{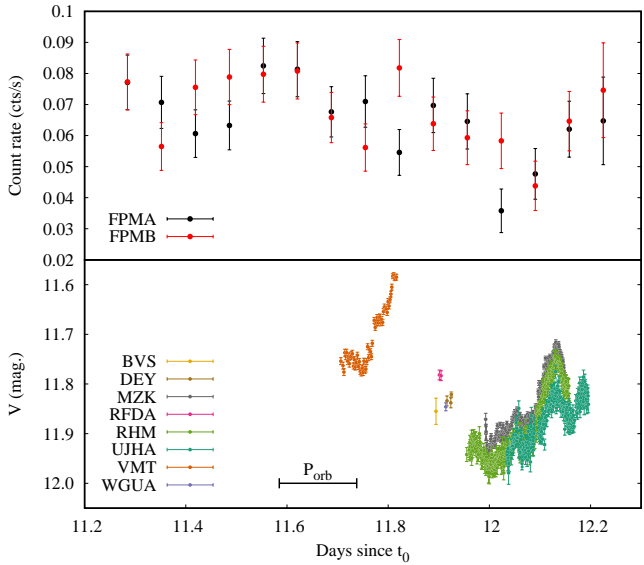}
    \caption{Top panel: the background-subtracted \nustarenergylow{}--\nustarenergyhigh{}\,keV {\em NuSTAR} lightcurve of \nova{}
(see \S~\ref{sec:nustarvar}).
Bottom panel: simultaneous optical $V$ band photometry by multiple 
observers identified by their AAVSO codes. 
% Try to avoid confusion with NuSTAR orbital period
The horizontal bar indicates the duration of \nova{} orbital period.}
    \label{fig:nustarlc}
\end{figure}

\section{Observations and analysis}
\label{sec:obs}

\subsection{{\em NuSTAR} hard X-ray observations}
\label{sec:nustarobs}

{\em NuSTAR} is a focusing hard X-ray telescope operating in the 3--79\,keV
energy range \citep{2013ApJ...770..103H,2015ApJS..220....8M}.
{\em NuSTAR} observed \nova{} between 
\nustarobsstartutc{} (\tnustarep) 
and \nustarobsstoputc{}\,UT 
(ObsID~\nustarobsid; PI:~Sokolovsky) for a total exposure of \nustarexpks{}\,ks. 
For the analysis we used \textsc{nupipeline} and \textsc{nuproducts} scripts from
\textsc{HEASoft\,\nustarheasoftver{}} \citep{2014ascl.soft08004N}, together
with the calibration files from the \textsc{CALDB} version \textsc{\nustarcaldbver{}}.
Following the same analysis procedure as \cite{2020MNRAS.497.2569S,2022MNRAS.514.2239S}, 
we utilized a circular extraction region with  radius of 30$\arcsec$ centred on the X-ray image of the nova using \textsc{ds9} \citep{2003ASPC..295..489J} 
independently for the two focal plane modules: FPMA and FPMB. The background was extracted from five circular regions 
of the same radius placed on the same Cadmium-Zinc-Telluride \citep[CZT;][]{2011hxra.book.....A} chip as the nova image.
For the following analysis we restricted the energy range to \nustarenergylow{}--\nustarenergyhigh{}\,keV.

\subsubsection{{\em NuSTAR} lightcurve}
\label{sec:nustarvar}

Fig.~\ref{fig:nustarlc} presents the \nustarenergylow{}--\nustarenergyhigh{}\,keV lightcurves of \nova{}
obtained during the {\em NuSTAR} observation described in \S~\ref{sec:nustarspec}. 
The lightcurves were background-subtracted and binned to 
\nustarorbitalperiodsec{}\,s (corresponding to the {\em NuSTAR}
orbital period at the time of the observations) 
resulting in one count rate measurement per orbit. 
Following \cite{2010AJ....139.1269D} and \cite{2017MNRAS.464..274S} we
perform a $\chi^2$ test to evaluate the significance of the count rate variations.
The probability of the observed scatter of count rate measurements arising from random noise
(while the true count rate is constant) is found to be very low (\nustarlcchiprob{}), 
allowing us to reject the constant count rate hypothesis.
The detected variations are happening on time-scales from one {\em NuSTAR} orbit to the total duration of the observation. 
The mean background-subtracted count rate is \nustarmeancountrateperfpm{}\,ct\,s$^{-1}$ per focal plane module.

We searched for periodicities in photon arrival times using 
the \textsc{patpc}\footnote{\url{https://github.com/kirxkirx/patpc}} code 
\citep{2022arXiv220610625S} constructing the `$H_m$-periodogram' 
\citep{1989A&A...221..180D,2010A&A...517L...9D,2011ApJ...732...38K}.
No periodicities were identified in the trial period range of 1 to 1000\,s
that would satisfy the following criteria:
\begin{enumerate}
\item be significant at $p<0.05$ level (\S~\ref{sec:intro});
\item be present in both FPMA and FPMB data;
\item not be a multiple of the {\em NuSTAR} orbital period.
\end{enumerate}
We repeated the search restricting the analysis to the lowest-energy
3.0--3.5\,keV events with the same null result. 
Finally, we compute the 
$H_m$ value for the spin and orbital periods reported by \cite{2022ApJ...940L..56P} and
find the associated single-trial probabilities to be $p \gg 0.05$ 
(no significant periodicity in {\em NuSTAR} data).

We note that irregular variability is present
in the AAVSO optical photometry obtained simultaneously with the {\em NuSTAR}
observations (Fig.~\ref{fig:nustarlc}). 
This variability is distinct from the overall optical brightness decline 
and is happening on a time-scale longer than the orbital period. 
The spin variations were first detected after the {\em NuSTAR} epoch \citep{2022ApJ...940L..56P}. 
The physical origin of these irregular brightness variations is uncertain.

\subsubsection{{\em NuSTAR} spectroscopy}
\label{sec:nustarspec}

\begin{table*}
        \centering
        \caption{{\em NuSTAR} spectral modelling}
        \label{tab:nustarspecmodels}
        \begin{tabular}{@{~}c@{~~}c@{~~}c@{~~}c@{~~}c@{~~}c@{~~}c@{~~}c@{~}} 
                \hline
\texttt{vphabs} $N_\mathrm{H}$  & k$T$  & $\Gamma$ & ${\rm C}/{\rm C}_{\odot}$, ${\rm N}/{\rm N}_{\odot}$, ${\rm O}/{\rm O}_{\odot}$, ${\rm Ne}/{\rm Ne}_{\odot}$,      & \nustarenergylow{}--\nustarenergyhigh{}\,keV Flux                    & unabs. \nustarenergylow{}--\nustarenergyhigh{}\,keV Flux             & $p$ & $\chi^2$/d.o.f.\\
($10^{22}$\,cm$^{-2}$)          & (keV) &          & ${\rm Mg}/{\rm Mg}_{\odot}$, ${\rm Si}/{\rm Si}_{\odot}$, ${\rm S}/{\rm S}_{\odot}$, ${\rm Fe}/{\rm Fe}_{\odot}$  & $\log_{10}$(erg\,cm$^{-2}$\,s$^{-1}$)  & $\log_{10}$(erg\,cm$^{-2}$\,s$^{-1}$)  &     &             \\ 
                \hline

% const_phabs_vphabs_pow__solar
\multicolumn{8}{c}{{\bf solar} abundances \texttt{constant*phabs*vphabs*powerlaw}} \\
$5.0 \pm 2.1 $ & $  $ & $3.2 \pm 0.1 $ & $1.0*, 1.0*, 1.0*, 1.0*, 1.0*, 1.0*, 1.0*, 1.0* $ & $-11.65 \pm 0.01 $ & $-11.59 \pm 0.03 $ & 0.50 & 62.24/63 \\

% const_phabs_vphabs_vapec__solar
\multicolumn{8}{c}{{\bf solar} abundances \texttt{constant*phabs*vphabs*vapec}} \\
$0.0^\dagger$ & $5.9 \pm 0.3 $ & $ $ & $1.0*, 1.0*, 1.0*, 1.0*, 1.0*, 1.0*, 1.0*, 1.0* $ & $-11.72 \pm 0.01 $ & $-11.72 \pm 0.01 $ & 0.00 & 219.60/63 \\

% const_phabs_vphabs_vapec__freeC
%\multicolumn{9}{c}{free C \texttt{constant*phabs*vphabs*vapec}} \\
%$0.0 \pm -1.0 $ & $4.1 \pm 0.2 $ & $ $ & $999.6 \pm 1353.6, 1.0*, 1.0*, 1.0*, 1.0*, 1.0*, 1.0*, 1.0* $ & $-11.69 \pm 0.01 $ & $-11.71 \pm 0.01 $ & 0.27 & 68.40 & 62 \\

% const_phabs_vphabs_vapec__freeFe
%\multicolumn{9}{c}{free Fe \texttt{constant*phabs*vphabs*vapec}} \\
%$0.0 \pm -1.0 $ & $4.4 \pm 0.2 $ & $ $ & $1.0*, 1.0*, 1.0*, 1.0*, 1.0*, 1.0*, 1.0*, 0.0 \pm 0.1 $ & $-11.69 \pm 0.01 $ & $-11.70 \pm 0.01 $ & 0.46 & 62.34 & 62 \\

% const_phabs_vphabs_vapec__freeNO
\multicolumn{8}{c}{{\bf free NO} abundances \texttt{constant*phabs*vphabs*vapec}} \\
$0.0^\dagger$ & $3.9 \pm 0.2 $ & $ $ & $1.0*, 1000.0^\dagger, 1000.0*, 1.0*, 1.0*, 1.0*, 1.0*, 1.0* $ & $-11.69 \pm 0.01 $ & $-11.71 \pm 0.01 $ & 0.30 & 67.33/62 \\

% const_phabs_vphabs_vapec__freeN
%\multicolumn{9}{c}{free N \texttt{constant*phabs*vphabs*vapec}} \\
%$0.0 \pm -1.0 $ & $4.0 \pm 0.2 $ & $ $ & $1.0*, 1000.0 \pm 806.9, 1.0*, 1.0*, 1.0*, 1.0*, 1.0*, 1.0* $ & $-11.70 \pm 0.01 $ & $-11.71 \pm 0.01 $ & 0.10 & 76.54 & 62 \\

% const_phabs_vphabs_vapec__freeO
%\multicolumn{9}{c}{free O \texttt{constant*phabs*vphabs*vapec}} \\
%$0.0 \pm -1.0 $ & $3.9 \pm 0.2 $ & $ $ & $1.0*, 1.0*, 997.9 \pm 4179.4, 1.0*, 1.0*, 1.0*, 1.0*, 1.0* $ & $-11.69 \pm 0.01 $ & $-11.71 \pm 0.01 $ & 0.30 & 67.44 & 62 \\

% const_phabs_vphabs_vapec__freeSi
%\multicolumn{9}{c}{free Si \texttt{constant*phabs*vphabs*vapec}} \\
%$0.0 \pm -1.0 $ & $3.5 \pm 0.2 $ & $ $ & $1.0*, 1.0*, 1.0*, 1.0*, 1.0*, 999.9 \pm 2717.8, 1.0*, 1.0* $ & $-11.70 \pm 0.01 $ & $-11.71 \pm 0.01 $ & 0.18 & 72.18 & 62 \\

% const_phabs_vphabs_vapec__freeS
%\multicolumn{9}{c}{free S \texttt{constant*phabs*vphabs*vapec}} \\
%$0.0 \pm -1.0 $ & $4.3 \pm 0.2 $ & $ $ & $1.0*, 1.0*, 1.0*, 1.0*, 1.0*, 1.0*, 290.6 \pm 101.2, 1.0* $ & $-11.67 \pm 0.01 $ & $-11.67 \pm 0.01 $ & 0.00 & 98.88 & 62 \\

% const_phabs_vphabs_vapec__NCar
\multicolumn{8}{c}{{\bf preferred model} --- {\bf V906\,Car} abundances \texttt{constant*phabs*vphabs*vapec}} \\
$0.0^\dagger$ & $4.1 \pm 0.2 $ & $ $ & $1.0*, 345.0*, 29.0*, 2.2*, 0.6*, 1.1*, 1.0*, 0.1* $ & $-11.69 \pm 0.01 $ & $-11.70 \pm 0.01 $ & 0.39 & 65.52/63 \\

% const_phabs_vphabs_vapec__V838Her
\multicolumn{8}{c}{{\bf V838\,Her} abundances \texttt{constant*phabs*vphabs*vapec}} \\
$0.0^\dagger$ & $3.5 \pm 0.2 $ & $ $ & $7.5*, 37.9*, 1.9*, 52.5*, 1.4*, 7.2*, 32.8*, 1.5* $ & $-11.71 \pm 0.01 $ & $-11.72 \pm 0.01 $ & 0.00 & 131.52/63 \\
                \hline
        \end{tabular}
\begin{flushleft}
The parameters that were kept fixed for the model fit are marked with the \** symbol. 
The $^\dagger$ symbol marks the limit of the search range reached during
the fitting procedure: the best-fitting value is set equal to the limit value and no fitting uncertainty is reported.
{\bf Column designation:}
Col.~1~-- intrinsic absorbing column (in excess of the total Galactic value);
Col.~2~-- temperature of the thermal component;
Col.~3~-- photon index of the power law component;
Col.~4~-- abundances of selected elements by number relative to the solar values of \cite{2009ARA&A..47..481A};
Col.~5~-- the logarithm of the integrated \nustarenergylow{}--\nustarenergyhigh{}\,keV flux under the model;
Col.~6~-- logarithm of the unabsorbed \nustarenergylow{}--\nustarenergyhigh{}\,keV flux;
Col.~7~-- chance occurrence (null hypothesis) probability;
Col.~8~-- $\chi^2$ value divided by the number of degrees of freedom.

\end{flushleft}

\end{table*}

\begin{figure}
\begin{center}
\includegraphics[width=0.48\textwidth,clip=true,trim=0cm 0cm 0cm 0cm,angle=0]{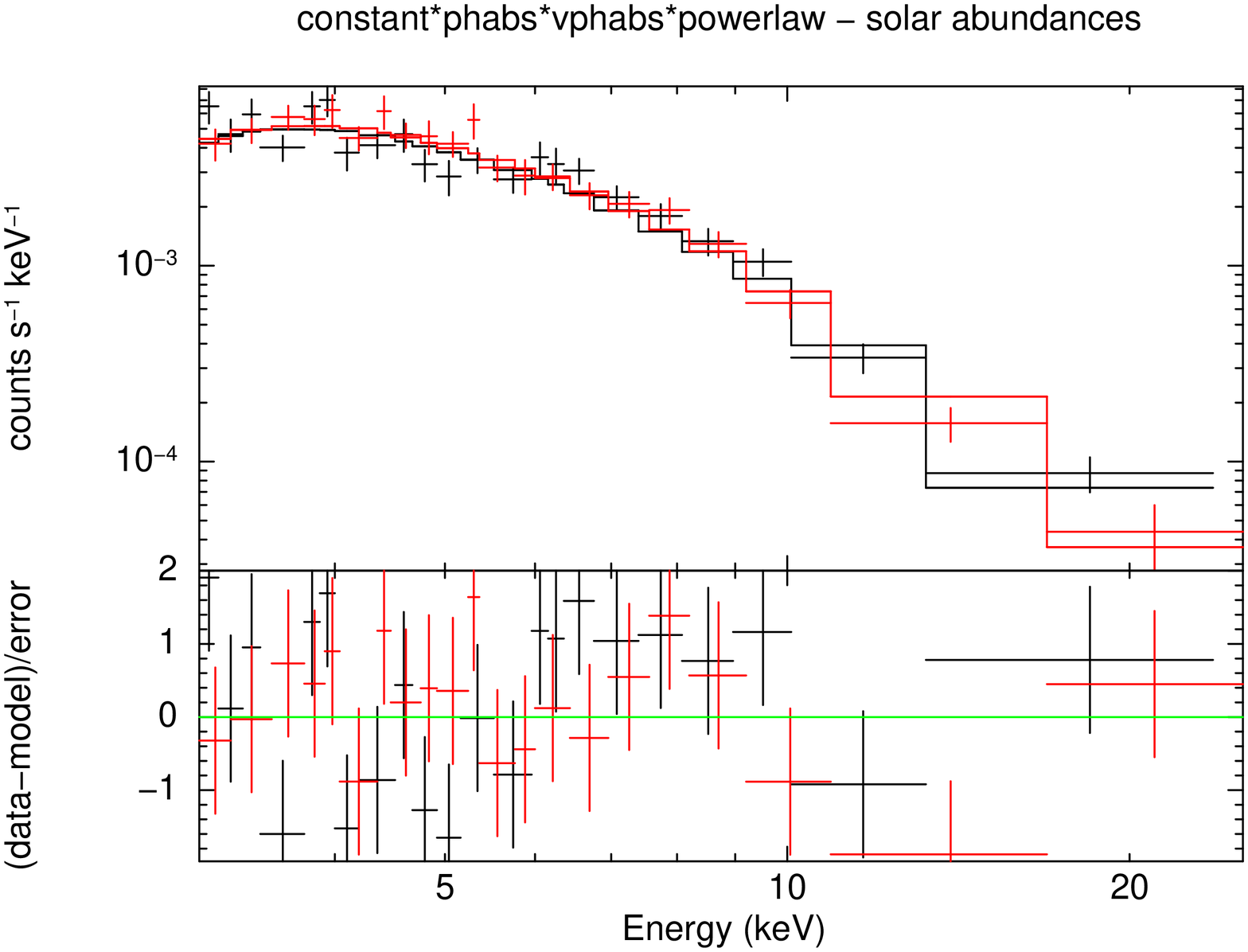}\\
~~\\
\includegraphics[width=0.48\textwidth,clip=true,trim=0cm 0cm 0cm 0cm,angle=0]{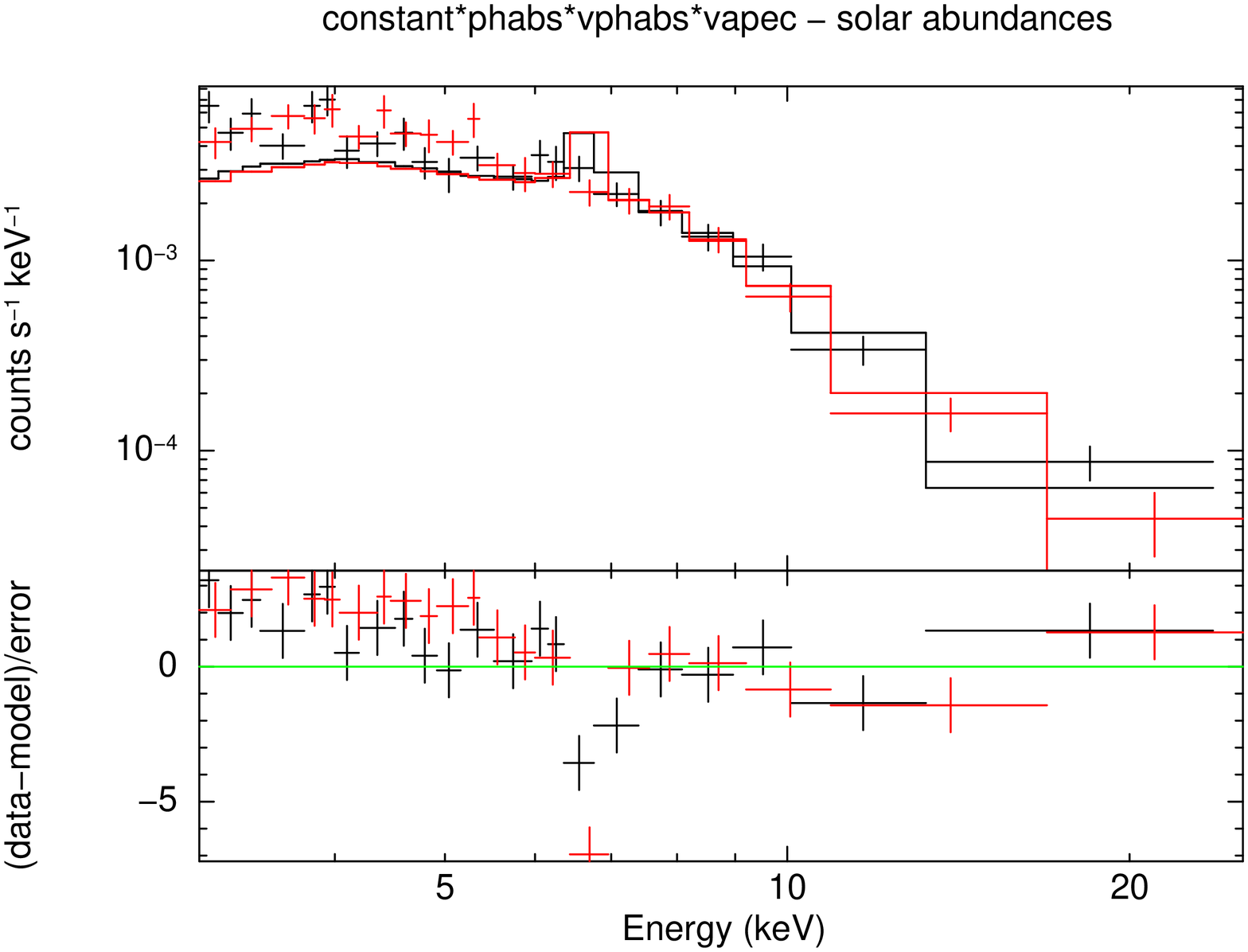}\\
~~\\
\includegraphics[width=0.48\textwidth,clip=true,trim=0cm 0cm 0cm 0cm,angle=0]{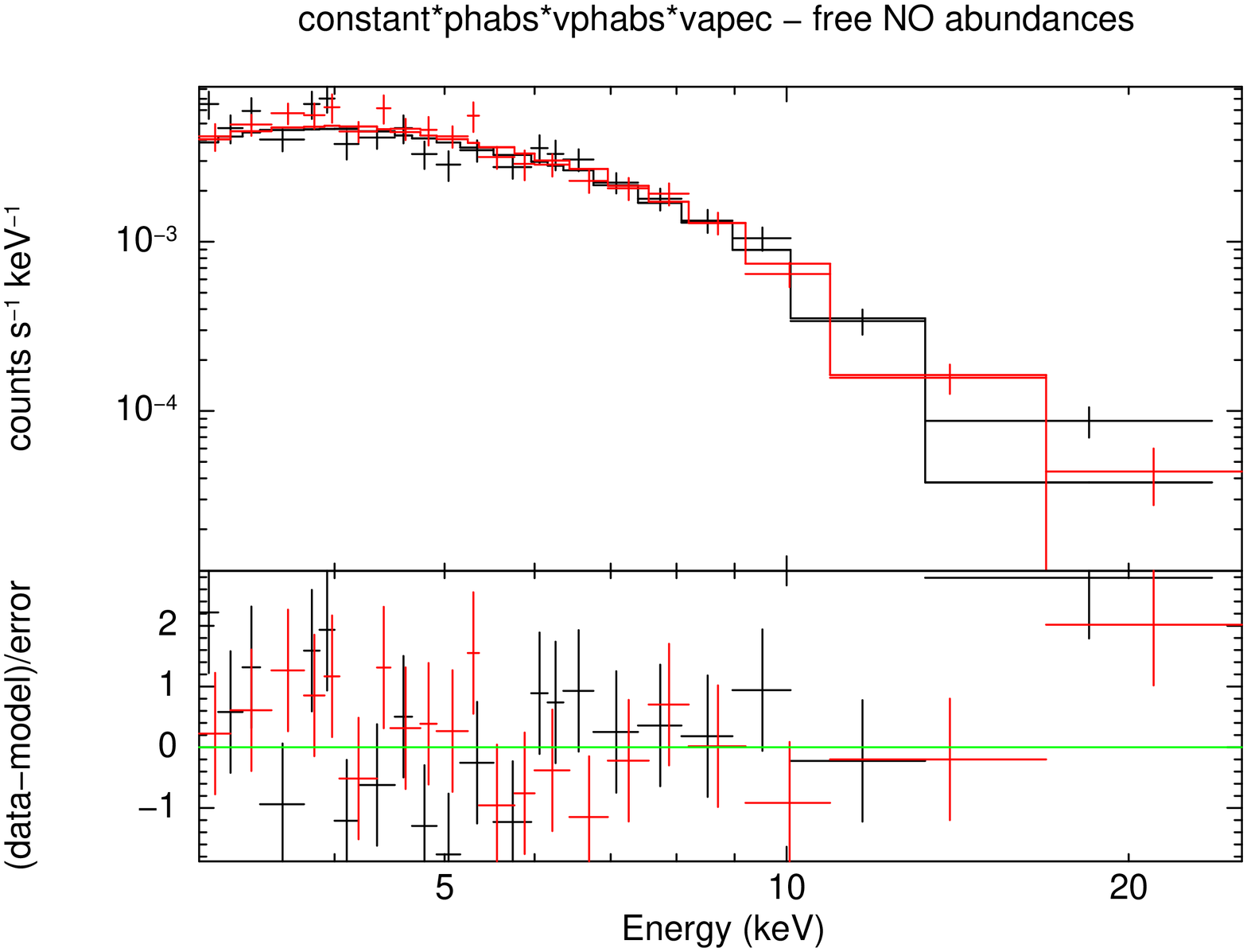}
\end{center}
\caption{Observed {\em NuSTAR} spectra compared with the models from 
Table~\ref{tab:nustarspecmodels}: 
power law emission with solar abundance absorber (top), 
\textsc{APEC} thermal plasma emission with solar abundances for both the
emitter and absorber (middle), 
and the \textsc{APEC} model, with NO abundances tied together and left free to vary (bottom). 
Black and red represent spectra obtained with the two {\em NuSTAR}
telescopes: FPMA and FPMB. For each model, the top sub-panel shows the spectrum and the model, while
the bottom sub-panel shows the difference between the spectrum and the model in
the units of uncertainty associated with each data bin.}
\label{fig:nuspec}
\end{figure}

The {\em NuSTAR} FPMA and FPMB spectra (Fig.~\ref{fig:nuspec}) were binned to
have at least \nustargroupmincounts{} counts per bin and were fit jointly
using the models listed in Table~\ref{tab:nustarspecmodels}.
All the considered models include a \texttt{constant} component to account for
the imperfect (and variable) cross-calibration of FPMA and FPMB, a \texttt{phabs} component
that accounts for the photoelectric absorption \citep{1992ApJ...400..699B} in the Galaxy along the line 
of sight by solar abundance material (\citealt{2009ARA&A..47..481A}; the equivalent hydrogen column
density, $N_{\rm H}$, is fixed to the value listed in \S~\ref{sec:thisnova}); and a
\texttt{vphabs} component that accounts for the possible intrinsic absorption within the nova shell 
($N_{\rm H}$ and the abundances associated with this model component are varied).

The simplest X-ray emission model we consider is \texttt{powerlaw}.
Allowing both the photon index and the intrinsic photoelectric absorption to
vary, one can account for the observed curvature of the spectrum and obtain 
a good fit even with the solar abundance absorber, see Table~\ref{tab:nustarspecmodels}
and Fig.~\ref{fig:nuspec}. 
However, as discussed by \cite{2022MNRAS.514.2239S}, 
the non-thermal X-ray emission mechanisms expected to operate in a nova 
should all produce hard photon spectra. 
\cite{2018ApJ...852...62V} predict that the low-energy extension of the GeV emission 
should have a $\Gamma = 1.2$ to $1.0$ in the {\em NuSTAR} band.
The other possible non-thermal mechanism -- Comptonization of the radioactive MeV lines 
-- should produce even harder spectra with $\Gamma \lesssim 0$ below 30\,keV \citep[see fig.~1--4 of][]{1998MNRAS.296..913G}. 
The observed soft photon index $\Gamma = 3.2 \pm 0.1$ contradicts these predictions.

The power law provides a convenient empirical description of the data, so 
we use it to compute the monochromatic flux 
%(in SED units; \S~\ref{sec:thispaper}) 
(in SED units; \S~\ref{sec:intro}) 
at 20\,keV 
(where the absorption is negligible, simplifying the computations) using equation~(4) of \cite{2022MNRAS.514.2239S}: 
$\nu F_\nu = 3.6 \times 10^{-10}$\,erg\,cm$^{-2}$\,s$^{-1}$. 
In the absence of an obvious physical mechanism that would produce non-thermal emission 
with a soft power law spectrum, we favour thermal emission models.

The GeV (\S~\ref{sec:latobs}) and non-thermal radio (\S~\ref{sec:radiodiscussion})   
emission reveal the presence of shock-accelerated particles within the ejecta of \nova{}. 
Shocks may heat plasma to X-ray temperatures \citep[e.g.][]{1967pswh.book.....Z,1997pism.book.....D}. 
Thermal emission of shock-heated plasma is the standard explanation for the $\gtrsim$1\,keV 
X-rays observed from novae (\S~\ref{sec:intro}). 
%(\S~\ref{sec:innovashock}). 
Therefore, we attempt a fit with a collisionally-ionized plasma emission model \citep[\texttt{vapec;}][]{2005AIPC..774..405B}. 
The model includes both the bremsstrahlung (free-free) continuum and line emission from specific elements.

We found no acceptable fit (\S~\ref{sec:intro}) to the data with the elemental abundances of the emitting plasma and 
the absorber fixed to the solar values
 (Table~\ref{tab:nustarspecmodels}).
Specifically, the \texttt{apec} model predicts a strong Fe~K$\alpha$ emission feature at 6.7\,keV 
that is clearly not present in the data (Fig.~\ref{fig:nuspec}). 
While fitting the solar-abundance \texttt{apec} model, 
\textsc{XSPEC} is trying to suppress the Fe~K$\alpha$ emission by increasing 
the temperature (compared to the non-solar abundances fits), essentially
trading off the reduced residuals in the 6-7\,keV region for the increased
residuals in the low energy regions.

One may change the elemental abundances in the \texttt{vapec} model to suppress 
the Fe~K$\alpha$ emission relative to the bremsstrahlung continuum. There
are two ways to do this:
\begin{enumerate}
\item decrease the Fe abundance to suppress the emission;
\item increase abundances of other heavy elements that, being ionized, will
shed more free electrons, enhancing the bremsstrahlung continuum (and
swamping the Fe emission).
\end{enumerate}

Nova ejecta are known to be overabundant in CNO elements
\citep{1985ESOC...21..225W,1986ApJ...308..721T,1998PASP..110....3G,2001MNRAS.320..103S,2005ApJ...624..914V,2012ApJ...755...37H}. 
The obvious explanation for such overabundance is that 
nova ejecta contain material ablated from the white dwarf with 
CO or ONeMg composition \citep[e.g.][and references therein]{2018ApJ...860..110S,2021JApA...42...13D}. 
Nitrogen is usually the most abundant of the CNO elements as the accreted material
is mixed with the white dwarf material and processed through the incomplete CNO cycle \citep{1972ApJ...176..169S,1986ApJ...308..721T} 
that changes the relative abundance of the CNO elements. 
The cycle is most likely to be interrupted by ejection while in the $^{14}$N bottleneck 
\citep[e.g.][]{2004A&A...420..625I,2006PhLB..634..483L}.
The exact composition of the ejecta depends on the degree 
of mixing between the accreted envelope and the white dwarf
\citep{2011Natur.478..490C,2016A&A...595A..28C,2018A&A...619A.121C,2013ApJ...762....8D,2022A&A...660A..53G} 
as well as the white dwarf composition.
To account for this, we let the abundances of the emitting plasma and the absorber intrinsic to the nova
ejecta vary. We also consider abundances for the emitter and absorber fixed to those found in another nova where they
are well constrained, and consider solar abundance for comparison.
We assume that the emitting and absorbing material both originate in the nova
ejecta and have the same elemental abundances. The Galactic absorber along
the line of sight is assumed to have solar abundances and is represented by
a separate model component \texttt{phabs}, as described earlier.

All the emission features of elements C, N, O, Ne, Mg,
Si or S, as well as their absorption K edges, are outside the {\em NuSTAR} band.
Therefore, the {\em NuSTAR} spectrum allows us to constrain only the total 
number of these medium-Z elements, not the individual abundances.
To facilitate direct comparison with the results of \cite{2020MNRAS.497.2569S,2022MNRAS.514.2239S}, we tie together
the abundances of N and O and let them vary while keeping the abundances of all other elements (including C and Fe) fixed to the solar values.
While the abundances of N and O are clearly super-solar, the exact values
are not well constrained (Table~\ref{tab:nustarspecmodels}). 
Good fits can be obtained even if we allow the abundance of any one of the C, N, O, Si elements vary while keeping all other abundances fixed to solar. This illustrates that the individual contributions of these elements cannot be really distinguished based on our {\em NuSTAR} spectrum and only their overall contribution is constrained.

To pick some specific illustrative values, we consider a model with the
abundances of N, O (along with Ne, Mg, and Si, and in addition tied-together Fe, Co, and Ni) 
fixed to the ones derived from {\em XMM-Newton} grating spectroscopy of 
a brighter nova V906\,Car \citep{2020MNRAS.497.2569S}.
The absorbed thermal plasma model with the V906\,Car
abundances provides an excellent fit to the {\em NuSTAR} spectrum of \nova{} 
(Table~\ref{tab:nustarspecmodels}; Fig.~\ref{fig:nuswspec}).

\cite{2021ATel14746....1W} classify \nova{} as a neon nova based on strong
forbidden Ne emission revealed by their optical spectroscopy. 
As nova V906\,Car erupted on a CO white dwarf \citep{2020MNRAS.497.2569S}, 
we also tried to fit the {\em NuSTAR} spectra with the abundances of V838\,Her -- a fast neon 
nova with well-determined elemental composition \citep{2007ApJ...657..453S}. 
This model, however, did not result in a good fit: %, as 
it overpredicted the Fe~K$\alpha$ 6.7\,keV emission---the same problem that led us to reject the solar abundances fit.

In all variations of the absorbed thermal emission model, the intrinsic
absorbing column is consistent with zero. The Galactic absorption is
sufficient to describe the curvature of the {\em NuSTAR} spectrum.

%
%
%%-----------------------------Figure Start------------------------------
%%-----------------------------Figure End--------------------------------
%

\subsection{{\em Swift} X-ray and UV observations}
\label{sec:swiftobs}

\begin{figure}
        \includegraphics[width=0.48\textwidth]{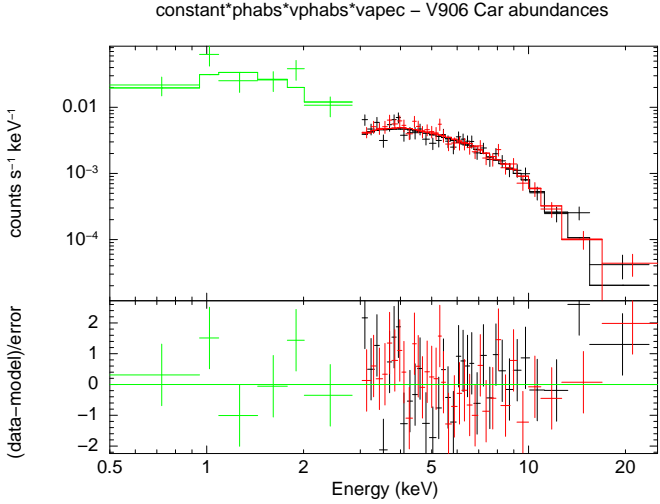}
    \caption{The quasi-simultaneous {\em NuSTAR}/FPMA and FPMB (colour-coded
as black and red, respectively) and {\em Swift}/XRT (green) spectra of \nova{}
compared to our preferred model---the single-temperature absorbed thermal plasma 
with the elemental abundances set to match those of nova V906\,Car \citep{2020MNRAS.497.2569S}.}
    \label{fig:nuswspec}
\end{figure}

{\em The Neil Gehrels Swift Observatory} \citep{2004ApJ...611.1005G}
combines multiple instruments 
including the X-ray Telescope \citep[XRT;][]{2005SSRv..120..165B} operating
in the 0.3--10\,keV energy range and 
the UV/Optical Telescope \citep[UVOT;][]{2005SSRv..120...95R}
on a space-based platform capable of fast repointing.
A detailed discussion of the {\em Swift}/XRT lightcurve of \nova{} is presented
by \cite{2021ApJ...922L..42D}. Here we analyse the {\em Swift} observation
performed on 2021-06-24 (ObsID~00014375014; PI~Orio). 
The nova was observed for a total exposure of 1\,ks split between 
two pointings around 02:49--02:53 and 12:25--12:37\,UT,
the first one overlapping with the {\em NuSTAR} observation described in
\S~\ref{sec:nustarobs}. In order to minimize the optical loading
we limited the analysis to \texttt{grade~0} events.
\nova{} is clearly detected in the 0.5--10.0\,keV
band by {\em Swift}/XRT at $0.091 \pm0.012$\,ct\,s$^{-1}$. The XRT spectrum 
presented in Fig.~\ref{fig:nuswspec} is consistent with the same absorbed single-temperature model  
describing the {\em NuSTAR} observations, if we allow for a constant offset
between the XRT and {\em NuSTAR} data. The ${\rm \texttt{constant}} = 1.5 \pm 0.3$ offset
accounts for the source variability (cf.~Fig.~\ref{fig:nustarlc}) and
the XRT to {\em NuSTAR} cross-calibration uncertainty (expected to be below
10\,per~cent; \citealt{2017AJ....153....2M}).
We also used \textsc{patpc} (\S~\ref{sec:nustarvar}) to test for the
presence of X-ray modulation at the {\it spin} period in {\em Swift}/XRT events. 
The resulting single-trial $H_m$ value corresponds to a chance occurrence probability 
of $p = 0.09$, which we consider a non-detection (\S~\ref{sec:intro}). 
An individual {\em Swift} pointing is shorter than the {\it orbital} period.
Investigations of X-ray orbital periodicity based on multiple {\em Swift} and {\em NICER} 
pointings are presented by \cite{2021ApJ...922L..42D}, \cite{2022ApJ...932...45O}, and \cite{2022MNRAS.517L..97L}.

The simultaneous ultraviolet photometry with {\em Swift}/UVOT resulted in the Vega system magnitudes 
${\rm uvw2} = 12.10 \pm 0.02$,
${\rm uvm2} = 12.87 \pm 0.03$,
${\rm uvw1} = 11.59 \pm 0.02$,
corresponding to a blackbody temperature of $T_{\rm UVOT} = 27000 \pm 4000$\,K. 
To compute $T_{\rm UVOT}$ we applied the reddening correction (\S~\ref{sec:thisnova})
using \cite{1989ApJ...345..245C} extinction law and the UVOT magnitude-to-flux conversion
of \cite{2008MNRAS.383..627P}.
%The quoted photometric uncertainties are statistical only, while 
%the \texttt{uvotsource} task reports an additional 0.03\,mag. systematic 
%uncertainty. 
%The source is bright enough that the coincidence-loss (pile-up) 
%effect in the UVOT detector may be important. The ${\rm uvw2}$ and ${\rm uvw1}$
%UVOT filters are known to suffer from a red leak, 
%% Wood be nice to cite this URL, but I'm desperately trying to conserve space
%%\footnote{\url{https://swift.gsfc.nasa.gov/analysis/uvot_digest/redleak.html}}, 
%while ${\rm uvm2}$ is affected to a lesser extent. Keeping in mind the above-mentioned 
%additional sources of uncertainty, applying the reddening correction 
%(corresponding to the $E(B-V)$ listed in \S~\ref{sec:thisnova}) according to 
%the \cite{1989ApJ...345..245C} extinction law and performing  
%the UVOT magnitude-to-flux conversion following \cite{2008MNRAS.383..627P}, 
%we conclude that the UVOT magnitudes roughly correspond to a blackbody 
%temperature of $T_{\rm UVOT} = 27000 \pm 4000$\,K.

%An ultraviolet source is detected at the position of the transient 
%with the following UVOT magnitudes (Vega system): 
%   JD        Band   Mag.  Err.
% 2459389.82  UVW2  12.10  0.02
% 2459389.82  UVM2  12.88  0.03
% 2459389.62  UVW1  11.59  0.03

\subsection{\fermilat{} $\gamma$-ray observations}
\label{sec:latobs}

The \fermilat{} \citep{2009ApJ...697.1071A} is a $\gamma$-ray instrument sensitive to the 30\,MeV-2\,TeV energy range.
We used \textsc{fermitools} version 2.0.8\footnote{\url{https://github.com/fermi-lat/Fermitools-conda/}} \citep{2019ascl.soft05011F} to perform a binned Maximum Likelihood analysis \citep{1996ApJ...461..396M} of \fermilat{} data on \nova{}.
We selected Pass~8 data (\texttt{P8R3}; \citealt{2013arXiv1303.3514A,2018arXiv181011394B};
with the associated \texttt{P8R3\_V3} instrument response functions) \texttt{SOURCE} class events,
in the energy range 0.1--300\,GeV, with maximum zenith angle of $90^\circ$, and with reconstructed positions within $15^\circ$ 
of (R.A., Dec.) = ($285\fdg0$, $16\fdg5$). The centre of the region of interest (ROI) was offset slightly ($0\fdg71$) from the optical position of \nova{} 
in order not to place the nova at the corner of spatial four bins (the bins were $0.1^\circ$ on a side). 
The events were filtered to include only times when the observatory was in normal science operations and the data were flagged as good.
The events were spatially binned in a $21\fdg2\times21\fdg2$ square region (sized to fit within the $15^\circ$ radius circular selection) and were binned in $\log({\rm energy})$ in 35 bins of equal size.
Energy dispersion (finite energy resolution of LAT) correction was enabled for all of our likelihood analyses, though the correction was disabled for the isotropic diffuse emission component\footnote{\url{https://fermi.gsfc.nasa.gov/ssc/data/analysis/documentation/Pass8_edisp_usage.html}}.

We constructed a spatial and spectral model of the region by including all point and extended sources from the third data release of the {\em Fermi} LAT fourth source catalogue (4FGL-DR3; \citealt{2020ApJS..247...33A}) within $25^\circ$ of the ROI centre. The model includes sources that are outside the field of view defined by the photon arrival direction selection at the previous step. 
% I'm referring to Pierre's analysis in the following sentence
We checked that the spectral analysis results do not depend critically on the exact choice of the model source and photon selection radii. 
The model also included components for the Galactic (using the spectral-spatial template \texttt{gll\_iem\_v07.fits}) and isotropic (using the file function \texttt{iso\_P8R3\_SOURCE\_V3\_v1.txt}) diffuse emission\footnote{Both files are available for download at \url{https://fermi.gsfc.nasa.gov/ssc/data/access/lat/BackgroundModels.html}}.
For our initial analysis, the spectral parameters of point sources detected in 4FGL-DR3 with $\geq15\sigma$ average significance and within $6^\circ$ of the ROI centre were allowed to vary in the fit, as well as the normalizations of the diffuse components.  Additionally, the normalization parameters of sources flagged as variable in 4FGL-DR3 were allowed to vary if
they were within $8^\circ$ of the ROI centre. 
A point source was added at the optical position of \nova{} having the \texttt{PowerLaw2} spectral model\footnote{\url{https://fermi.gsfc.nasa.gov/ssc/data/analysis/scitools/source_models.html\#PowerLaw2}}, 
with the \texttt{Integral} parameter free to vary, the photon index ($\Gamma$) fixed to 
a value of 2.2 (typical for $\gamma$-ray novae; \S~\ref{sec:specshape}), 
% Franckowiak has harder indexes because of the exponential cut-off model
the Lower-Limit (Upper-Limit) parameter fixed at 0.1\,GeV (300\,GeV).

\begin{figure*}
%                               left  bottom right  top
\includegraphics[width=0.48\textwidth,clip=true,trim=0cm 0cm 0cm 0cm,angle=0]{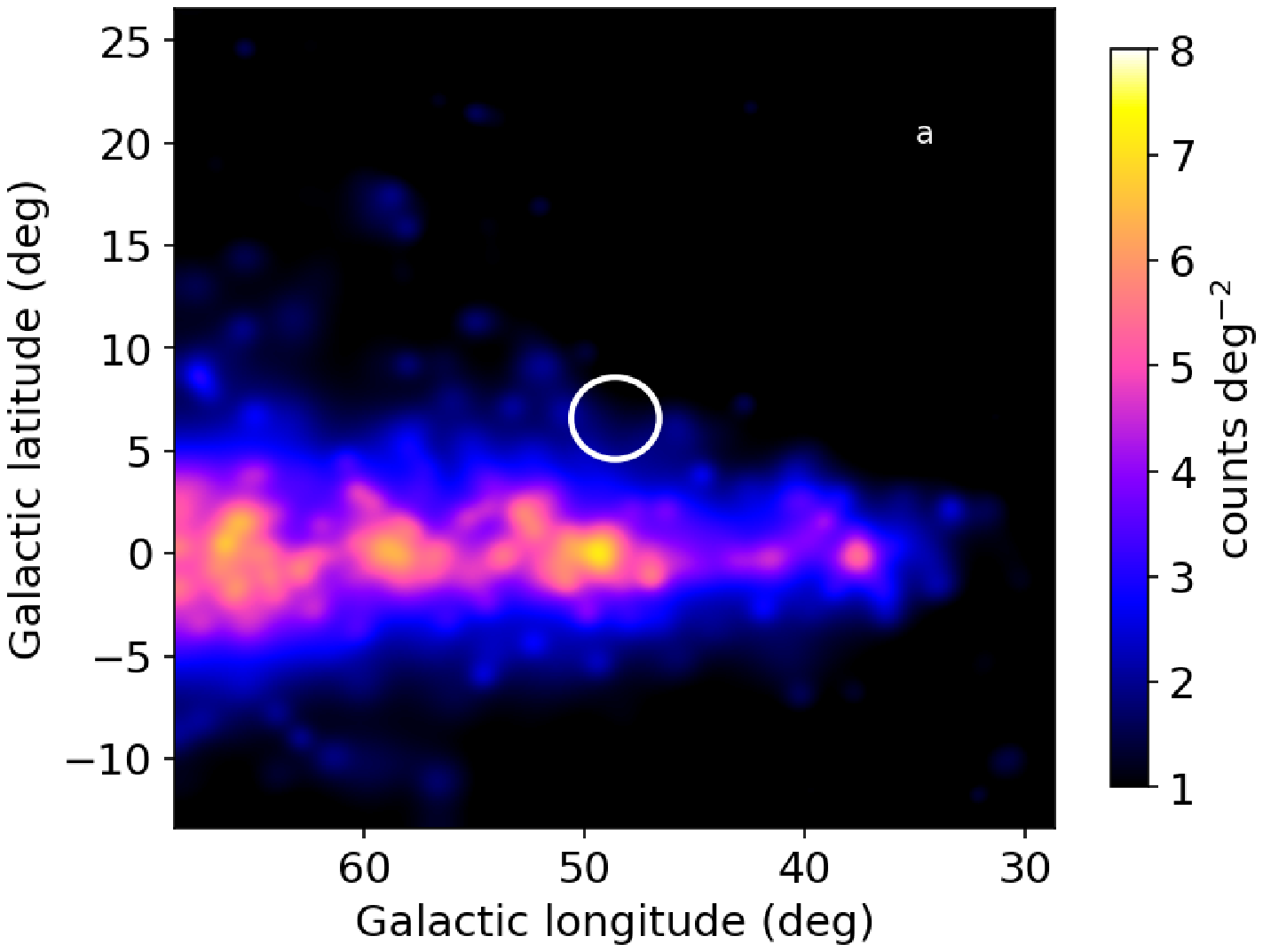}~~
\includegraphics[width=0.48\textwidth,clip=true,trim=0cm 0cm 0cm 0cm,angle=0]{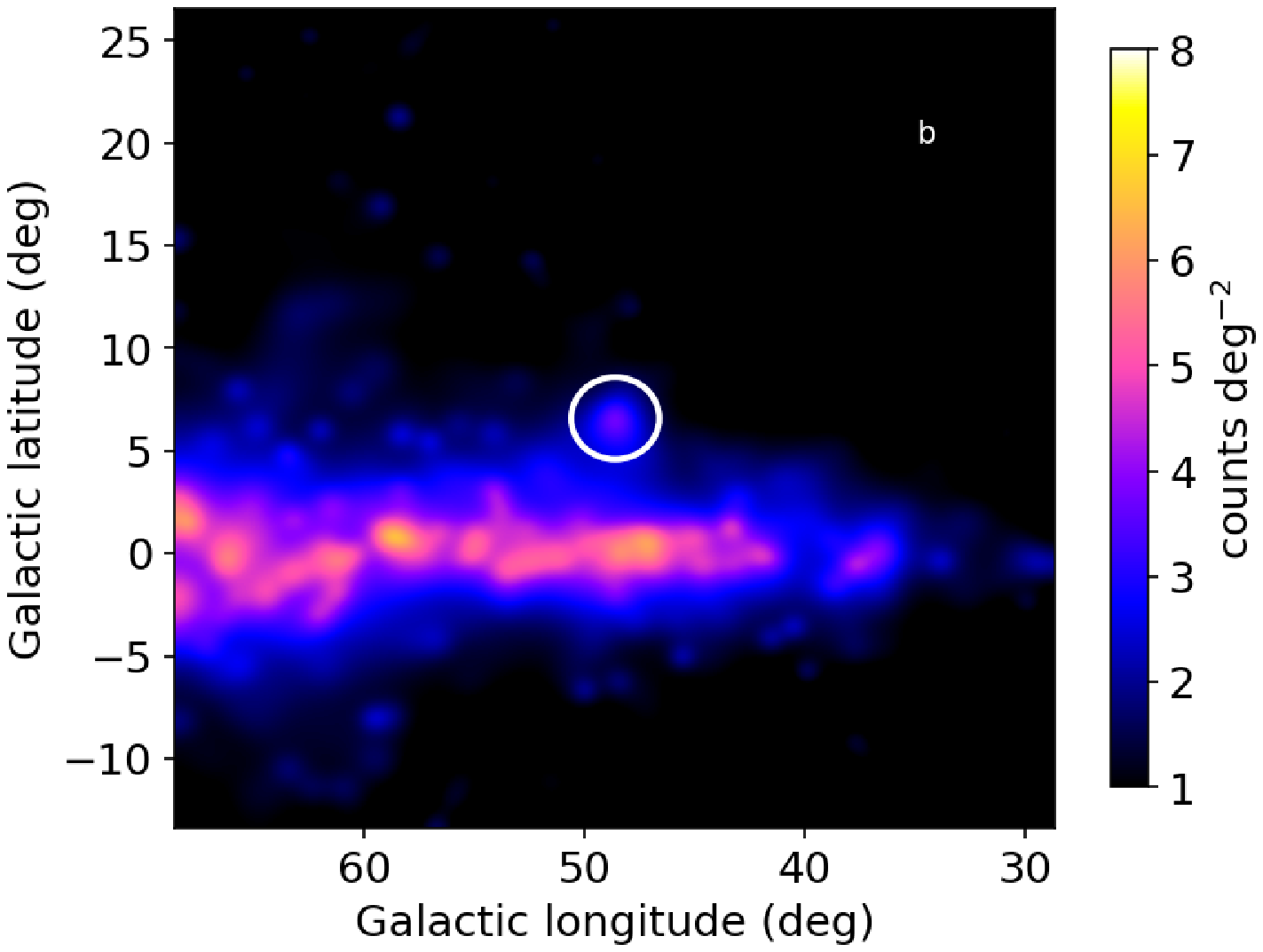}
\caption{The \fermilat{} smoothed 0.1--2\,GeV count images centered on \nova{}. The left image (a) covers the time interval
2021-06-10~10:34 to 2021-06-11 08:34\,UT before the eruption. 
The right image (b) covers the 18\,h interval when the $\gamma$-ray emission was detected.
The white circle marks the optical position of the nova.}
    \label{fig:latimg}
\end{figure*}

\begin{figure}
%                               left  bottom right  top
\includegraphics[width=1.0\linewidth,clip=true,trim=0.4cm 0.5cm 2.5cm 2.0cm,angle=0]{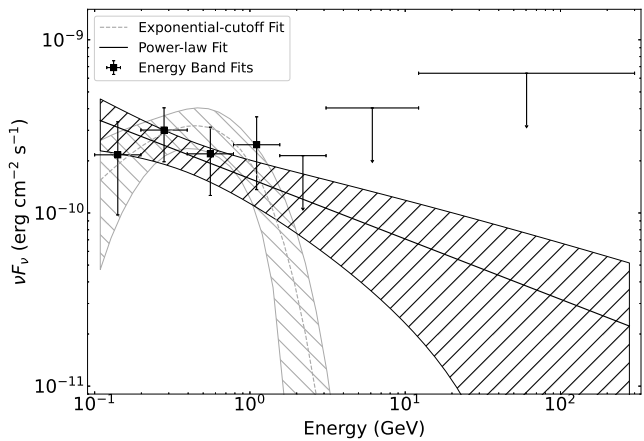}
\caption{The \fermilat{} spectral energy distribution of \nova{}, compared to the power law
(solid black line) and power law with an exponential cut-off (dashed grey line) models. The models were fit to the 0.1--300\,GeV photon data using the maximum likelihood technique. 
The filled regions correspond to the $1\sigma$ uncertainty range for the power law (black, forward slash fill) and cutoff (grey, back slash fill) models.} 
    \label{fig:latspec}
\end{figure}

To refine the free parameters in our model, 
we analysed one year of data prior to the outburst (2020-06-01 to 2021-06-01). 
After an initial fit, examination of the spatial residuals suggested that the normalization of the point source 4FGL\,J1930.3$+$0911 needed to be free to vary. 
Additionally, the point source at the position of \nova{} was not significantly detected and was removed from the model before refitting.

We then analysed data spanning 40 days from $t_0-10$\,d to $t_0+30$\,d. 
We started from the model best-fitting the one-year pre-eruption data, as described above.
We added a point source at the optical position of \nova{}, 
fixed the spectral parameters of the Galactic diffuse emission, 
and fixed all but the normalization parameters for sources that were free to vary in the previous analysis.
Finally, we fixed the normalizations of the faint sources that were detected with 
the test statistic ${\rm TS}<9$ \citep{1996ApJ...461..396M} in the 40-day data to their respective 1-year values.
The new source at the nova position was detected with 
${\rm TS}=5$ during this 40-day time interval, which is just less than $2\sigma$ for two degrees of freedom.

Using the 40-day model as a starting point, we constructed a lightcurve with 6\,hour bins (Fig.~\ref{fig:latlc}) spanning $t_0 \pm 2$\,d.
We assumed a power-law spectrum for the nova with free normalization and photon index.
The nova was detected in three consecutive 6\,h bins, with the first detection at $t_0+6$\,hours having ${\rm TS}=31$.
We consider as detections the 6\,h bins with ${\rm TS}>=6$ ($2\sigma$ detection for two degrees of freedom) 
and at least 4 predicted counts, while calculating the 95\,per~cent upper limits for the other bins (Fig.~\ref{fig:latlc}).
% Tyrel writes: yes, they are 95% upper limits.  I use the Bayesian approach (IntegralUpperLimit module) if TS<=0 OR Npred<4, otherwise I use the frequentist approach (UpperLimits module).
According to \cite{2022MNRAS.517L..97L}, this is the shortest-duration $\gamma$-ray nova ever observed.

During the 18\,h time span covered by the detections, \nova{} was identified with ${\rm TS} = 49$ (Fig.~\ref{fig:latimg}), having 
$\Gamma = 2.3 \pm 0.2$. % (Fig.~\ref{fig:latspec}). 
The 0.1--300\,GeV photon flux is $(1.6 \pm 0.4) \times 10^{-6}$\,photons\,cm$^{-2}$\,s$^{-1}$, which is equivalent 
to an integrated energy flux of $(9.5 \pm 2.7) \times 10^{-10}$\,erg\,cm$^{-2}$\,s$^{-1}$.
%
%
%  18-hour "high time":
%    PowerLaw2 model:
%      Integral = (1.6 +/- 0.4) e-6 cm^-2 s^-1
%      Index = 2.3 +/- 0.2
%      LowerLimit == 100 MeV
%      UpperLimit == 300000 MeV
%      TS = 49
%      F (0.1-300 GeV) = (1.6 +/- 0.4) e-6 cm^-2 s^-1
%      G (0.1-300 GeV) = (9.5 +/- 2.7) e-10 erg cm^-2 s^-1
%    PowerLawExpCutoff2 model:
%      Prefactor =  (1.5 +/- 2.3) e-9 cm^-2 s^-1 MeV^-1
%      Index1 = 0.9 +/- 0.9
%      Expfactor = (2.5 +/- 1.6) e-3
%      Index2 == 1
%      TS = 52
%      F (0.1-300 GeV) = (1.3 +/- 0.4) e-6 cm^-2 s^-1
%      G (0.1-300 GeV) = (6.7 +/- 1.5) e10 erg cm^-2 s^-1
%
%
Using \texttt{gtfindsrc} on this 18\,hour interval, we found a best-fitting position for the
$\gamma$-ray point source of 
(R.A., Dec.) = ($284\fdg17$,$16\fdg86$), 
offset from the nova optical position by $0\fdg20$, 
well within the 95\,per~cent confidence-level containment radius of $0\fdg26$.
Starting with the full-energy range 18-hour model, we constructed the SED of \nova{} 
by performing fitting in individual energy bands. The monochromatic flux values in Fig.~\ref{fig:latspec}
are shown for the bins where the nova was detected with ${\rm TS}\geq4$ and had at least four predicted counts, 
otherwise a 95\,per~cent confidence-level upper limit is reported.

No significant improvement (maximum $\Delta {\rm TS}=3$) was found by changing the source spectrum model 
to a curved log-parabola or power law with an exponential cut-off or hadronic model. 
For the hadronic model the integrated 0.1--300\,GeV flux is $(1.4 \pm 0.3) \times 10^{-6}$\,photons\,cm$^{-2}$\,s$^{-1}$ 
for a best fitting slope of the power law proton spectrum $3.0^{+0.6}_{-0.3}$. 
The hadronic model uses the method of \cite{2006ApJ...647..692K} to calculate the $\gamma$-ray spectrum 
due to the decay of neutral pions produced in proton-proton collisions. 
This model was applied to novae earlier by \cite{2010Sci...329..817A}, \cite{2014Sci...345..554A},
and \cite{2022ApJ...935...44C}.

While we cannot prefer an exponential cut-off over a simple power law on
the basis of the available photon data, we still perform the exponential cut-off
fit to facilitate comparison with previously detected LAT novae. 
For \nova{} the exponential cut-off is
found at $0.4 \pm 0.3$\,GeV, and the photon index $\Gamma = 0.9 \pm 0.9$ is
harder than what is predicted by the simple power law model.
For the exponential cut-off power-law model, the integrated 0.1--300\,GeV photon flux is 
$(1.3 \pm 0.4) \times 10^{-6}$\,photons\,cm$^{-2}$\,s$^{-1}$ equivalent to an
integrated energy flux of 
$(6.7 \pm 1.5) \times 10^{-10}$\,erg\,cm$^{-2}$\,s$^{-1}$.
The corresponding monochromatic flux 
%(SED point; \S~\ref{sec:thispaper}) 
(SED point; \S~\ref{sec:intro}) 
at 100\,MeV is 
$\nu F_\nu = 1.5 \times 10^{-10}$\,erg\,cm$^{-2}$\,s$^{-1}$ where we used 
equation~(3) of \cite{2022MNRAS.514.2239S} to convert the \textsc{fermitools}
%\texttt{PLSuperExpCutoff} or \texttt{PLSuperExpCutoff2} 
model parameters to $\nu F_\nu$.

% Short paragraph to discuss the gamma-ray to optical luminosity ratio (PJ)
For comparison with previous $\gamma$-ray novae, we computed the average $\gamma$-ray to optical luminosity ratio of 
\nova{} between day $t_0$+6\,hours and day $t_0$+24\,hours. The optical 
flux changes by a factor of two over that 18\,h time interval passing its peak. 
To estimate average optical flux over the time interval of the detected $\gamma$-ray emission 
we fitted a fireball model to the Evryscope (g-band), MLO-ASC and AAVSO ($V$-band and visual magnitudes) data assuming a blackbody temperature of
8000\,K. The fireball model calculates the lightcurve emitted by an expanding ionized ejecta \citep[e.g. see section 5 of][and references therein]{2017MNRAS.469.4341M}.
The resulting $\gamma$-ray to optical luminosity ratio ranges from 
$(2.1 \pm 0.4) \times 10^{-3}$ to $(2.7 \pm 0.5) \times 10^{-3}$ 
for a $\gamma$-ray spectral distribution described by an exponential cut-off power-law and a hadronic model, respectively. 
These ratios are similar to the ones obtained with other $\gamma$-ray novae 
\citep{2015MNRAS.450.2739M,2017NatAs...1..697L,2020NatAs...4..776A,2022ApJ...935...44C}. 

The analysis of \fermilat{} data over the time interval matching the {\em NuSTAR} observation (\S~\ref{sec:nustarobs}) 
results in non-detection with a 95\,per~cent upper limit on the photon flux of $< 4 \times 10^{-7}$\,photons\,cm$^{-2}$\,s$^{-1}$
(0.1--300\,GeV integrated energy flux less than $2.2 \times 10^{-10}$\,erg\,cm$^{-2}$\,s$^{-1}$.).

Finally, we tested for the presence of orbital and spin periodicities \citep{2022ApJ...940L..56P} 
in the $\gamma$-ray photon arrival times. For the test we selected 19 events within $5^\circ$ of \nova{} 
(recorded during the time it has been detected by \fermilat{}) and 
assigned to the nova with a \textsc{gtsrcprob} probability of at least 68\,per\,cent. 
% yes, per cent instead of % is the MNRAS house style
The event times were corrected to the Solar system barycentre with \textsc{gtbary}. 
We used \textsc{patpc} (\S~\ref{sec:nustarvar}) to compute the $H_m$ values 
for the orbital and spin periods and the corresponding probability of obtaining this $H_m$ value by chance. 
The events were found to be consistent with arriving at random phases for both trial periods (no periodicity found). 
This is in line with the results of \cite{2022MNRAS.517L..97L} who report no significant periodicity in the $\gamma$-ray data.

\subsection{VLA radio observations}
\label{sec:vla}

The National Radio Astronomy Observatory's Karl G. Jansky Very Large Array 
consists of 27 25\,m-diameter radio telescopes combined to form a connected interferometer
% 299792458/4/1e9 - 4m is 0.07 GHz
operating in a 0.07--50\,GHz frequency range \citep{1980ApJS...44..151T,2011ApJ...739L...1P}, 
see \cite{2017isra.book.....T} for a detailed discussion of radio interferometry techniques.
We observed \nova{} with the VLA 
% as of now - defined at first use in Intro
%Karl G. Jansky Very Large Array (VLA) 
at 15 epochs between 2021-06-15 ($t_0+3.2$\,d) and 2022-07-26 ($t_0+409.1$\,d). 
Most observations were performed while the array was in its C configuration (baseline range 0.035--3.4\,km). 
At each epoch a 1h45\,min observation was split between $S$, $C$, $K_u$ and $K_a$ bands. 
The data at each band were further split in two adjacent sub-bands to improve spectral resolution. 
The images were reconstructed independently at the following central 
frequencies: 2.6, 3.4, 5.1, 7.0, 13.7, 16.5, 31.1, 34.9\,GHz.
The VLA observing log is presented in Table~\ref{tab:vlaobslog}.

\begin{table}
\begin{center}
\caption{VLA observing log}
\label{tab:vlaobslog}
\begin{tabular}{cccccc}
\hline\hline
Epoch & Days & Date     & ID      & VLA   & Prim. \\
      & since $t_0$ &   &         & config. & calib.  \\
\hline
 1 &   3.2 & 2021-06-15 & SD1113  &  C & 3C\,48 \\
 2 &   4.2 & 2021-06-16 & SD1113  &  C & 3C\,48 \\
 3 &   5.2 & 2021-06-17 & SD1113  &  C & 3C\,48 \\
 4 &   9.2 & 2021-06-21 & SD1113  &  C & 3C\,48 \\
 5 &  10.1 & 2021-06-22 & SD1113  &  C & 3C\,48 \\
 6 &  13.1 & 2021-06-25 & SD1113  &  C & 3C\,286 \\
 7 &  15.2 & 2021-06-27 & SD1113  &  C & 3C\,48 \\
 8 &  31.0 & 2021-07-13 & SD1113  &  C & 3C\,286 \\
 9 &  44.1 & 2021-07-26 & SD1113  &  C & 3C\,48 \\
10 &  74.0 & 2021-08-25 & SD1113  &  C & 3C\,48 \\
11 &  83.0 & 2021-09-03 & SD1113  &  C & 3C\,48 \\
12 &  91.0 & 2021-09-11 & SD1113  &  C & 3C\,48 \\
13 & 142.8 & 2021-11-01 & 21B-351 &  B & 3C\,48 \\
14 & 315.4 & 2022-04-23 & 22A-169 &  A & 3C\,48 \\
15 & 409.1 & 2022-07-26 & 22A-169 &  D & 3C\,48 \\
\hline
\end{tabular}
\begin{flushleft}
\end{flushleft}
\end{center}
\end{table}

We used the quasar J1857$+$1624 (GB6\,B1855$+$1620; located $0\fdg48$ 
from \nova{}) as the complex gain 
calibrator. 3C\,48 was used to set the absolute flux density scale with the
exception of the observations on 2021-06-25 and 2021-07-13 when 3C\,286 was
used (Table~\ref{tab:vlaobslog}; \citealt{2017ApJS..230....7P}; c.f.~\citealt{1987A&A...183...38T}). 
For each observing epoch we produced two versions of the VLA schedule optimized 
for different local sidereal time ranges to facilitate dynamic scheduling. 
One version had 3C\,48 scans at the end of the experiment while the other started 
with the scans on 3C\,286. 
Following \cite{2021ApJS..257...49C} we expect the absolute flux density uncertainty to be
5~per~cent (10~per~cent) at frequencies below (above) 10\,GHz. As we rely on phase transfer
from the phase calibrator and do not attempt self-calibration (as the target
is rather weak), there might be an additional uncertainty as large as a few
tens of per~cent at high frequencies associated with imperfect phase
transfer in 
%sub-optimal weather conditions.
bad weather.

% Do you expect to see "We used standard CASA procedures for calibration and
% imaging"? No, it was much-much worse than that.

We relied on the remotely accessible computing resources of the NRAO's New Mexico Array Science Center \texttt{nmpost} cluster
for VLA data calibration and imaging. Specifically, we used \textsc{CASA 6.1.2} \citep{2007ASPC..376..127M} 
with the VLA \texttt{pipeline 2020.1.0.40} %\footnote{\url{https://science.nrao.edu/facilities/vla/data-processing/pipeline}} 
for calibration, 
\textsc{CASA 4.7.2} for writing out single sub-band multi-source \texttt{FITS} files that were loaded to 
\textsc{AIPS 31DEC21} \citep{2003ASSL..285..109G} for indexing and splitting into single-source \texttt{FITS} 
files suitable for imaging in \textsc{difmap 2.5e} \citep{1994BAAS...26..987S,1997ASPC..125...77S}.

The \texttt{CLEAN} \citep{1974A&AS...15..417H} imaging
was performed in \textsc{difmap}, 
which was also used for manually flagging the data affected by RFI or poor
system performance 
% no coma after e.g. is the MNRAS house style
(e.g. a warm receiver at a specific antenna). 
To identify the bad data that survived automated flagging by the \textsc{CASA} 
pipeline, we used \textsc{difmap}'s \texttt{radplot} command to inspect 
the correlated flux densities in the Stokes V, Q and U parameters as a function of
baseline length. These plots allow one to easily identify unusually noisy or highly 
polarized data points commonly associated with corrupted data. We then plot
the Stokes I correlated flux density as a function of time for each pair of
antennas (\texttt{vplot} with \texttt{vflags="1f"}) and identify groups of
flagged visibility measurements that affect the same frequency channels at
multiple antennas (RFI) or multiple channels at one antenna (receiver
problems) and flag all the visibilities in the affected groups of channels.

We tried two imaging strategies. First we \texttt{CLEAN}'ed the naturally weighted data by manually 
putting \texttt{CLEAN} windows around regions with visible emission, followed
by a full-map \texttt{CLEAN} once no visible emission remains in the residual map. 
The second approach was to employ a version of Dan Homan's
automated multi-resolution \texttt{CLEAN}'ing script\footnote{\url{http://personal.denison.edu/~homand/final_clean_rms}}
that \texttt{CLEAN}'s the full map first at super-uniform (\texttt{uvweight 10,-1}),
then uniform (\texttt{uvweight 2,-1}) and natural weighting (\texttt{uvweight 0,-2}). 
The script does not rely on manually placed \texttt{CLEAN} windows and performs no self-calibration.
The rationale behind the multi-resolution \texttt{CLEAN}'ing was discussed
by \cite{1999PhDT.........8M}, and the script we utilize is the one used 
for analyzing Very Long Baseline Array data in the framework of the MOJAVE
project \citep{2009AJ....137.3718L}. The results of the manual \texttt{CLEAN}'ing
at the natural weighting and automated multi-scale \texttt{CLEAN}'ing were
found to be very similar, with the multi-scale procedure typically resulting
in a slightly lower image noise, but on rare occasions, exaggerating a
pattern of stripes crossing the image that result from amplitude calibration issues
(for the datasets affected by this problem, manual \texttt{CLEAN}'ing
reduced the amplitude of the stripes).

The image pixel size was chosen so that we have at least five pixels across the half-power beam width at the observing frequency. The nova flux density was measured simply by taking the image peak value near the nova position. We checked that the nova is consistent with being a point source even at 34.9\,GHz by fitting variable-width Gaussian source model to the $uv$-data using \textsc{difmap}. 
At frequencies/epochs where no emission at the nova position is visible we report an upper limit computed as 
\citep[e.g.][]{2023arXiv230109116N}
\begin{equation}
{\rm UL} = \max(\,0,\,{\rm image\_value\_at\_nova\_position}) + 3 \times {\rm image\_rms}.
\label{eq:radioUL}
\end{equation}
The first term accounts for the possible presence of sub-threshold flux from the source. The image is in the units of surface brightness (Jy/beam), so we assume the target source remains unresolved even when invisible to interpret the result as an upper limit on the total flux density.

\begin{figure}
%                                                        left  bottom right  top
        \includegraphics[width=1.0\linewidth,clip=true,trim=0cm 4cm 0cm 0cm,angle=0]{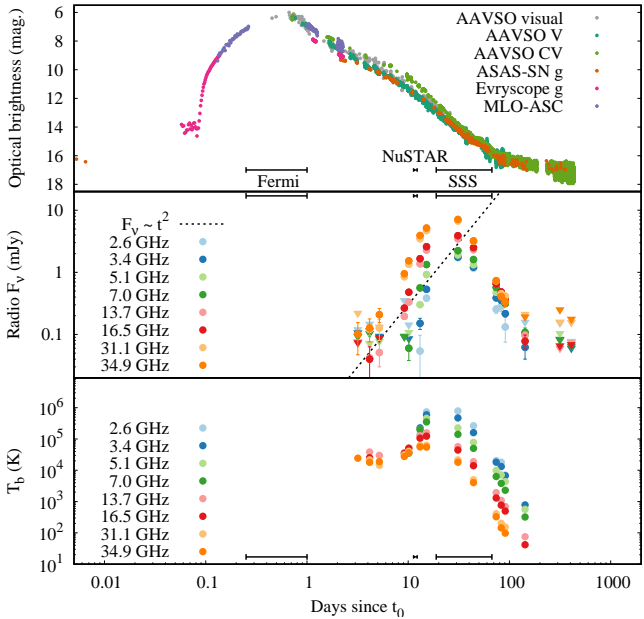}
    \caption{The optical (top panel) lightcurve of \nova{} compared to the
VLA radio lightcurve (middle panel) and the radio brightness temperature curve 
(bottom panel). The three horizontal bars indicate the time intervals when 
the $\gamma$-ray emission was detected by \fermilat{} (left bar),
the duration of the {\em NuSTAR} pointing (small bar in the middle) and 
the approximate duration of the SSS emission observed by {\em Swift}/XRT
(right bar; from \citealt{2021ApJ...922L..42D}). The dotted %dashed 
line indicates 
the rate at which a uniformly expanding optically thick thermal cloud would 
increase its flux density.}
    \label{fig:vlalc}
\end{figure}

\begin{figure*}
%                                                        left  bottom right  top
        \includegraphics[height=0.23\textwidth,clip=true,trim=0.0cm 0cm 0cm 0cm,angle=0]{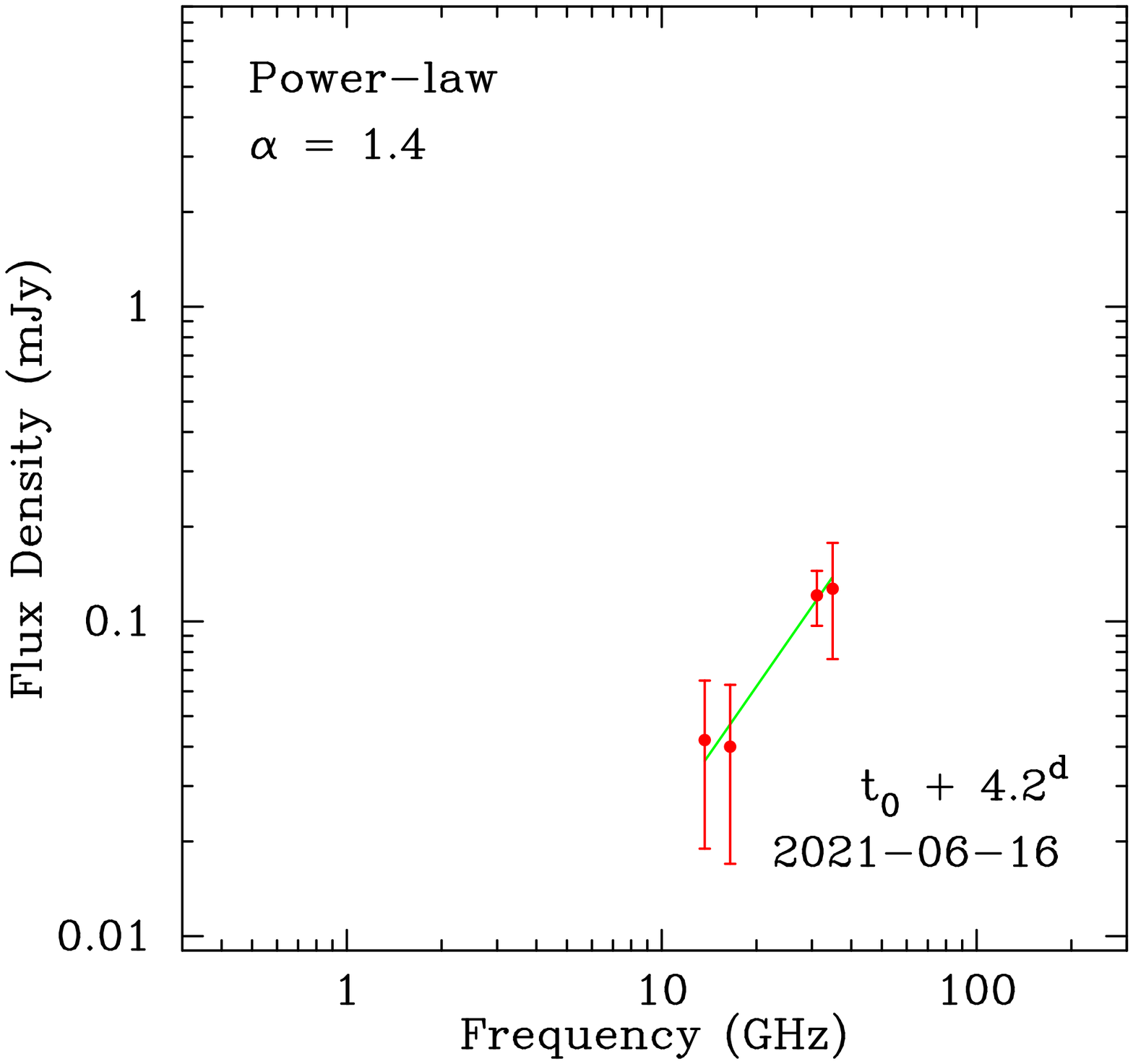}
        \includegraphics[height=0.23\textwidth,clip=true,trim=0.8cm 0cm 0cm 0cm,angle=0]{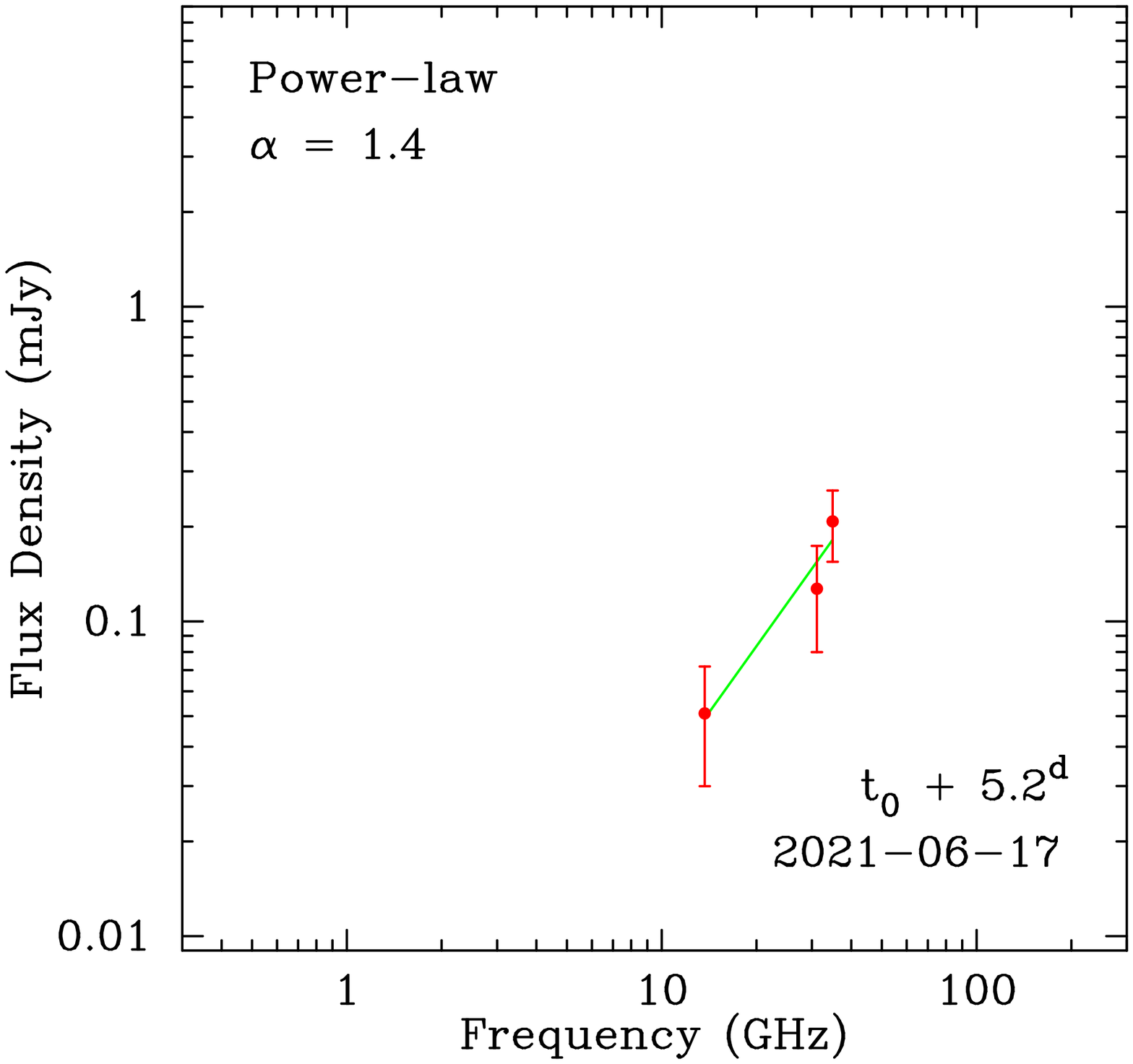}
        \includegraphics[height=0.23\textwidth,clip=true,trim=0.8cm 0cm 0cm 0cm,angle=0]{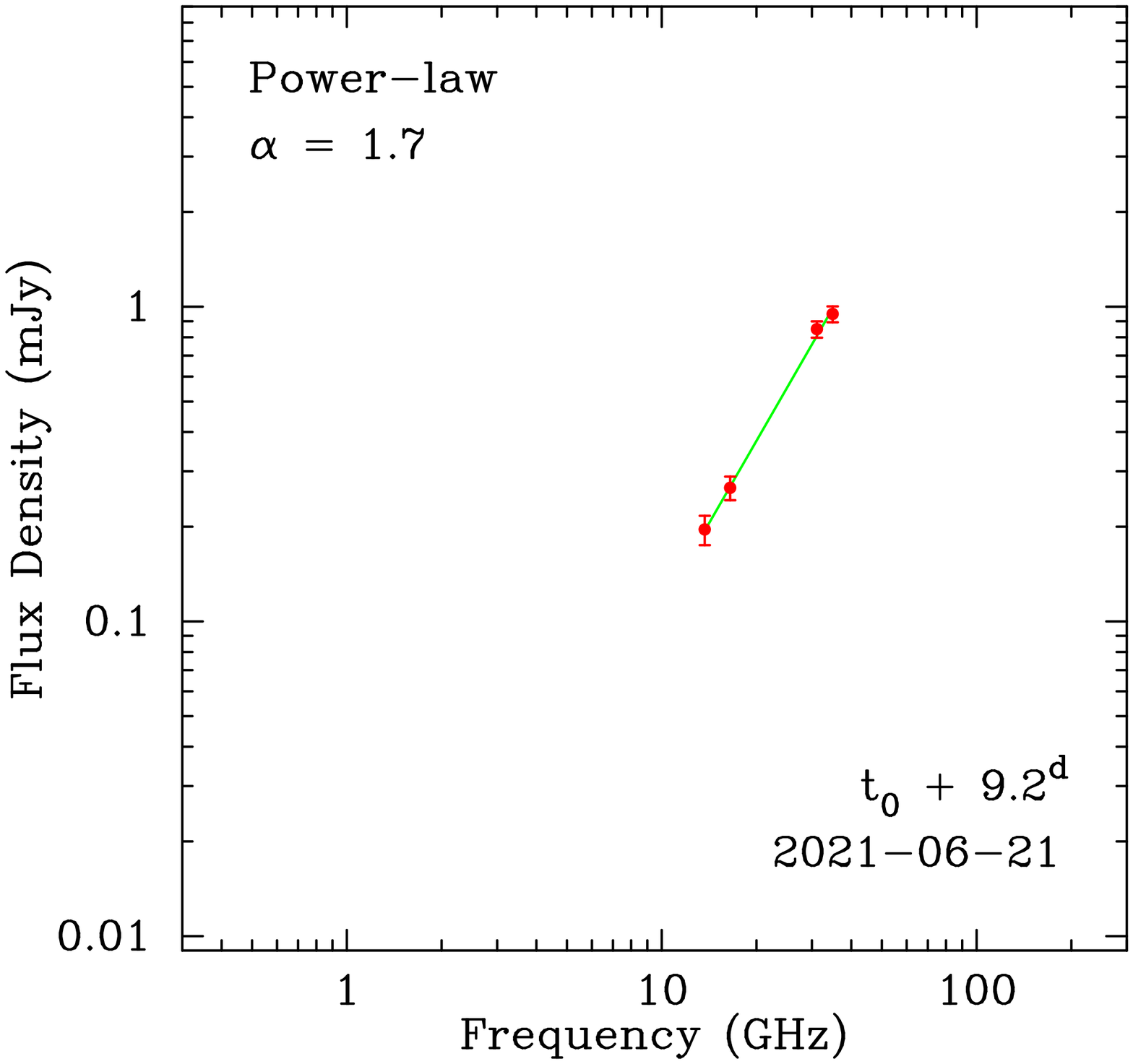}
        \includegraphics[height=0.23\textwidth,clip=true,trim=0.8cm 0cm 0cm 0cm,angle=0]{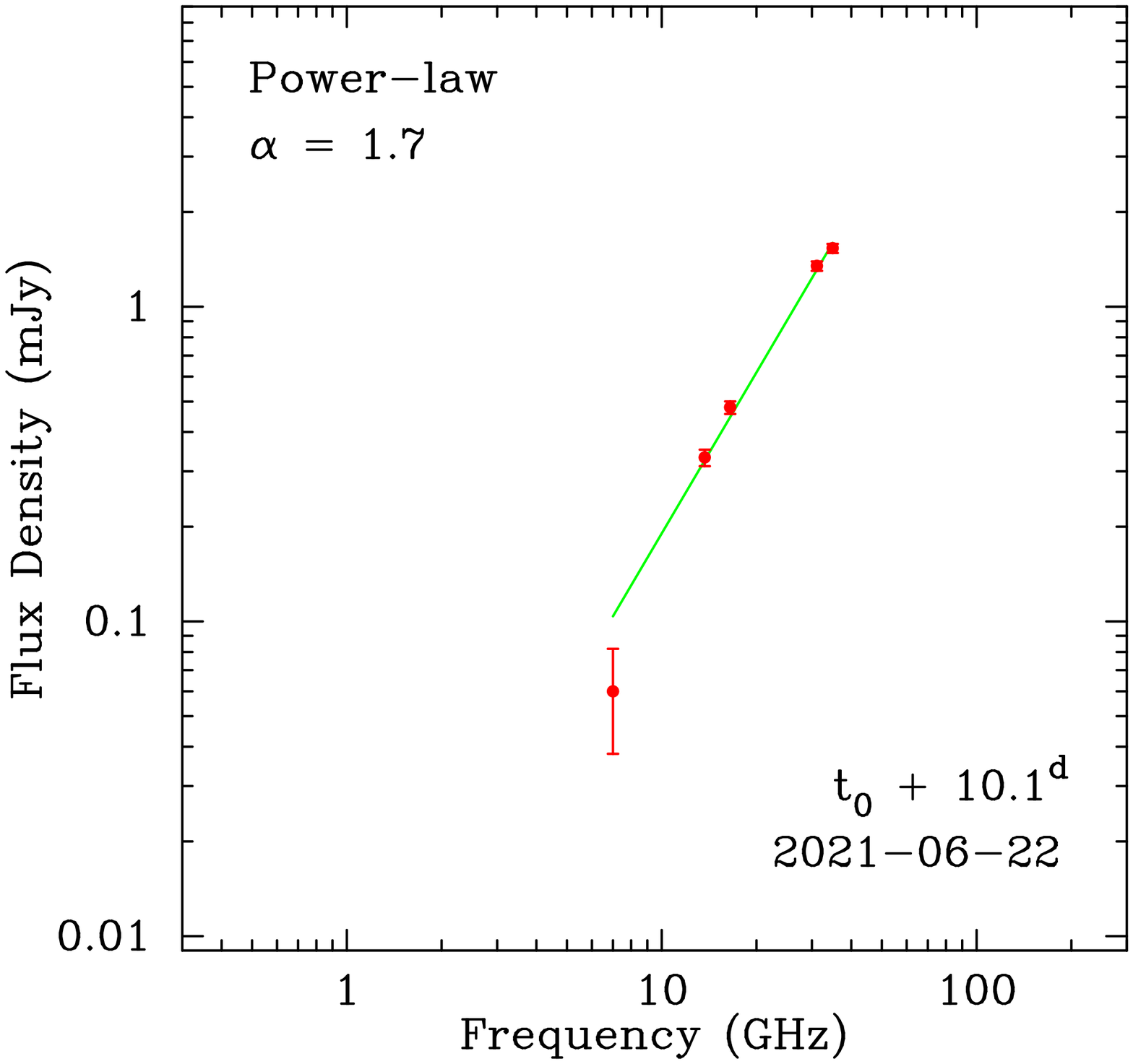}\\
\vspace{0.2cm}
%                                                        left  bottom right  top
        \includegraphics[height=0.23\textwidth,clip=true,trim=0.0cm 0cm 0cm 0cm,angle=0]{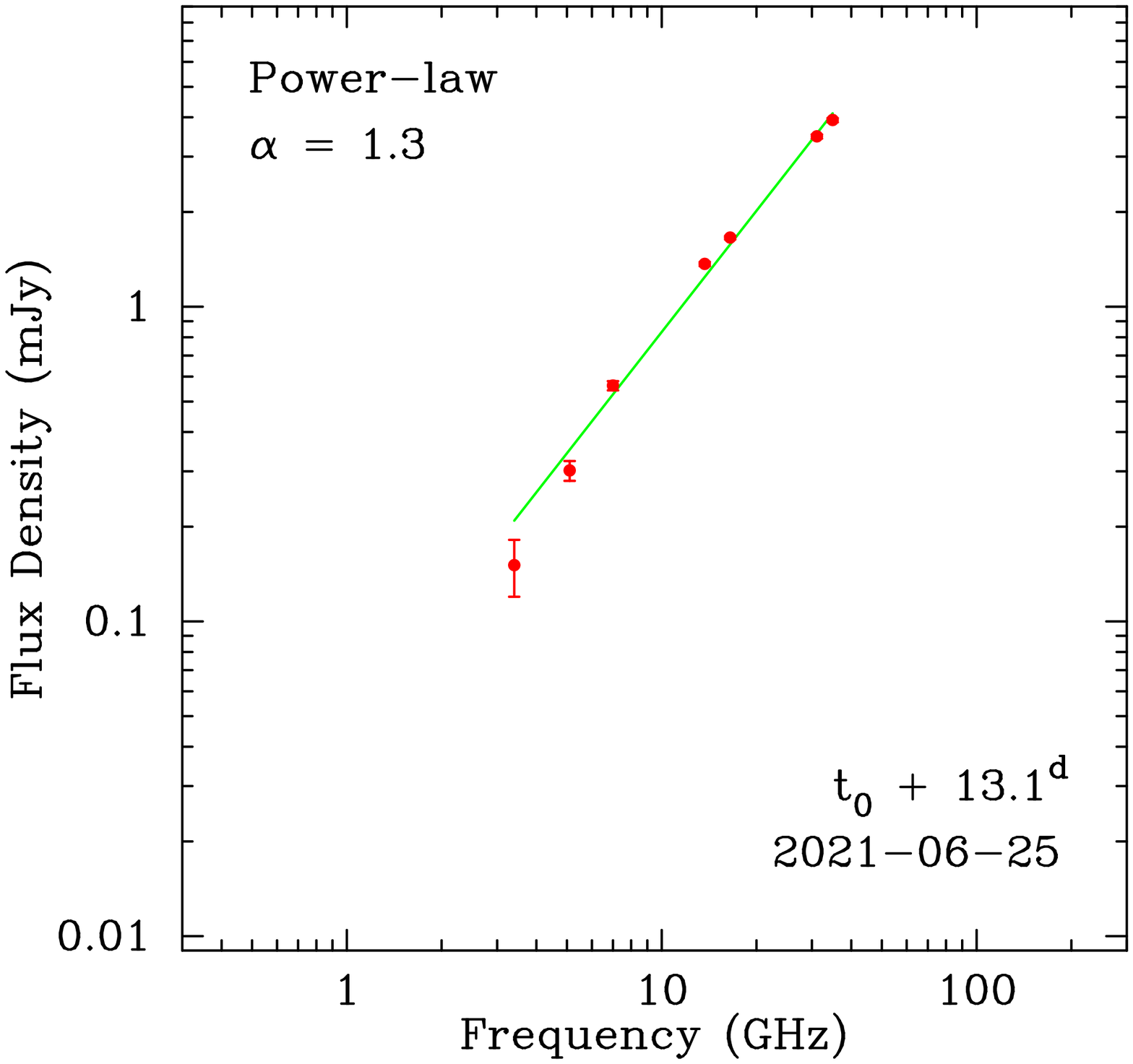}
        \includegraphics[height=0.23\textwidth,clip=true,trim=0.8cm 0cm 0cm 0cm,angle=0]{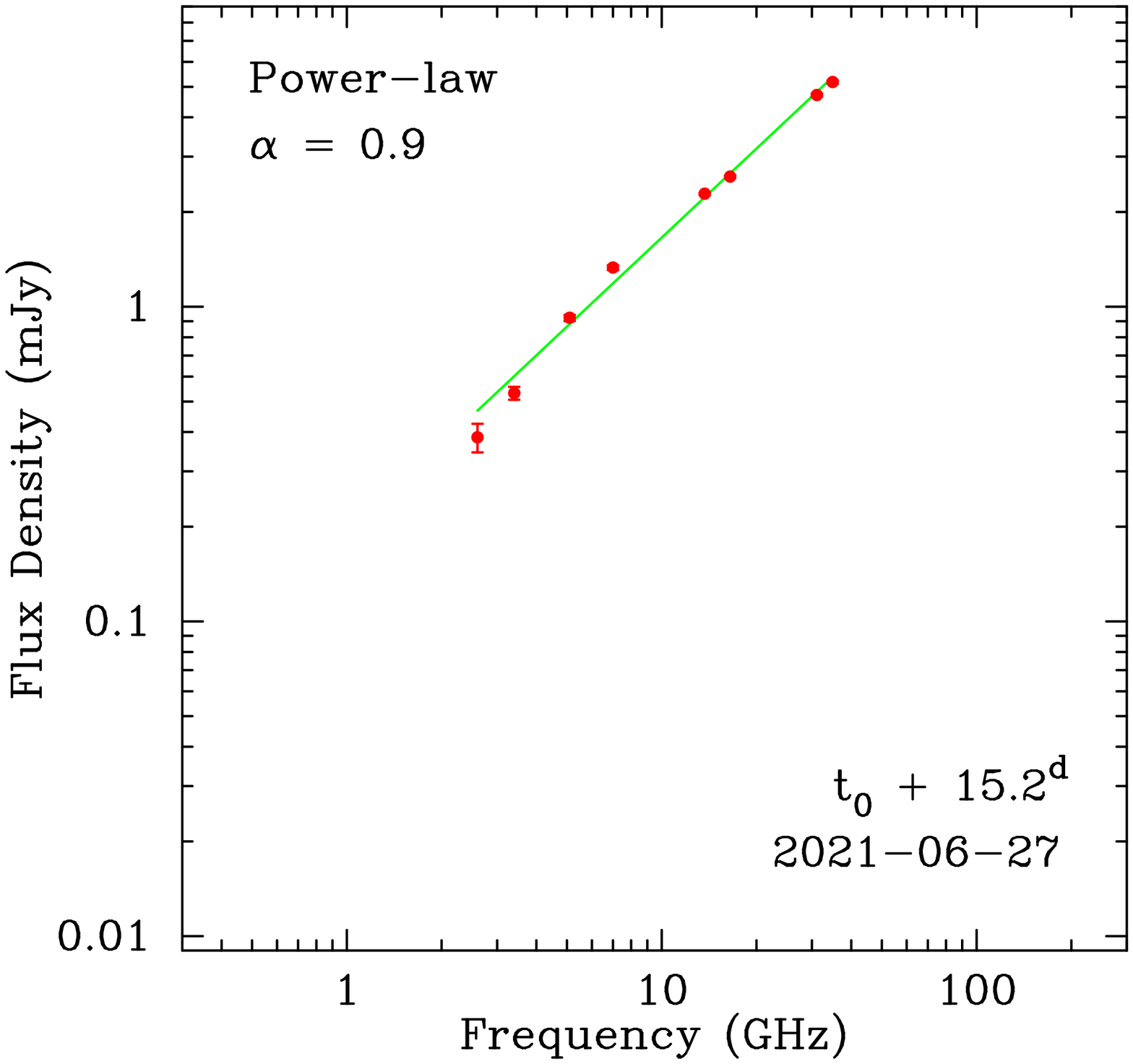}
        \includegraphics[height=0.23\textwidth,clip=true,trim=0.8cm 0cm 0cm 0cm,angle=0]{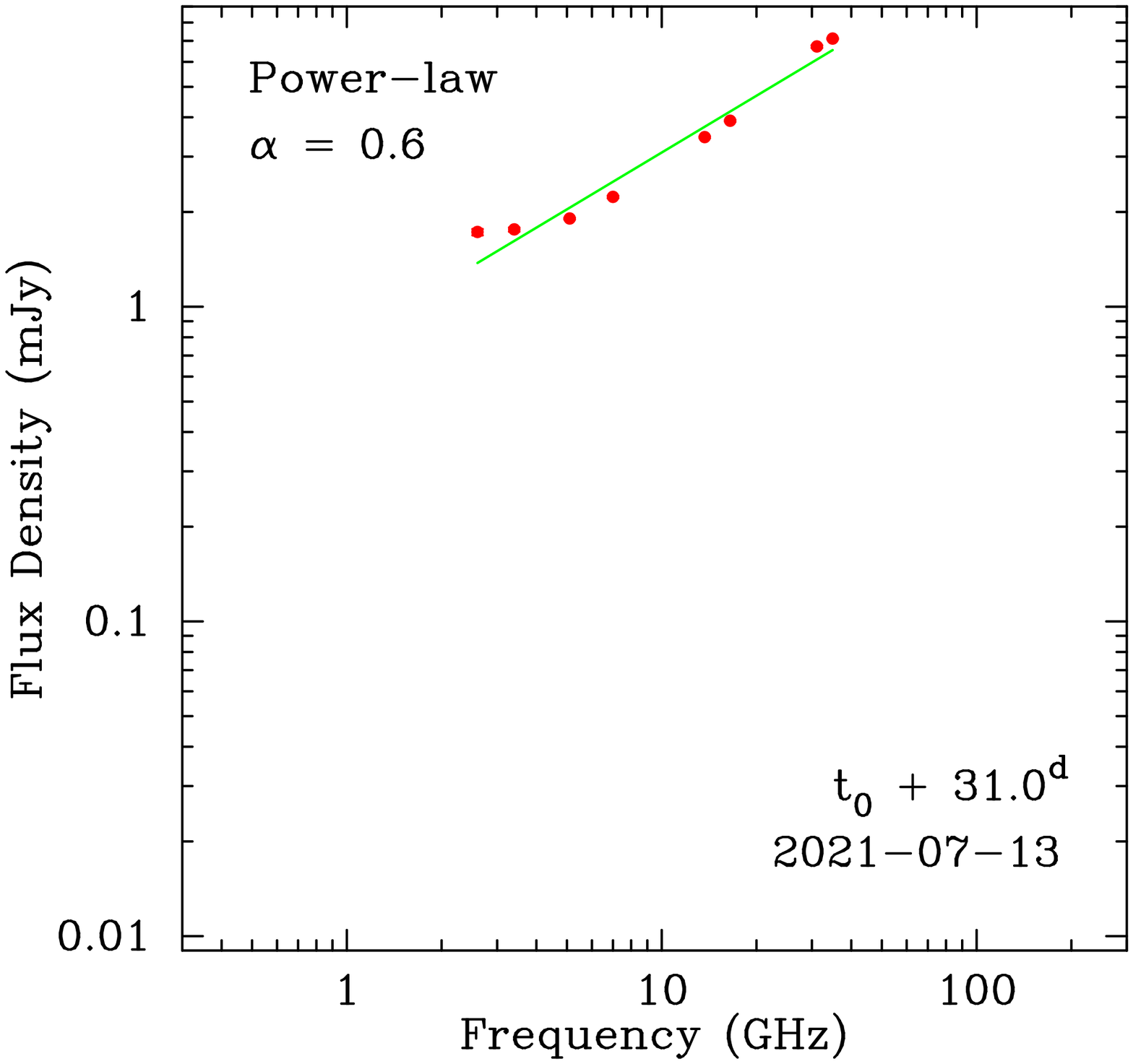}
        \includegraphics[height=0.23\textwidth,clip=true,trim=0.8cm 0cm 0cm 0cm,angle=0]{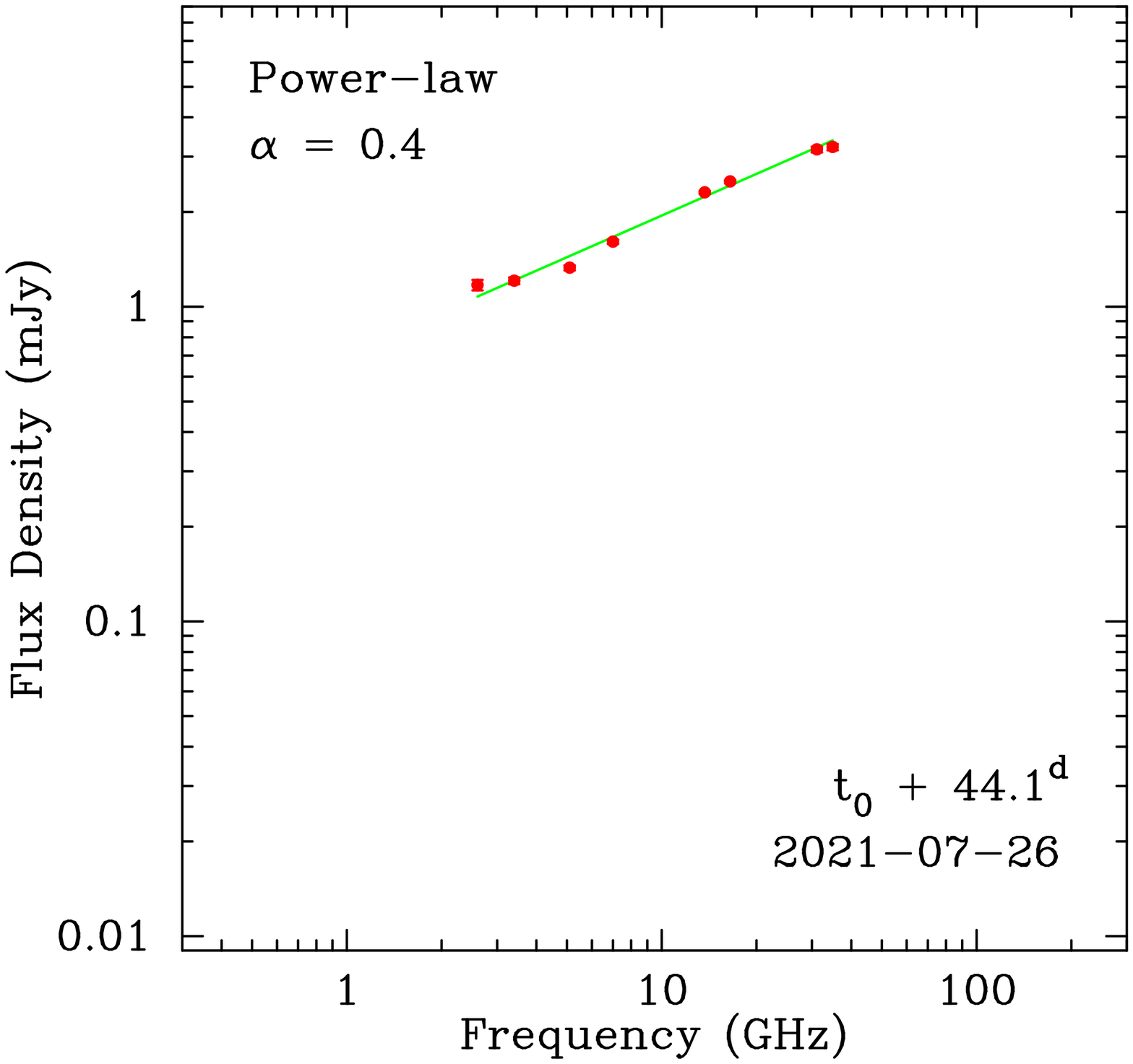}\\
\vspace{0.2cm}
%                                                        left  bottom right  top
        \includegraphics[height=0.23\textwidth,clip=true,trim=0.0cm 0cm 0cm 0cm,angle=0]{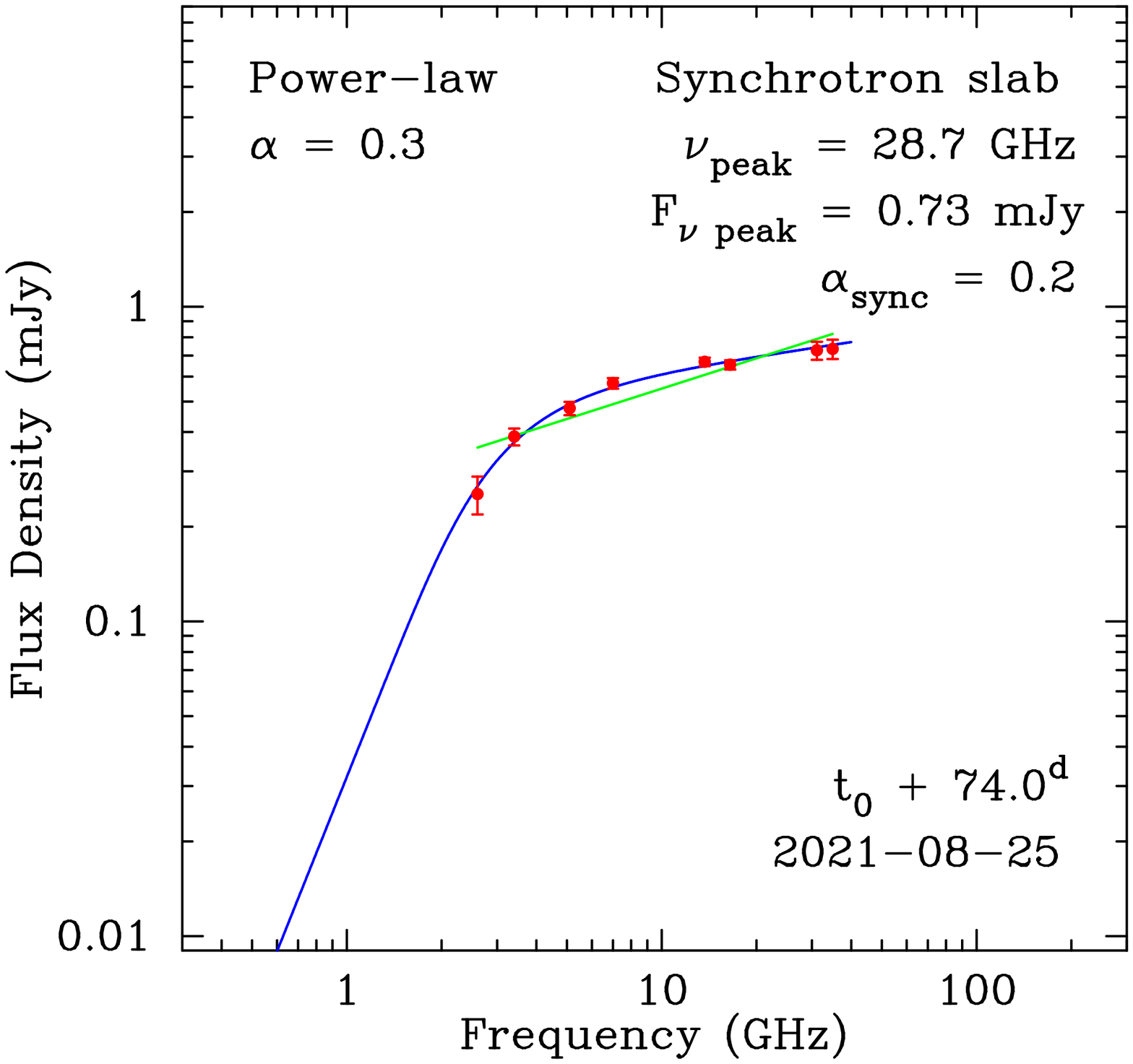}
        \includegraphics[height=0.23\textwidth,clip=true,trim=0.8cm 0cm 0cm 0cm,angle=0]{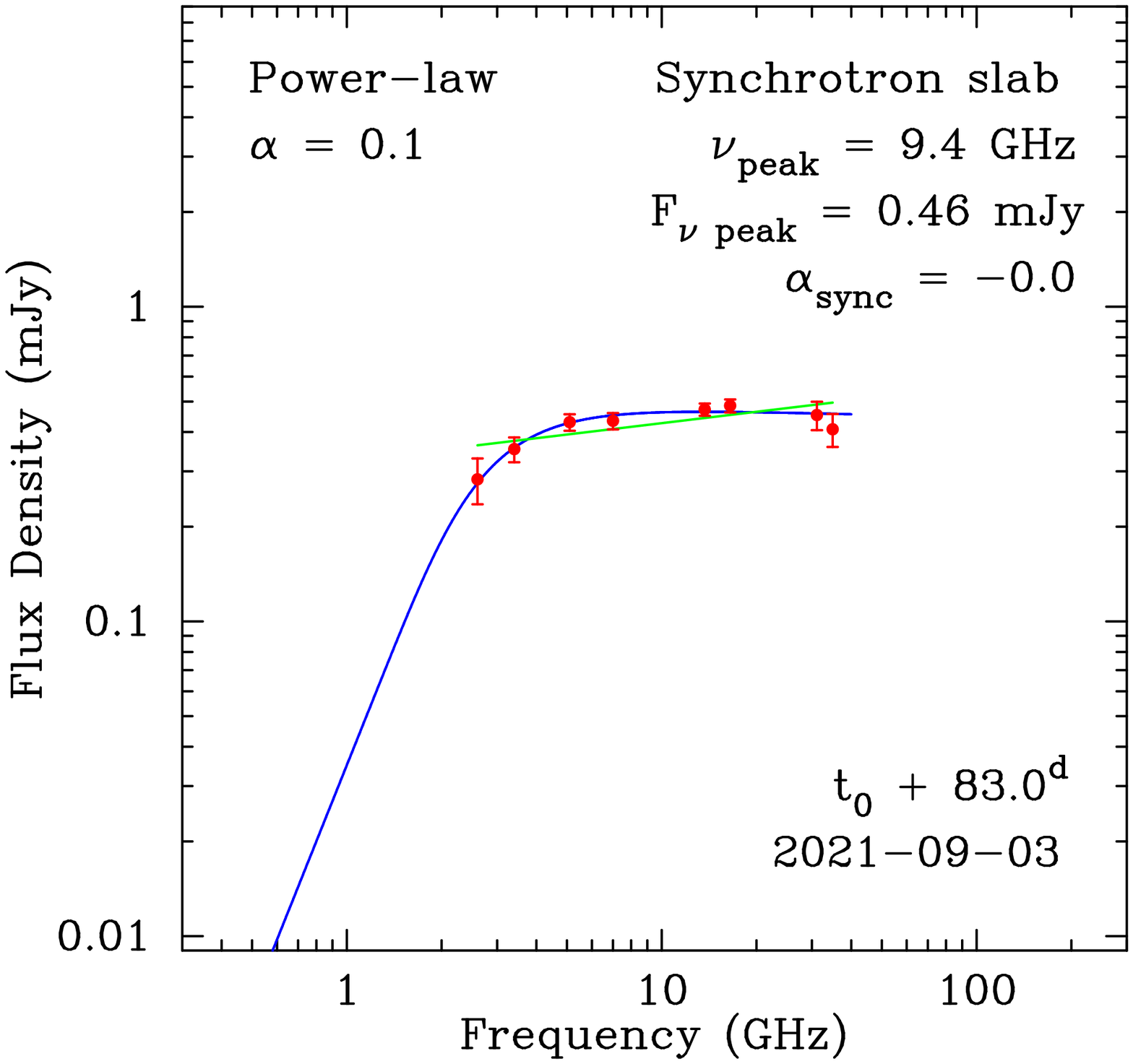}
        \includegraphics[height=0.23\textwidth,clip=true,trim=0.8cm 0cm 0cm 0cm,angle=0]{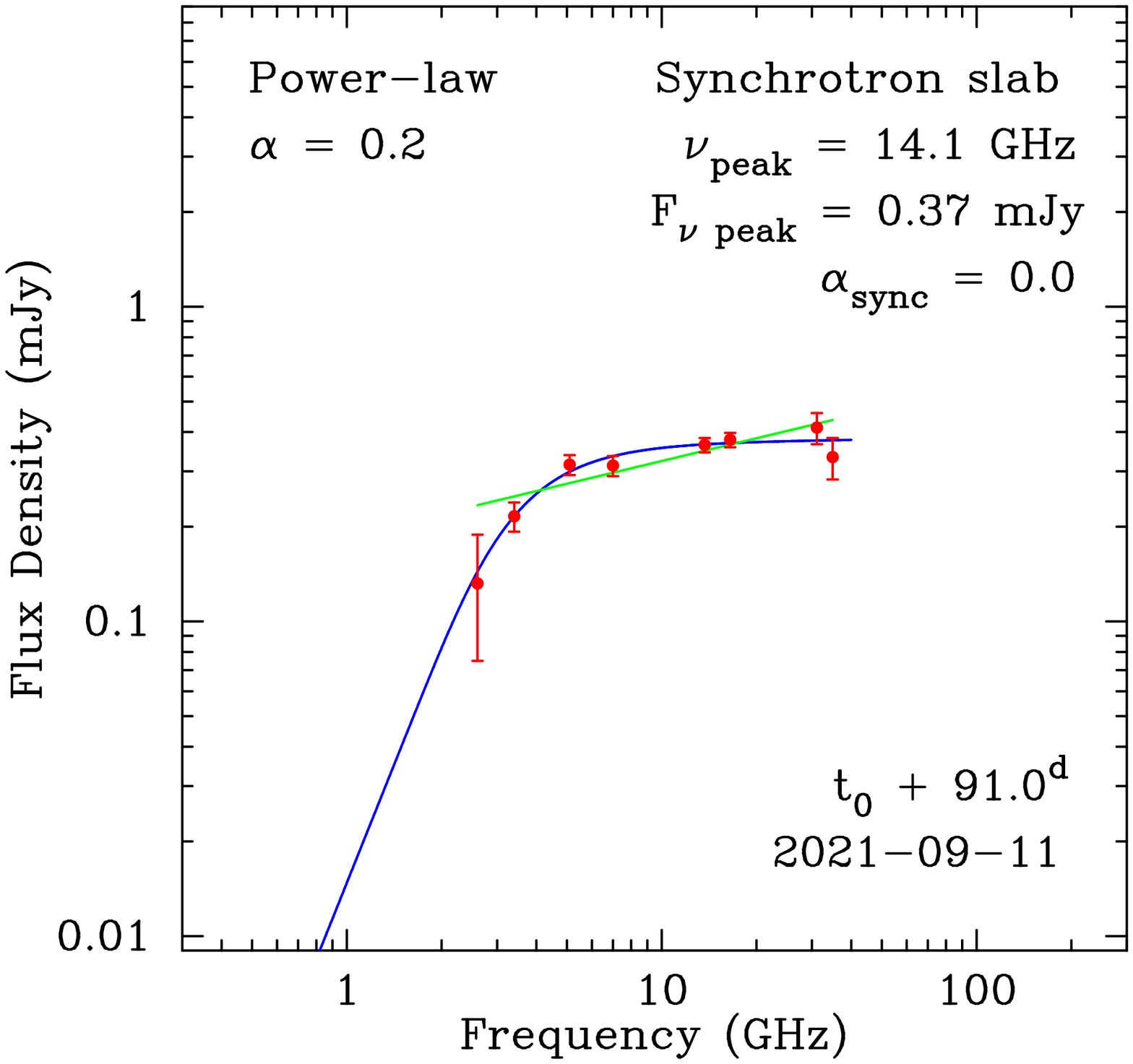}
        \includegraphics[height=0.23\textwidth,clip=true,trim=0.8cm 0cm 0cm 0cm,angle=0]{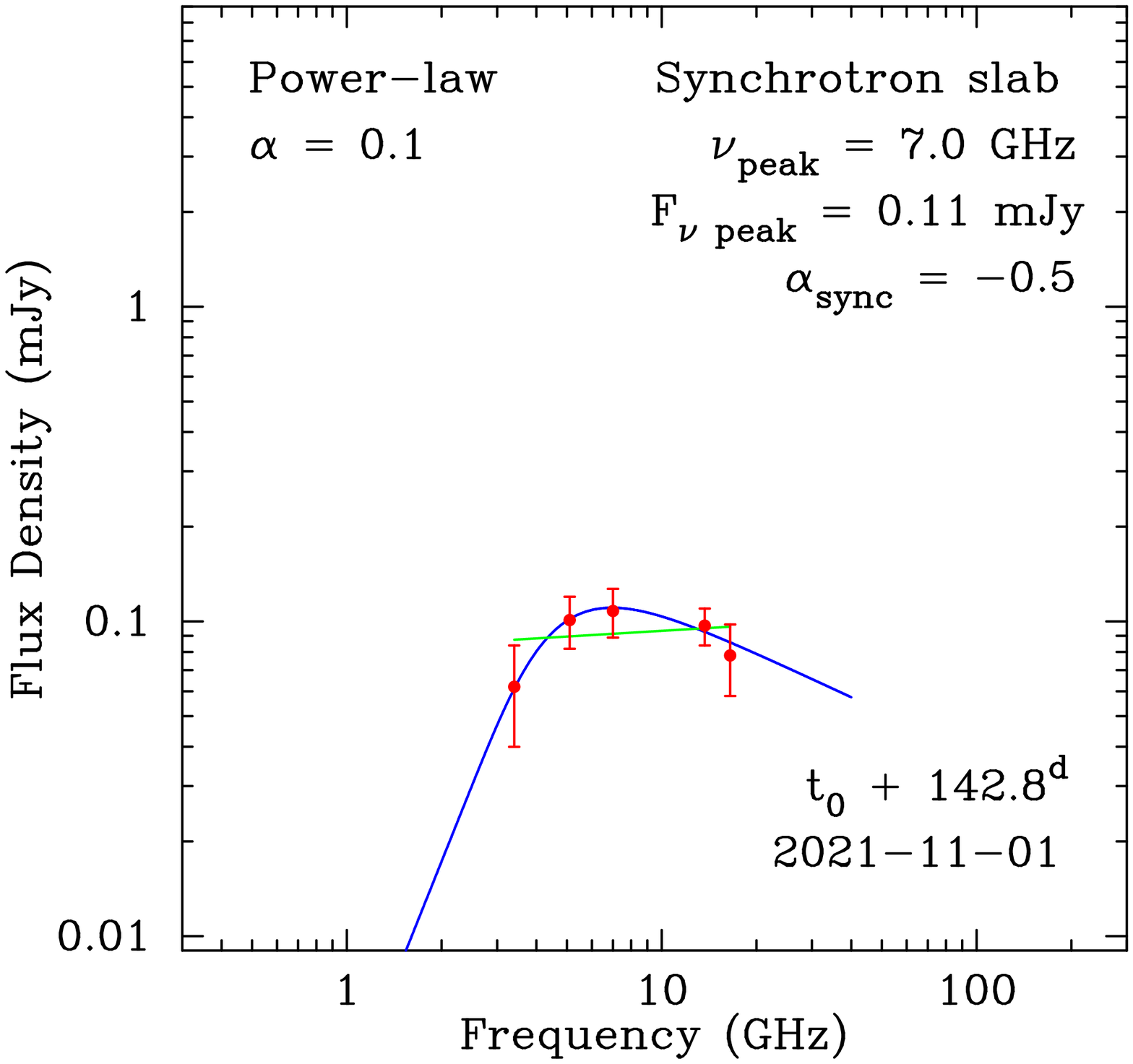}

    \caption{The evolution of the radio spectrum of \nova{}. The VLA flux
density measurements (red) are compared to the simple power-law fit (green
line) and a spectrum of a uniform synchrotron-emitting slab (blue curve). 
The uncertainty on the spectral index 
is $\sim0.3$ for the power law fits on 2021-06-16, 2021-06-17, 2021-11-01 and $\lesssim 0.1$ in all other cases. 
The synchrotron slab spectrum can approximate the observations only at late epochs. 
The spectrum shape is most likely determined by the non-uniform optical depth across the source.}
    \label{fig:vlaspec}
\end{figure*}

The VLA lightcurve of \nova{} is presented in Fig.~\ref{fig:vlalc} 
while the evolution of its radio spectrum is presented in Fig.~\ref{fig:vlaspec}. 
In the first observation at $t_0+3.2$\,d, the radio emission at the optical position 
of the nova is barely detected at 34.9\,GHz with a flux density of $0.10 \pm 0.05$\,mJy.
The next day ($t_0+4.2$\,d) brings a secure detection of the nova at four
high-frequency sub-bands revealing an inverted spectrum (Fig.~\ref{fig:vlaspec}). 
The flux densities continued increasing with time at all the observing frequencies,  peaking
around $t_0+31.0$\,d at the values ranging from $1.73 \pm 0.04$\,mJy
at 2.6\,GHz to $7.12 \pm 0.04$\,mJy at 34.9\,GHz (the quoted uncertainties
reflect the residual map noise and do not include the systematic effects
listed above). After the peak, the nova gradually fades, falling below the
detection limit at all bands after $t_0 + 142.8$\,d. The radio lightcurve appears
rather smooth with a single peak characterized by a fast rise and slower decay.
The peak and decline of the radio lightcurve coincide with the appearance of
SSS X-ray emission (Fig.~\ref{fig:vlalc}). 
As the eruption progresses, the radio spectrum gradually
flattens out, becoming somewhat curved at the latest epochs (Fig.~\ref{fig:vlaspec}).

\section{Discussion}
\label{sec:discussion}

\subsection{High-energy spectra}
\label{sec:specshape}

The featureless {\em NuSTAR} band emission, 
consistent with being produced by thermal plasma with non-solar abundances, 
is similar to that observed in classical novae previously detected by {\em NuSTAR}:
YZ\,Ret \citep{2022MNRAS.514.2239S} and V906\,Car \citep{2020MNRAS.497.2569S}. 
The {\em Swift}/XRT spectrum (obtained quasi-simultaneously with the {\em NuSTAR} pointing) 
is consistent with the thermal emission model attenuated by the Galactic absorption (\S~\ref{sec:swiftobs}). 
The first {\em NuSTAR}-detected classical nova V5855\,Sgr \citep{2019ApJ...872...86N} also
had a similar spectrum, but the photon statistics were too low to judge if
the emitting plasma abundances were super-solar. {\em NuSTAR} observations
of the other two classical novae, V339\,Del and V5668\,Sgr, resulted in
non-detections \citep{2018ApJ...852...62V}.

Two recurrent novae were also detected by {\em NuSTAR}: 
V745\,Sco \citep{2015MNRAS.448L..35O} and RS\,Oph
(\citealt{2021ATel14872....1L}, M.~Orio, 2022, private communication).
Unlike the classical novae, the spectra of both recurrent
novae show strong Fe~K$\alpha$ emission. This can be understood as V745\,Sco
and RS\,Oph both being `embedded novae' %--- they have 
with red giant donors. Much of the shocked material in embedded novae 
originates in the giant's wind that is likely to have nearly-solar
composition (but see \citealt{2019MNRAS.490.3691D} and \citealt{2022ApJ...938...34O}). 
On the contrary, in classical novae we expect no dense circumbinary material
\citep{2014ApJ...786...68H},
so shocks must be internal to ejecta. The nova ejecta includes a lot of 
white dwarf material, making its elemental abundances highly non-solar.

%     Gamma       Gamma_EC    EC(GeV)       Nova       Ref
%  2.37+/-0.09   1.84+/-0.23  2.3+/-1.0    V1369 Cen  2016ApJ...826..142C
%  2.42+/-0.13   2.27+/-0.25   9+/-14      V5668 Sgr  2016ApJ...826..142C
%  2.26+/-0.12                             V5855 Sgr  2019ApJ...872...86N
%  2.11+/-0.05   1.86+/-0.11  5.9+/-2.6    V5856 Sgr  2017NatAs...1..697L
%                1.92+/-0.16  7.7+/-4.7    V1324 Sco  2018A&A...609A.120F
%                1.50+/-0.28  1.3+/-0.5     V959 Mon  2018A&A...609A.120F
%                1.68+/-0.22  3.0+/-1.8     V339 Del  2018A&A...609A.120F
%                2.00+/-0.26  2.0+/-1.0    V1369 Cen  2018A&A...609A.120F
%  2.04+/-0.02   1.76+/-0.05  5.9+/-1.1     V906 Car  2020NatAs...4..776A
%  2.2+/-0.3                                V407 Lup  2021ApJ...910..134G
%  1.8+/-0.2                                V549 Vel  2021ApJ...910..134G,2020ApJ...905..114L
%  2.2+/-0.1                                V357 Mus  2021ApJ...910..134G
%  2.0+/-0.1                                V392 Per  2021ApJ...910..134G,2022arXiv220110644A 
%                1.59+/-0.16  1.9+/-0.6       YZ Ret  2022MNRAS.514.2239S
%
% Gamma
% MEDIAN= 2.200000
% weighted MEAN= 2.073729
% weighted MEAN_ERR= 0.045717
% weighted SD= 0.137150
% 5<sigma, 2>sigma, 2>2*sigma, 0>3*sigma
%
% Gamma_EC
% MEDIAN= 1.840000
% weighted MEAN= 1.785047
% weighted MEAN_ERR= 0.050524
% weighted SD= 0.151573
% 5<sigma, 3>sigma, 0>2*sigma, 1>3*sigma
%
% EC(GeV)
% MEDIAN= 3.000000
% weighted MEAN= 2.160681
% weighted MEAN_ERR= 0.526274
% weighted SD= 1.578822
% 5<sigma, 0>sigma, 2>2*sigma, 2>3*sigma

The typical nova GeV photon index value 
differs between the simple power-law and
exponential cut-off power-law models.
Summarizing the spectral fits for the 14 classical novae (excluding the
embedded nova V407\,Cyg) reported by 
\cite{2016ApJ...826..142C,2017NatAs...1..697L,2018A&A...609A.120F,2019ApJ...872...86N,2021ApJ...910..134G,2020ApJ...905..114L,2022MNRAS.514.2239S,2022arXiv220110644A}, 
for the simple power-law model the median photon index is $\Gamma = 2.20$ 
with a standard deviation of 0.14. Meanwhile for the exponential cut-off 
power-law, the median index is 1.84 with a standard deviation of 0.15 and 
the median cut-off energy is 3.0\,GeV (standard deviation 1.6\,GeV). 

The slope of the \fermilat{} spectrum of \nova{} (\S~\ref{sec:latobs}; Fig.~\ref{fig:latspec}) 
is consistent with the typical $\Gamma$ values cited above for the simple power-law model.
Though we do not find significant evidence for spectral curvature, 
a cut-off energy below 1\,GeV (if real) would be the smallest one to date among the {\em Fermi}-detected novae.
The cut-off would reflect the particle energy spectrum, expected at a photon energy of 0.1 of the maximum particle energy according to \cite{2016MNRAS.457.1786M}.
The shock velocity (estimated at a later epoch from the X-ray temperature, 
as described in \S~\ref{sec:shocklocation}), GeV luminosity (\S~\ref{sec:distlum}), 
and the low cut-off energy would place the \nova{} shock in a region of
surprisingly high density of $>10^{11}\,{\rm cm}^{-3}$
\citep[see Fig.~3 of][]{2016MNRAS.457.1786M}.
We stress that there is no evidence for the existence of a cut-off in 
the $\gamma$-ray spectrum of \nova{}. The simple power law model provides an acceptable fit. 
We consider the model that includes the cut-off because it was preferred for other, brighter $\gamma$-ray novae.

In \S~\ref{sec:xrayemittermass} we make a crude estimate of the density of
the X-ray emitting plasma 11 days after eruption to be $\sim10^{6}\,{\rm cm}^{-3}$.
This estimate assumes that the X-ray plasma uniformly fills the whole volume of the ejecta.
For a uniform expansion the volume should increase as the third power of
time, so 10 days earlier, when the $\gamma$-ray emission was detected, the
same plasma could have a factor of 1000 higher density. To reconcile the
X-ray density estimate with the high density suggested by the low-energy
cut-off in the $\gamma$-ray spectrum, we can assume that the X-ray emitting material 
should occupy less than 1\,per~cent of the ejecta volume. Considering that
the X-ray emitting plasma may be confined in a thin shell that may cover only
a fraction of a sphere, such an assumption does not seem impossible.
Alternatively, rather than being an intrinsic feature of particle energy 
spectrum, the GeV cut-off may result from opacity (\S~\ref{sec:comfastslow}).

\subsection{Shock location}
\label{sec:shocklocation}

Shocks in classical novae may result from 
parts of the nova shell being ejected at different times and 
at different speeds. 
There is evidence of multiple ejections 
in optical spectra of novae \citep{1987Msngr..50....8D,2020ApJ...905...62A}. 
The exact mechanism by which novae eject their envelope is debated
\citep[sec.~2.2 of][]{2021ARA&A..59..391C}. 
As outlined in 
%\S~\ref{sec:innovashock}, 
\S~\ref{sec:intro},
it has been 
proposed both on observational \citep{2014Natur.514..339C} and theoretical
\citep{1990LNP...369..342L,2022ApJ...938...31S} grounds that a nova
eruption includes two phases of mass loss driven by different mechanisms: 
\begin{enumerate}
\item The initial ejection of the common envelope formed by the expanded white
dwarf atmosphere that engulfs the binary. The envelope is ejected by the
binary motion and is concentrated towards the orbital plane of the binary
\citep{1980MNRAS.191..933M,1990ApJ...356..250L,1997MNRAS.284..137L}.
\item The fast radiation-driven wind from the hot nuclear-burning white dwarf
\citep{1994ApJ...437..802K,2011A&A...536A..97F}.
\end{enumerate}
A similar interacting wind model was proposed to explain the shapes of planetary nebulae
(\citealt{1982ApJ...258..280K}, \citealt{1989ApJ...339..268S}, but see \citealt{2002ARA&A..40..439B}). 
The fact that the slow outflow can confine the fast wind suggests that 
the slow outflow carries most of the ejected mass -- a conclusion confirmed by \cite{2022ApJ...938...31S}.
Confinement of the initially spherically symmetric fast wind by the dense 
orbital-plane-concentrated flow gives rise to the bipolar (dumbbell-shaped) morphology
often inferred for nova ejecta
(e.g.~\citealt{2013MNRAS.433.1991R,2014ApJ...792...57R,2016AstL...42...10T,2022MNRAS.511.1591T,2022ApJ...932...39N}. 
but see also counter-examples presented by \citealt{2022MNRAS.512.2003S}).
The interface between the slow orbital-plane-concentrated flow and the fast wind is a natural shock formation site.
Alternatively, the shock may be located close to
the white dwarf, as suggested by the report of 544.8\,s periodicity in the $\gamma$-ray
emission of V5856\,Sgr by \cite{2022ApJ...924L..17L}.

The absence of fast variability in the {\em NuSTAR} X-ray data on \nova{} 
dominated by shock-heated plasma (\S~\ref{sec:nustarvar}) suggests that the shocked
region is large and located far away from the white dwarf.
The shock velocity ($v_{\rm shock}$) can be related to the post-shock temperature:
\begin{equation}
{\rm k} T_{\rm shock} = \frac{3}{16} \mu m_p v_{\rm shock}^2
\label{eq:ktshock}
\end{equation}
(equation~[6.58] of \citealt{1997pism.book.....D}),
where $m_p$ is the proton mass, $k$ is the Boltzmann constant, and $\mu$ is the mean molecular weight. 
For a fully ionized gas with the abundances derived for V906\,Car by \cite{2020MNRAS.497.2569S},
$\mu=0.74$. Together with ${\rm k} T_{\rm shock} = 4$\,keV (Table~\ref{tab:nustarspecmodels}),
this corresponds to $v_{\rm shock} \simeq 1700$\,km\,s$^{-1}$.
Taking 10\,ks as the variability time-scale in the {\em NuSTAR}
band (Fig.~\ref{fig:nustarlc}), the corresponding length-scale of the shocked
region would be $10^{12}$\,cm or 0.1\,a.u. It is larger than the white dwarf
and the accretion disc (\citealt{2018A&A...613A...8F}; these could, in principle, be a site of shock formation).

Association of all the hard X-ray emission observed by {\em NuSTAR} with the
shock within the nova ejecta is natural in the context of other {\em NuSTAR}-observed novae, 
but is not a trivial conclusion given the IP nature of \nova{}. 
The IPs are known for optically thin, spin-modulated X-rays \citep{1989MNRAS.237..853N} 
produced in the post-shock region of their accretion columns just above the white dwarf surface.
In the case of \nova{}, however, the spin modulation is detected only in the optically-thick, supersoft component. 
Accretion powered X-rays of IPs are usually seen to be spin-modulated
\citep{2015ApJ...807L..30M}, although not always \citep[see e.g.][]{2022MNRAS.511.4582C}. 
This may be simply due to the well-known decrease in IP spin modulation amplitude with 
increasing photon energy \citep{1989MNRAS.237..853N}.

The reason for the X-ray modulation in \nova{} being confined to soft 
X-rays might be that, as the white dwarf atmosphere expanded following 
the nova eruption, 
the shock within the accretion column formed further away 
from the white dwarf resulting in lower velocity reached by 
the infalling material and hence a lower post-shock temperature.
Another possibility is enhanced Compton cooling of the accreting material 
\citep{2002apa..book.....F,2011ApJ...737....7N}  
facilitated by the dense radiation field near the surface of 
the hydrogen-burning white dwarf.

Depending on the relative masses of the accretion disc and nova ejecta, the
accretion disc may survive the eruption or be completely swept away
\citep{2010ApJ...720L.195D,2018A&A...613A...8F}.
The periodic variations were missing in both optical and X-ray bands near the nova peak
\citep{2021RNAAS...5..244H}, either as a result of temporarily arrested accretion
or due to the expanded photosphere engulfing the binary system outshining 
and obscuring the effects of accretion. 
The orbital and spin modulations emerged in the optical band on day 4 and day 12 after the
eruption, respectively \citep{2022ApJ...940L..56P}. If we interpret the
spin modulation as the result of accretion (like in the non-nova IPs), that
would mean the accretion has restarted by the time of our {\em NuSTAR}
observation.
The super-soft X-ray emission modulated with the white dwarf spin period
appeared around $t_0 + 18.9$\,d \citep{2021ATel14747....1P,2021ApJ...922L..42D}, 
after our {\em NuSTAR} observation. 

%%% How do we know Fe lines in Drake's spectrum are from donor??
From the X-ray grating spectroscopy by \cite{2021ApJ...922L..42D}, we know 
the donor star has non-zero Fe abundance. Therefore, we would expect to see strong 
Fe lines in the 6--7\,keV range (commonly found in IPs; 
\citealt{2018MNRAS.476..554S}, \citealt{2018ApJ...852L...8L},
\citealt{2022MNRAS.511.4582C}, \citealt{2022A&A...657A..12J}) had the shocked plasma been accreted from the donor.
One could also expect strong intrinsic absorption for accretion-powered X-rays of IPs.
In summary, the X-ray emission of \nova{} observed by {\em NuSTAR} is likely associated 
with a shock within the nova ejecta, not accretion on the magnetized white dwarf.

\subsection{Distance and luminosity}
\label{sec:distlum}

The progenitor of nova \nova{} has a negative parallax value reported in 
the {\em Gaia} Early Data Release~3, so the distance of $6.0^{+3.8}_{-2.8}$\,kpc 
listed by \cite{2021AJ....161..147B} reflects the prior constructed from a three-dimensional model of 
the Galaxy rather than the actual geometric distance measurement. 
Therefore, we have to rely on indirect distance indicators to determine 
the nova luminosity (results summarized in Table~\ref{tab:luminositiestable}).

\nova{} is an IP, and according to \cite{2022ApJ...940L..56P} it brightened from $V=20.5$. 
Using $A(V)=1.7$ mag, the dereddened quiescent magnitude is then $V_0 = 18.8$. In comparison, 
a normal IP with an orbital period of 3.67\,h (\S~\ref{sec:thisnova}) has a typical 
$M_V = 4.8$ (with a scatter of 1 magnitude; the majority of IPs are near the absolute 
$V$ magnitude -- orbital period relationship for dwarf novae in outburst of
\citealt{1987MNRAS.227...23W}; Mukai et al. in prep.). This implies a distance of $6.3^{+3.8}_{-2.4}$\,kpc. %, indicated. 
A rare low-luminosity IP (that would correspond to the dwarf-novae-in-quiescence relation of \citealt{1987MNRAS.227...23W}) with $P_{\rm orb}=3.67$\,h
would have $M_V \sim 8.5$, placing it at 1.1\,kpc. 
%
% Elias wrote specifically:
% "From Galactic reddening maps and E(B-V) = 0.55. I get distance between 5 and 6+ kpc
% There seems to be a good chunk of dust between 5 and 6 kpc"
%
Such a small distance is inconsistent with an estimate based on optical
extinction: adopting $E(B-V)=0.55$ mag reported by \cite{2021ATel14704....1M}, we use 
the 3D Galactic dust map of \cite{2016ApJ...818..130B} to estimate the
distance to \nova{}, $> 5$\,kpc.
%We consider the low-luminosity IP possibility less
%likely as they are more rare than the high-luminosity ones.
The $\sim 6.3$\,kpc distance is also consistent with the angular diameter of
the ejecta measured at $t_0+2$\,d and $t_0+3$\,d with the CHARA optical
interferometer (Gail Schaefer, private communication; \citealt{2005ApJ...628..453T,2020SPIE11446E..05S}).
We note that \cite{2021ApJ...922L..10W} estimated a smaller distance of 4.7\,kpc based on the purported relationship between
the absolute magnitude at maximum and the rate of the lightcurve decline for novae, 
\cite{2022MNRAS.517.6150S} reported the distance of $3.2^{+2.1}_{-0.8}$\,kpc
by combining the negative {\em Gaia} parallax with complex nova-specific priors, 
while \cite{2021ApJ...922L..42D} adopted a nominal distance of 5\,kpc in their analysis.
%Throughout this paper w
We prefer the 6.3\,kpc distance based on the expected IP host
magnitude, preliminary CHARA expansion parallax, and  
the lower limit from the 3D Galactic extinction model.

At 6.3\,kpc distance, the \nustarenergylow{}--\nustarenergyhigh{}\,keV
luminosity of \nova{} is \nustarlum{}. 
The {\em NuSTAR} band luminosity of \nova{} is comparable to that of V5855\,Sgr \citep{2019ApJ...872...86N} and V906\,Car \citep{2020MNRAS.497.2569S} 
and is an order of magnitude larger than the luminosity of YZ\,Ret \citep{2022MNRAS.514.2239S}.
% These things tend to vary
Only an order-of-magnitude comparison is appropriate here, as these
novae only have one (two for V906\,Car) {\em NuSTAR} snapshots, 
while the flux in the {\em NuSTAR} band is expected to vary with time.

%%%%%%%%%%%%%%%%%%% Optical luminosity %%%%%%%%%%%%%%%%%%%%%%%%%%

% I would like to have all the luminosity values only in the table, but I was told to also cite them in the text...
Adopting the brightest visual estimate 6.0\,mag. as the peak magnitude of \nova{} and assuming it to be equivalent to $V$ 
we follow the procedure detailed by \cite{2022MNRAS.514.2239S} to estimate the peak bolometric 
flux per unit area \citep{2015arXiv151006262M} 
\begin{equation}
f = 2.518 \times 10^{-5} \times 10^{-0.4 (V+{\rm BC}-A_V) } \,{\rm erg}\,{\rm cm}^{-2}\,{\rm s}^{-1},
\label{eq:bolometricflux}
\end{equation}
where BC is the bolometric correction (for the nova at peak we adopt ${\rm BC} = -0.03$). 
At the abovementioned distance this translates to a bolometric luminosity of \opticalpeaklum{}
(absolute magnitude $M = -9.7$). 

Taking $V=11.8$ during the {\em NuSTAR} observation 
(Fig.~\ref{fig:nustarlc}), applying a bolometric correction 
of $-2.36$ for the {\em Swift}/UVOT derived temperature (\S~\ref{sec:swiftobs}) 
according to table~3.1 of \cite{2007iap..book.....B}, 
and correcting for extinction (\S~\ref{sec:thisnova}), 
we estimate a bolometric luminosity of \opticalnustareplum{}. 
Due to the combined uncertainties in magnitude 
(typically 0.1\,mag. for visual and 0.02\,mag for CCD measurements, 
real variability over the {\em Fermi} and {\em NuSTAR} exposures), 
distance, bolometric correction and the magnitude-to-flux conversion, 
the combined uncertainty on the luminosity is expected to be at 
the tens of per~cent level.
We also compute monochromatic fluxes at 5500\,\AA (2.25\,eV):  
$\nu F_\nu = 3.8 \times 10^{-7}$\,erg\,cm$^{-2}$\,s$^{-1}$ (peak) and
$\nu F_\nu = 1.8 \times 10^{-9}$\,erg\,cm$^{-2}$\,s$^{-1}$ ({\em NuSTAR} epoch)
using the magnitude zero points from \cite{1998A&A...333..231B}.

\begin{table}
\begin{center}
\caption{\nova{} luminosity}
\label{tab:luminositiestable}
\begin{tabular}{rc}
\hline\hline
Band       & Luminosity \\
%           & erg\,s$^{-1}$      \\
\hline
\multicolumn{2}{l}{$\gamma$-ray bright epoch near the optical peak:} \\
18\,h integration 0.1--300\,GeV & \latpeaklum{} \\
18\,h average bolometric optical & $1.5 \times 10^{39}$\,erg\,s$^{-1}$ \\
        peak bolometric optical & \opticalpeaklum{} \\
\hline
\multicolumn{2}{l}{{\em NuSTAR} epoch at \tnustarep{}:} \\
                  0.1--300\,GeV & \latnustareplum{} \\
 \nustarenergylow{}--\nustarenergyhigh{}\,keV  & \nustarlum{} \\
                  0.3--78\,keV  & \nustarlumextrapolated{} \\
             bolometric optical & \opticalnustareplum{} \\
\hline
% I would love to include radio flux in this table, but over what frequency range should I integrate?
\hline
\end{tabular}
\begin{flushleft}
\end{flushleft}
\end{center}
\end{table}

% micronova
Overall, the eruption of \nova{} was well within the normal diversity of classical nova eruptions. 
This is at odds with the `micronova' scenario where the nuclear burning is confined to a small region of the white dwarf surface
near the magnetic poles (\citealt{2022Natur.604..447S,2022MNRAS.514L..11S}, see also \citealt{1988ApJ...330..264L} and \citealt{1989PASP..101....5S}). 
The magnetically confined nuclear burning region would produce the soft X-ray spin modulation observed 
in \nova{} by \cite{2021ApJ...922L..42D,2022MNRAS.517L..97L}.
A `micronova' has a peak optical luminosity of $10^{34}$\,erg\,s$^{-1}$ and eruptions lasting $\sim 10$\,h \citep{2022Natur.604..447S}.
\nova{} was among the fastest known novae and may well be at the extreme of the distribution in other ways, too, 
but it definitely was not a micronova. 
Had the magnetic field of an IP been capable of confining accreted matter, 
the eruption of \nova{} would have been much less energetic. 
Also, 3 out of the 6 brightest novae of the 20th century are firmly established to be magnetic 
- DQ\,Her \citep{1956ApJ...123...68W}, GK\,Per \citep{1988MNRAS.231..783N}, and V1500\,Cyg
\citep{1988ApJ...332..282S} --- and there are quite a few others that were proposed to be 
magnetic, at various levels of trustworthiness. Normal nova eruptions seem to routinely occur on magnetic white dwarfs.

\subsection{The origin of radio emission}
\label{sec:radiodiscussion}

\nova{} remains unresolved in all our VLA observations listed in Table~\ref{tab:vlaobslog}. 
However, by knowing the time of the eruption ($t_0$), distance, and the nova shell expansion velocity 
(from optical spectroscopy), we can estimate the shell's angular size. Then, together 
with the observed total flux density, we can constrain the surface brightness of the 
% note - no coma after i.e. and e.g. as this is MNRAS (its style is opposite to AAS journals in this respect)
radio emission --- i.e. the brightness temperature, $T_b$. 
The brightness temperature of thermal emission cannot exceed the physical temperature of
the emitting body (equation~(\ref{eq:TbT})) 
while the synchrotron $T_b$ can reach $10^{11}$\,K \citep{1969ApJ...155L..71K,1994ApJ...426...51R}. 
By comparing the estimated $T_b$ with a knowledge of the emitter's physical temperature, 
one can determine if the observed emission is consistent with being thermal. 
Following e.g.~\cite{2021MNRAS.501.1394N}
we calculate 
\begin{equation}
 T_b = 1222 \times \bigg(\frac{\nu}{\rm 1\,GHz}\bigg)^{-2} \times \bigg(\frac{F_\nu}{\rm 1\,mJy}\bigg) \times \bigg(\frac{\theta}{\rm 1\,arcsec}\bigg)^{-2}\,{\rm K}
 \label{eq:brightnessTemperature}
\end{equation}
where $\nu$ is the observing frequency, $F_\nu$ is the spectral flux density
and $\theta$ is the FWHM of a source that is assumed to have a Gaussian shape. 
The difference between the coefficients in eqn~(\ref{eq:brightnessTemperature}) and equation~(\ref{eq:radioflux})
arises from the different meaning of $\theta$. 
While in eqn~(\ref{eq:brightnessTemperature}) 
we use the $\theta$ definition common in radio astronomy observations,
in equation~(\ref{eq:radioflux}) $\theta$ is the angular diameter of a circle
having the surface area $\ln 2 \times$ that of a Gaussian source with ${\rm FWHM} = \theta$.
%
% ln 2 = 0.6931
This $\ln 2$ factor is unimportant for the following order-of-magnitude discussion of $T_b$.

Adopting an expansion velocity of 3500\,km\,s$^{-1}$ from \cite{2021ATel14710....1A}
and the distance from \S~\ref{sec:distlum}, 
we use eqn~(\ref{eq:brightnessTemperature}) to calculate the lower limit on $T_b$ at each VLA epoch. The lowest $T_b$ values are achieved when the observed radio emission covers the whole expanding ejecta, rather than a few compact knots.

% Early radio detections might be thermal
The first few detections of \nova{} at 13.7 to 34.9\,GHz have the estimated lower limit
$T_b > {\rm a few} \times 10^4$\,K (Fig.~\ref{fig:vlalc}),
consistent with the expected effective temperature of nova ejecta photoionized by
the central nuclear-burning white dwarf \citep{2015ApJ...803...76C}.
While the white dwarf atmosphere is very hot 
($10^5$--$10^6$\,K; \citealt{2013ApJ...777..136W,2015ApJ...803...76C}) producing 
super-soft X-ray emission, the photoionized ejecta are cooled by forbidden line emission 
down to an equilibrium temperature of $\sim 10^4$\,K \citep{1997pism.book.....D,2014A&A...561A..10P}.

The estimated $T_b$ lower limit values rise steeply around $t_0+10.1$\,d and reach $\gtrsim 10^5$\,K 
by $t_0+13.1$\,d at and below 16.5\,GHz (Fig.~\ref{fig:vlalc}). 
The high peak $T_b$ values suggest a non-thermal origin of the radio emission near its peak.
In \S~\ref{sec:thermalradiofromxrayplasma} we discuss why the fraction of
the nova shell shock-heated to $\sim 10^7$\,K (that we observe with {\em NuSTAR})
cannot be responsible for a significant fraction of the radio emission 
(and absorption) observed with the VLA.

The final argument supporting a non-thermal origin of the radio emission peak is
the fast rise of the radio flux density. The dashed line in the middle panel of Fig.~\ref{fig:vlalc}
indicates $F_\nu \propto t^2$, the rate at which the flux density of a uniformly
expanding, constant temperature, optically thick cloud should increase \citep[e.g.][]{Seaquist_Bode_2008}. 

The conclusion about the likely synchrotron origin of most of \nova{}'s
radio emission does not contradict the e-EVN 1.6\,GHz upper limit of 
36\,$\mu$Jy/beam on 2021-06-22 ($t_0 + 10$\,d) reported by \cite{2021ATel14758....1P}.
The total flux density observed by the VLA on that day (Fig.~\ref{fig:vlaspec})
extrapolated to 1.6\,GHz is well below the reported limit. One could
speculate that had the e-EVN observation been conducted at least a few days
later, it could have resulted in a positive detection.

The radio spectrum of \nova{} evolves from steeply inverted ($\alpha = 1.4-1.7$) to flat with a sign of curvature. 
The optically thin synchrotron spectrum cannot have a slope greater than $\alpha = +1/3$
for any energy distribution of the emitting electrons as the synchrotron spectrum of 
a single electron has no region rising faster than $F_\nu \propto \nu^{+1/3}$ 
\citep{1965ARA&A...3..297G,1966AuJPh..19..195K,1979tpa..book.....G}.
The optically thin thermal free-free (bremsstrahlung) emission slope is $F_\nu \propto \nu^{-0.1}$ \citep{1970ranp.book.....P}.
Therefore, at early times, the shape of the spectrum is
determined by optical depth effects.
The spectral slope, however, never reaches the canonical $\alpha = +2.5$ of 
a source experiencing synchrotron self-absorption \citep{1970ranp.book.....P}
or $\alpha = +2$ of an optically-thick thermal source \citep{2013LNP...873.....G}. 
The optically thick spectral slope that is more shallow than the canonical
values may arise from the source being inhomogeneous. For a thermal source,
the slope of $\alpha = +2$ is expected in the idealized case of an infinitely steep drop-off 
in density at the outermost edge of the source. A power-law drop-off in
density with distance from the source centre, $n \propto r^{-8}$, would
account for $\alpha = +1.7$ for a thermal bremsstrahlung source.

According to \cite{1966AuJPh..19..195K}, free-free absorption of optically thin synchrotron emission
with $F_\nu = A \nu^\alpha$ will result in an exponential drop at low
frequencies, $F_\nu = A \nu^\alpha e^{-(\nu_{\tau=1}/\nu)^2}$ 
in the case of an external screen 
(an absorber located between the emitter and observer), 
and a modified power-law spectrum
$F_\nu = \frac{A}{\nu_{\tau=1}^2} \nu^{\alpha+2}$ if the absorber is mixed
with the synchrotron-emitting plasma. Here $\nu_{\tau=1}$ is the frequency
at which optical depth is equal to unity, $A$ is a constant, and it is
assumed that we can neglect the thermal emission of the absorbing plasma.
As with the `cosmic conspiracy' producing flat
spectra of cores in extragalactic radio sources \citep{1979ApJ...232...34B,1980Natur.288...12M,1981ApJ...243..700K}, 
the likely explanation of the \nova{} radio spectrum slope is that there is a gradient of physical 
properties across the radio emitting region. 
The change in slope reflects the changes of these properties in time or over 
the source,  as the outer regions of 
the source become transparent and the inner source regions become visible.

% Late epochs are synchrotron
At late epochs, the $T_b$ estimates are consistent with 
thermal emission (Fig.~\ref{fig:vlalc}). One may wonder if the synchrotron
emission is dominating the peak of the radio lightcurve but is overtaken by
thermal emission later.
Two considerations disfavor this possibility.
First, the decline from the maximum $T_b$ appears smooth and gradual, with no abrupt
change in the decline rate that could suggest change in the dominating
emission mechanism (Fig.~\ref{fig:vlalc}).
Second, the latest VLA epoch with a positive detections of \nova{} 
(2021-11-01, $t_0+142.8$; Fig.~\ref{fig:vlaspec}) reveals a curved spectrum with a
high-frequency spectral index of $\alpha_{\rm sync} = -0.5 \pm 0.1$, which is $<-0.1$ --- indicative of synchrotron emission.

If we take the radio spectrum peak parameters from this last-detection epoch 
(Fig.~\ref{fig:vlaspec}), the angular diameter at this epoch (90\,mas, 
estimated earlier for the $T_b$ calculation), and naively apply equation~(2) of 
\cite{1983ApJ...264..296M} describing a uniform synchrotron cloud, 
we end up with an unrealistically high magnetic field
needed to produce the observed spectral turnover via synchrotron 
self-absorption. This supports the conclusion of \cite{2016MNRAS.463..394V} that 
it is the free-free opacity that is dominating the radio spectral evolution in novae,
irrespective of the emission mechanism (thermal or synchrotron; see
also \citealt{2016MNRAS.456L..49K} for a discussion of embedded novae).
The synchrotron self-absorption 
spectrum fits to the late VLA epochs in Fig.~\ref{fig:vlaspec}
appear to be just a convenient peaked function that happens to fit the data
well with no physical meaning. 

There is another mechanism that may attenuate the synchrotron spectrum at low
frequencies in the presence of a thermal plasma in the emitting region: 
the Razin-Tsytovich effect \citep{1965ARA&A...3..297G,1966AuJPh..19..195K,1979rpa..book.....R}. 
It has been proposed as the reason for the inverted cm-band spectrum of the nova QU\,Vul by \cite{1987A&A...183...38T}. 
The condition $\nu_{\rm RT peak} = 20 n_e/B$ (where the peak
frequency $\nu_{\rm RT peak}$ is in Hz, the electron density $n_e$ is in cm$^{-3}$ 
and the magnetic field strength $B$ is in Gauss) for 
$\nu_{\rm RT peak} = 7$\,GHz (the lowest radio spectrum peak frequency
observed in \nova{} on $t_0+142.8$\,d; Fig.~\ref{fig:vlaspec}) implies a density $n_e = 3.5\times10^8 B$.
This may not be unreasonable for a shocked nova shell \citep[][c.f.~\S~\ref{sec:thermalradiofromxrayplasma}]{1987A&A...183...38T,2016MNRAS.463..394V} 
depending on the assumed magnetic field strength. 
The spectral index resulting from the Razin-Tsytovich effect is 
frequency-dependent and is ultimately steeper than that of a self-absorbed
source \citep[according to figure~1 of][]{1966AuJPh..19..195K}, 
in contrast to the relatively shallow spectral indices of \nova{}. 

Synchrotron emission in novae has been identified before on the basis of 
high estimated $T_b$ of  
V809\,Cep \citep{2022MNRAS.515.3028B}, 
V392\,Per, V357\,Mus, V5855\,Sgr, V5668\,Sgr, V2672\,Oph, V2491\,Cyg,
V838\,Her, V1500\,Cyg \citep{2021ApJS..257...49C},
helium nova V445\,Pup \citep{2021MNRAS.501.1394N}, 
V906\,Car \citep{2020NatAs...4..776A}, V1324\,Sco \citep{2018ApJ...852..108F},
V1723\,Aql \citep{2016MNRAS.457..887W}, 5589\,Sgr \citep{2016MNRAS.460.2687W},
V959\,Mon \citep{2014Natur.514..339C} and QU\,Vul \citep{1987A&A...183...38T}, 
and 
via direct VLBI imaging of V959\,Mon \citep{2014evn..confE..53Y}, \nova{} 
itself (J.~Linford and M.~Williams, private communication) 
and the embedded (red giant donor) novae 
RS\,Oph \citep{1989MNRAS.237...81T,2006Natur.442..279O,2008ApJ...688..559R,2008ApJ...685L.137S,2009MNRAS.395.1533E,2022A&A...666L...6M,2023arXiv230110552D},
V407\,Cyg \citep{2020A&A...638A.130G}, and V1535\,Sco \citep{2017ApJ...842...73L}.
Synchrotron emission was also reported in embedded novae
V745\,Sco \citep{2016MNRAS.456L..49K}
and 
V3890\,Sgr \citep{2023arXiv230109116N}.
A special case is the long-lived synchrotron-emitting remnant of 
the 1901 nova GK\,Per that apparently erupted within a planetary nebula
\citep{1989ApJ...344..805S,2016A&A...595A..64H}.

% Sync spec never rises faster than nu^1/3 for any distribution of electron energies 1966AuJPh..19..195K

\subsection{Thermal radio emission from the X-ray emitting plasma}
\label{sec:thermalradiofromxrayplasma}

The observed spectrum of a free-free emitting cloud is determined by its
temperature, distance and a combination of the electron density and volume
of the cloud. There is a degeneracy between the electron density and volume,
so a combination of the two values known as the emission measure (EM) is
often used in calculations. There are two definitions of EM \citep[e.g.][]{2016MNRAS.457..887W}, 
the one used in X-ray astronomy:
\begin{equation}
{\rm (EM)}_{\rm X-ray} = \int n_e n_i dV,
\label{eq:emxray}
\end{equation}
where $n_e$ is the electron number density, $n_i$ is the number density of
ions and $V$ is the volume occupied by the emitting particles. The
\textsc{apec} plasma emission model normalization, $\mathcal{E}$, provides the volume emission measure of the
plasma scaled by the distance \citep{2011hxra.book.....A}:
\begin{equation}
\mathcal{E} = \frac{10^{-14}}{4 \pi D_{\rm cm}^2} {\rm (EM)}_{\rm X-ray},
\label{eq:emnorm}
\end{equation}
where $D_{\rm cm}$ is distance to the emitting cloud in cm.

\begin{figure}
\begin{center} 
        \includegraphics[width=0.75\linewidth,clip=true,trim=0cm 0cm 0cm 0cm,angle=0]{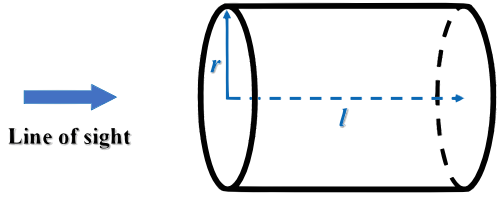}
\end{center}
\caption{The assumed cylindrical slab geometry of the cloud.}
    \label{fig:slab}
\end{figure}

Another definition of the emission measure is commonly used in radio
astronomy \citep[e.g.][]{1970ranp.book.....P}:
\begin{equation}
{\rm (EM)}_{\rm radio} = \int n_e n_i dl,
\label{eq:emxray}
\end{equation}
where the integration is done along the line of sight crossing the cloud, $l$,
not the cloud volume as in the X-ray definition. To relate ${\rm (EM)}_{\rm X-ray}$
to ${\rm (EM)}_{\rm radio}$, we have to assume some specific geometry of the
cloud. For simplicity, we assume the cloud to be a uniform cylindrical slab
(the shape of a ice hockey puck; Fig.~\ref{fig:slab}) of radius $r$ and depth $l$ along the line of sight:
\begin{equation}
{\rm (EM)}_{\rm X-ray} = \pi r^2 {\rm (EM)}_{\rm radio}.
\label{eq:emxrayemradio}
\end{equation}
Note that ${\rm (EM)}_{\rm X-ray}$ is often expressed in units of cm$^{-3}$, 
while ${\rm (EM)}_{\rm radio}$ is in pc\,cm$^{-6}$ -- the value of ${\rm (EM)}_{\rm radio}$
obtained from equation~(\ref{eq:emxrayemradio}) needs to be divided by the number 
of centimetres in a parsec to be expressed in the commonly used units.
In reality, the emitting region geometry may resemble a sector of a spherical shell with density varying with
radius, resulting in an additional factor of a few in the equation relating the
two EM values.

% https://www.cv.nrao.edu/~sransom/web/A5.html

From the ${\rm (EM)}_{\rm radio}$ and temperature of the emitting cloud,
we can calculate the free-free optical depth in the radio band using the approximate relation of
\cite{1967ApJ...147..471M}:
\begin{equation}
\tau \approx 3.28 \times 10^{-7} \left( \frac{T}{10^4\,{\rm K}} \right)^{-1.35}  \left( \frac{\nu}{1\,{\rm GHz}} \right)^{-2.1}  \left( \frac{ {\rm (EM)}_{\rm radio} }{1\,{\rm pc}\,{\rm cm}^{-6}} \right)
\label{eq:radioopticaldepth}
\end{equation}
For the ${\rm (EM)}_{\rm X-ray}$ and temperature (both derived from the {\em NuSTAR}
spectrum, \S~\ref{sec:nustarspec}) 
and assuming linear expansion of $r$ starting at $t_0$ with the optical
spectroscopy-derived velocity (\S~\ref{sec:radiodiscussion}), the resulting ${\rm (EM)}_{\rm radio} = 3.8 \times 10^8$\,pc\,cm$^{-6}$ for the assumed cylindrical geometry (Fig.~\ref{fig:slab}).
The implied optical depth is in the range $\tau = 10^{-3}$ -- $10^{-6}$ for $\nu = 1$ to 30\,GHz.
This suggests that the X-ray emitting plasma 
cannot be responsible for the {\it absorption} of cm-band radio waves. 
The absorption may be provided by another cooler component of the plasma located in the foreground of the radio emitting region.

To find out if the X-ray emitting plasma is responsible for the observed
radio {\it emission} we use the convenient equation~(1.7) of \cite{2016PhDT........30W} to calculate the radio flux density
\begin{equation}
% Sokoloski's version:
%F_\nu = \frac{T_b}{1200\,{\rm K}} \bigg(\frac{\nu}{1\,{\rm GHz}}\bigg)^2 \bigg(\frac{\theta}{1\,{\rm arcsec}}\bigg)^2\,{\rm mJy},
% Weston's version double-checked by Sokolovsky:
% k_cgs=1.380649e-16;c_cgs=2.99792458e10; 1/( 2*pi*k_cgs/c_cgs^2/(2*206264.8)^2*(10^9)^2*1e23*1000 )
% here 2 is so we are not confusing radius with diameter
F_\nu = \frac{T_b}{1763\,{\rm K}} \bigg(\frac{\nu}{1\,{\rm GHz}}\bigg)^2 \bigg(\frac{\theta}{1\,{\rm arcsec}}\bigg)^2\,{\rm mJy},
\label{eq:radioflux}
\end{equation}
where 
\begin{equation}
T_b = (1 - e^{-\tau}) T
\label{eq:TbT}
\end{equation}
is the brightness temperature related to the physical (electron) temperature, $T$, 
through the optical depth, $\tau$, that we derive from
equation~(\ref{eq:radioopticaldepth}), see e.g.~\cite{1979rpa..book.....R} or
\cite{SnellKurtzMarr}. 
The predicted $F_\nu \approx 2\,\mu$Jy at cm wavelengths -- well below the sensitivity limit of our snapshot VLA observations.

\subsection{Ejecta mass constraints}
\label{sec:ejmassconstr}

Fast novae are associated with massive white dwarfs and low ejecta masses. 
At a given accretion rate, the higher surface gravity and smaller surface area
of a massive white dwarf facilitate earlier ignition of the thermonuclear
runaway, resulting in a smaller amount of ejected material. 
If the fractional change of $10^{-4}$ between \nova{}'s pre- and post-eruption white
dwarf spin period is attributed to the mass loss via a magnetized wind, this
would require mass loss of $10^{-5}$ to $10^{-4}\, {\rm M}_\odot$ according
to \cite{2021ApJ...922L..42D}. The authors also point out that a large
ejecta mass of $10^{-4}\, {\rm M}_\odot$ was reported for another fast nova,
V838\,Her, by \cite{1996MNRAS.282..563V}. 

In the following, we constrain the ejecta mass of \nova{} using the information 
about the X-ray emission, X-ray absorption and thermal radio emission.
The amount of X-ray emission probes the shock-heated plasma while 
the X-ray absorption and radio emission independently probe 
the colder photoionized fraction of the ejecta. All these estimates point to 
a low ejecta mass (and hence alternative explanations for the spin
period change; \S~\ref{sec:thisnova}); however the estimates are highly model-dependent.

\subsubsection{Minimum ejecta mass from thermal radio emission}
\label{sec:radiomass}

The majority of novae produce thermal radio emission \citep{2021ApJS..257...49C}.
Emission from the hot white dwarf photo-ionizes nova ejecta causing it to produce free-free radio emission 
as it expands. How and when the free-free emission in the radio band changes from optically thick to optically thin 
is determined by the ejecta mass \citep{2015ApJ...803...76C}.
We can put constraints on the mass of the ejecta in \nova{} if a fraction of 
its radio emission is thermal in origin \citep[e.g.][]{1996ASPC...93..174H,2016MNRAS.457..887W}.

In the radio lightcurves of novae, where the thermal and synchrotron peaks can be separated, 
the synchrotron peak typically precedes the thermal peak \citep{2016MNRAS.457..887W,2018ApJ...852..108F,2021ApJS..257...49C}.
In contrast to this tendency, the $T_b$ history (Fig.~\ref{fig:vlalc})
suggests that
thermal radio emission may have dominated the radio lightcurve
of \nova{} at {\em early} times, before $t_0+10$\,d.

By $t_0+10$\,d the radio spectrum appears slightly curved
(Fig.~\ref{fig:vlaspec}), suggesting the optical depth might be starting to drop.
From equation~(\ref{eq:radioopticaldepth}) and the condition $\tau > 1$ at 13\,GHz, we derive 
${\rm (EM)}_{\rm radio} > 2.1 \times 10^{27}$\,cm$^{-5}$ assuming $T=10^4$\,K
(unshocked photoionized ejecta).

The thermal radio lightcurves of novae are often described with the `Hubble flow' model
\citep{1979AJ.....84.1619H,1996ASPC...93..174H,2005MNRAS.362..469H}; see  Appendix~A
of \cite{2018ApJ...852..108F}. The ejecta are assumed to be in a spherical shell with 
an inner radius expanding with velocity $v_{\rm min}$ and an outer radius expanding with $v_{\rm max}$
resulting in $n \propto r^{-2}$ density profile. 
Assuming the shell ejected at $t_0$ with $v_{\rm max} = 5000$\,km\,s$^{-1}$ 
\citep[the maximum velocity reported by][]{2021ATel14710....1A} and 
$v_{\rm min} = 0.1 v_{\rm max}$, we find that the ${\rm (EM)}_{\rm radio}$
constraint corresponds to an ejecta mass of $>10^{-7}\, {\rm M}_\odot$
(if the 13\,GHz emission at $t_0+10$ is mostly thermal). 

\subsubsection{Upper limit on the ejecta mass from the absence of intrinsic X-ray absorption}
\label{sec:noxrayabsmass}

Following \cite{2020MNRAS.497.2569S} and \cite{2021MNRAS.500.2798N} we
estimate the mass of unshocked ejecta by assuming the X-rays originating deep 
within the ejecta are absorbed by the spherical `Hubble flow' shell. 
Taking the same model parameters as for the radio-emission-based mass
estimation in \S~\ref{sec:radiomass} and assuming that an intrinsic
absorption of $N_\mathrm{H} = 10^{21}\,{\rm cm}^{-2}$ would have been
detectable in our {\em NuSTAR} plus {\em Swift}/XRT observations 
(\S~\ref{sec:nustarspec} and \ref{sec:swiftobs}) we obtain an upper limit on
the ejecta mass of a~few~$\times 10^{-7} {\rm M}_\odot$, close to 
the thermal radio lower limit. 

The absence of detectable intrinsic absorption is inconsistent with
a larger ejecta mass of $\times 10^{-4} {\rm M}_\odot$ that would
produce an absorbing column of $N_\mathrm{H} = 5 \times 10^{23}\,{\rm cm}^{-2}$
at day 12 with the above model parameters. 
The caveat here is that the X-ray upper limit on the unshocked ejecta mass 
strongly depends on the assumptions of spherical symmetry of the absorber and 
the minimum expansion velocity of the `Hubble flow' model (the slower moving
material provides most absorption as it remains dense for a longer time).

\subsubsection{The mass of X-ray emitting plasma}
\label{sec:xrayemittermass}

In order to estimate the density and total mass of the X-ray emitting material we 
assume the simple cylindrical shape of the emitting region (as in \S~\ref{sec:thermalradiofromxrayplasma})
with the height equal to its radius 
\citep[geometry used by][]{1989MNRAS.237...81T}. % yes, they use this assumption 
Under this assumption, the observed ${\rm (EM)}_{\rm X-ray}$ value 
corresponds to the density of $n \sim 10^6$\,cm$^{-3}$. Multiplying this by
the mean molecular weight appropriate for the nova ejecta
(\S~\ref{sec:shocklocation}), the proton mass 
and the cylinder volume we 
get the total mass of the X-ray emitting plasma to be $10^{-7} {\rm M}_\odot$. 
We stress that the above values of mass and density are nothing more than 
an indication of where in the parameter space the true values might lay.
The actual values depend critically on the geometry and density distribution
of the shocked ejecta. Specifically, we assumed that the plasma is
distributed uniformly across the cylindrical volume (has a filling factor of
unity). This mass estimate refers to the shock-heated material
which is, presumably, a small part of the ejecta.

\subsection{\nova{} and our understanding of novae}
\label{sec:comfastslow}

It has long been argued on the basis of optical lightcurves and spectral
evolution that there is no fundamental difference between fast and slow
novae \citep{1939PA.....47..410M,1939PA.....47..481M,1939PA.....47..538M}.
The observation that \nova{}, being one of the fastest novae ever seen in 
the Galaxy (\S~\ref{sec:thisnova}), produces $\gamma$-ray and X-ray emitting shocks 
similar to the ones found in slower novae suggests this is also true for the
structure of nova ejecta. The structures within the ejecta responsible for 
the shock formation, likely the slow equatorial outflow and fast omnidirectional
wind \citep[][\S~\ref{sec:shocklocation}]{2014Natur.514..339C,2022ApJ...938...31S}, must be present in \nova{}.
It does not appear to be the case that a fast nova is dominated by the fast
wind while the ejecta of a slow nova is mostly the slow equatorial outflow.
Instead, both structures must be present in the fast-evolving \nova{}, 
while the time-scale of their interaction is compressed compared to most
other novae.

The extreme properties of \nova{} challenge our understanding 
of nova outflows and specifically the `slow torus -- fast bipolar wind'
scenario (\S~\ref{sec:intro}, \ref{sec:shocklocation}).
%(\S~\ref{sec:innovashock}, \ref{sec:shocklocation}). 
%
It is tempting to attribute the shock velocity 
$v_{\rm shock} \simeq 1700$\,km\,s$^{-1}$ (corresponding to ${\rm k} T_{\rm shock} = 4$\,keV)
to the shock between the early 3500\,km\,s$^{-1}$ and late 5000\,km\,s$^{-1}$ outflows seen 
in optical spectra by \cite{2021ATel14710....1A}. Even if the shock is
between the 3500\,km\,s$^{-1}$ outflow and a slower ejecta not seen in spectroscopy, 
that would still imply the slow torus velocity of $>1000$\,km\,s$^{-1}$. 
This is considerably faster than the expected orbital velocity of the binary companion, 
suggesting that it is unlikely that the common envelope interaction was the mechanism
responsible for the slow torus ejection in \nova{} \citep{2021ARA&A..59..391C}.
Another challenge for the common envelope ejection origin of the shocked material is
the six-hour delay between the start of the eruption ($t_0$) and the onset 
of $\gamma$-ray emission (\S~\ref{sec:latobs}, Fig.~\ref{fig:latlc}).
It would be surprising if less than two orbital revolutions (3.67\,h period) are sufficient to eject the common envelope. 

An onset of $\gamma$-rays delayed by a few days from the optical rise is commonly observed 
in slower novae \citep[][]{2014Sci...345..554A,2016ApJ...826..142C,2017MNRAS.469.4341M}. 
While such a delay is naturally expected in the two-flow scenario, the
alternative possibility is that if $\gamma$-rays are produced simultaneously
with the optical rise, they may have a hard time escaping the system due to 
photo-nuclear or $\gamma\gamma$ pair production opacity
\citep[][]{2016MNRAS.457.1786M,2018MNRAS.475.1521M,2020ApJ...904....4F,2022Sci...376...77A}.

The correlated variations in optical and $\gamma$-ray flux observed in
V5856\,Sgr by \cite{2017NatAs...1..697L} and V906\,Car by \cite{2020NatAs...4..776A}
revealed that a significant fraction of nova optical light might be
shock-powered. \cite{2017MNRAS.469.4341M} suggested that in addition to the
main optical lightcurve peak associated with the greatest expansion of the photosphere 
(common to all novae) there might be a separate peak in the lightcurves of GeV-bright novae
associated with optical emission of the $\gamma$-ray-producing shocks.
The lightcurve of \nova{} has a single peak. However a kink in the optical 
lightcurve is hinted around the time the $\gamma$-ray emission ended 
($t_0 + 1$\,d; the rate of decline has decreased). One may speculate 
that shocks could have contributed to the optical light before $t_0 + 1$\,d
producing an additional bump right on top of the common `fireball' lightcurve peak. 

Detection of the GeV $\gamma$-rays before the optical peak (Fig.~\ref{fig:latlc})
is at odds with the prediction of the `continuously changing velocity wind'
model of \cite{2022ApJ...939....1H}. However, one may overcome this
contradiction if the optical peak is prolonged by the contribution of
shock-powered optical light.

\section{Conclusions}
\label{sec:conclusions}

We conducted a joint analysis of $\gamma$-ray (\fermilat{}), X-ray ({\em NuSTAR}, {\em Swift}/XRT), optical (AAVSO, Evryscope,
ASAS-SN), and radio (VLA) observations of an exceptionally fast Galactic nova \nova{}.

\begin{enumerate}
\item \nova{} was clearly detected by \fermilat{}, but only for the
duration of 18\,h near the optical peak. There is a delay of about 6\,h 
between the onset of optical and detectable $\gamma$-ray emission. 
The shape and the cut-off energy of the $\gamma$-ray spectrum are poorly constrained taking into account the limited statistics.
\item The {\em NuSTAR} spectrum of \nova{} is consistent with 
having been produced 
by shock-heated plasma with non-solar elemental abundances. It is remarkably
similar to the spectra of three classical novae previously detected by {\em NuSTAR}.
The lack of periodic variability in the hard X-ray flux at the spin period of the white dwarf 
suggests that the {\em NuSTAR}-detected X-rays from
\nova{} are associated with a shock within the
nova ejecta, not accretion on the magnetized white dwarf.
\item Given the strong similarity between the high-energy properties of \nova{} and those of other classical novae in the days to weeks after eruption, it appears that neither the exceptionally high speed
of this nova, nor the intermediate polar nature of the host system affect the shock development within the ejecta.
\item We interpret the radio emission of \nova{} as being shock-powered synchrotron emission attenuated by free-free absorption. 
Unlike many other novae, \nova{} displayed weak thermal radio emission that
contributed before the synchrotron emission reached its peak.
\item The radio emission (\S~\ref{sec:radiomass}) and X-ray emission and absorption
(\S~\ref{sec:noxrayabsmass}) point to a low ejecta mass of $\sim 10^{-7} {\rm M}_\odot$, 
however the different ejecta mass estimation techniques do not necessarily probe the same parts of the ejecta.
\item Being an exceptionally fast nova, \nova{} might serve as a stress test for 
the `slow torus -- fast bipolar wind' scenario (outlined in \S~\ref{sec:shocklocation} and \ref{sec:comfastslow}) 
of shock formation in novae. For this scenario to hold in \nova{}, common-envelope action must have been able to eject 
the envelope very quickly and the fast flow must have begun before the detection of $\gamma$-rays within 6 hours of $t_0$.
\end{enumerate}

\section*{Data availability}

The raw \fermilat{}, {\em NuSTAR}, {\em Swift}, AAVSO, ASAS-SN and VLA data are
available at the respective archives. The readers are encouraged to contact the authors for processed data and analysis details. 
The data files and plotting scripts corresponding 
to Fig.~\ref{fig:latlc}, \ref{fig:nustarlc}, \ref{fig:nuspec},
\ref{fig:nuswspec}, \ref{fig:vlalc}, \ref{fig:vlaspec} and Table~\ref{tab:nustarspecmodels}
may be found at 
\url{https://kirx.net/~kirx/V1674_Her__public_data/}
and 
\url{http://scan.sai.msu.ru/~kirx/data/V1674_Her__public_data/}

\section*{Acknowledgements}
%%The Acknowledgements section is not numbered. Here you can thank helpful
%%colleagues, acknowledge funding agencies, telescopes and facilities used etc.
%%Try to keep it short.
%%
%% The standard AAVSO ack for MW papers https://www.aavso.org/data-usage-guidelines
We acknowledge with thanks the variable star observations from the AAVSO International Database contributed by observers worldwide and used in this research. 
We thank  
Elizabeth O. Waagen and Dr. Brian K. Kloppenborg for their assistance in 
communicating with the AAVSO observers and 
Dr.~Nikolai~N.~Samus for the help in gathering information about the
circumstances surrounding the discovery of \nova{}. 
%%

% NuSTAR ack
This paper made use of data from the {\em NuSTAR} mission, a project led by the California Institute of Technology, 
managed by the Jet Propulsion Laboratory, funded by the National Aeronautics and Space Administration.
This research has made use of the {\em NuSTAR} Data Analysis Software (NuSTARDAS) jointly developed by 
the ASI Science Data Center (ASDC, Italy) and the California Institute of Technology (USA).

% Swift ack
We acknowledge the use of public data from the {\em Swift} data archive.

% Standard LAT ack
The \textit{Fermi} LAT Collaboration acknowledges generous ongoing support
from a number of agencies and institutes that have supported both the
development and the operation of the LAT as well as scientific data analysis.
These include the National Aeronautics and Space Administration and the
Department of Energy in the United States, the Commissariat \`a l'Energie Atomique
and the Centre National de la Recherche Scientifique / Institut National de Physique
Nucl\'eaire et de Physique des Particules in France, the Agenzia Spaziale Italiana
and the Istituto Nazionale di Fisica Nucleare in Italy, the Ministry of Education,
Culture, Sports, Science and Technology (MEXT), High Energy Accelerator Research
Organization (KEK) and Japan Aerospace Exploration Agency (JAXA) in Japan, and
the K.~A.~Wallenberg Foundation, the Swedish Research Council and the
Swedish National Space Board in Sweden.
 
Additional support for science analysis during the operations phase is gratefully
acknowledged from the Istituto Nazionale di Astrofisica in Italy and the Centre
National d'\'Etudes Spatiales in France. This work performed in part under DOE
Contract DE-AC02-76SF00515.

% VLA ack
The National Radio Astronomy Observatory is a facility of the National Science Foundation operated under cooperative agreement by Associated Universities, Inc.

%%
% MSU team funding
This material is based upon work supported by the National Science Foundation under Grants~No.~AST-1751874 and AST-1816100. 
We acknowledge support for this work from NASA projects 
{\em Fermi}~80NSSC20K1535,
%131113, 
{\em NuSTAR}~80NSSC21K0277, %8136, 
{\em XMM-Newton}~80NSSC21K0277, %90327 and 80NSSC21K0715, 
and {\em Swift}~80NSSC21K0173. %1821098.
BDM acknowledges support from NASA under Grant No.~80NSSC22K0807.
KLP acknowledges support from the UK Space Agency.
IV acknowledges support by the ETAg grant PRG1006 and by EU through the ERDF CoE grant TK133.
% Sara Buson
This work was supported by the European Research Council, ERC Starting grant \emph{MessMapp}, S.B. Principal Investigator, under contract no.~949555.
% Teddy & Tyrel
Research at the Naval Research Laboratory is supported by NASA DPR S-15633-Y.
%% Software
This work made use of \textsc{Astropy}: %\footnote{\url{http://www.astropy.org}} 
a community-developed core Python package 
and an ecosystem of tools and resources for astronomy \citep{2013A&A...558A..33A,2018AJ....156..123A,2022ApJ...935..167A}. 
This research made use of \textsc{Photutils}, an \textsc{Astropy} package for 
detection and photometry of astronomical sources \citep{larry_bradley_2022_6825092}. 
%%%%%%%%%%%%%%%%%%%%%%%%%%%%%%%%%%%%%%%%%%%%%%%%%%%
%
%
%
%
%%%%%%%%%%%%%%%%%%%%%%%%%%%%%%%%%%%%%%%%%%%%%%%%%%%
%
%%%%%%%%%%%%%%%%%%%%% REFERENCES %%%%%%%%%%%%%%%%%%


%\bibliographystyle{mnras}
\bibliographystyle{mnras_vanHack}
\bibliography{nher} % if your bibtex file is called example.bib

%
%% Alternatively you could enter them by hand, like this:
%% This method is tedious and prone to error if you have lots of references
%%\begin{thebibliography}{99}
%%\bibitem[\protect\citeauthoryear{Author}{2012}]{Author2012}
%
%%Author A.~N., 2013, Journal of Improbable Astronomy, 1, 1
%%\bibitem[\protect\citeauthoryear{Others}{2013}]{Others2013}
%%Others S., 2012, Journal of Interesting Stuff, 17, 198
%%\end{thebibliography}
%
%%%%%%%%%%%%%%%%%%%%%%%%%%%%%%%%%%%%%%%%%%%%%%%%%%%
%
%%%%%%%%%%%%%%%%%% APPENDICES %%%%%%%%%%%%%%%%%%%%%
%
%%If you want to present additional material which would interrupt the flow of the main paper,
%%it can be placed in an Appendix which appears after the list of references.
%
%\appendix

% Don't change these lines
\bsp % typesetting comment
\label{lastpage}
\end{document}